\documentclass[
  twocolumn,english,aps,pra,
  superscriptaddress,amsmath,amssymb,floatfix,nofootinbib,longbibliography
]{revtex4-2}

\usepackage{amsthm}
\usepackage{amsfonts}
\usepackage{siunitx}
\usepackage{amsmath}
\usepackage{amssymb}
\usepackage{float}
\makeatletter
\let\newfloat\newfloat@ltx
\makeatother
\usepackage{algpseudocode}
\usepackage{algorithm}
\usepackage{graphicx}
\usepackage{verbatim}
\usepackage[colorlinks]{hyperref}
\usepackage{tikz}
\usepackage{pgfplots}
\pgfplotsset{compat=1.18}
\usepackage{braket}
\usepackage{xcolor}
\usepackage{physics}
\usepackage{amssymb} 
\usepackage{graphicx}
\usepackage{dcolumn}
\usepackage{bm}
\usepackage{mathtools}
\usepackage{hyperref}
\usepackage{mathrsfs}
\usepackage{subcaption}
\usepackage{titlesec}
\definecolor{linkcolor}{RGB}{0,83,166}
\hypersetup{
  colorlinks = true,
  allcolors = {linkcolor}
}

\begin{document}
\title{Partition Function Estimation Using Analog Quantum Processors}

\author{Thinh Le}
\affiliation{Theoretical Division, Los Alamos National Laboratory}
\affiliation{Department of Computer Science, North Carolina State University, Raleigh, USA 27606.}

\author{Elijah Pelofske}
\email[]{epelofske@lanl.gov}
\affiliation{Information Systems \& Modeling, Los Alamos National Laboratory}

\begin{abstract}

We evaluate using programmable superconducting flux qubit D-Wave quantum annealers to approximate the partition function of Ising models. We propose the use of two distinct quantum annealer sampling methods: chains of Monte Carlo-like reverse quantum anneals, and standard linear-ramp quantum annealing. The control parameters used to attenuate the quality of the simulations are the effective analog energy scale of the J coupling, the total annealing time, and for the case of reverse annealing the anneal-pause. The core estimation technique is to sample across the energy spectrum of the classical Hamiltonian of interest, and therefore obtain a density of states estimate for each energy level, which in turn can be used to compute an estimate of the partition function with some sampling error. This estimation technique is powerful because once the distribution is sampled it allows thermodynamic quantity computation at arbitrary temperatures. On a $25$ spin $\pm J$ hardware graph native Ising model we find parameter regimes of the D-Wave processors that provide comparable result quality to two standard classical Monte Carlo methods, Multiple Histogram Reweighting and Wang-Landau. Remarkably, we find that fast quench-like anneals can quickly generate ensemble distributions that are very good estimates of the true partition function of the classical Ising model; on a Pegasus graph-structured QPU we report a logarithmic relative error of $7.6 \times 10^{-6}$, from $171,000$ samples generated using $0.2$ seconds of QPU time with an anneal time of $8$ nanoseconds per sample which is interestingly within the closed system dynamics timescale of the superconducting qubits.

\end{abstract}

\maketitle

%%%%%%%%%%%%%%%%%%%%%%%%%%%%%%%%%%%%%%%%%%%%%%%%%%%%%%%
%%%%%%%%%%%%%%%%%%%%%%%%%%%%%%%%%%%%%%%%%%%%%%%%%%%%%%%
\section{Introduction}\label{section:introduction}
%%%%%%%%%%%%%%%%%%%%%%%%%%%%%%%%%%%%%%%%%%%%%%%%%%%%%%%
%%%%%%%%%%%%%%%%%%%%%%%%%%%%%%%%%%%%%%%%%%%%%%%%%%%%%%%
\raggedbottom

The partition function, typically denoted as $Z$, is an incredibly important quantity in statistical mechanics. Its accurate computation is central to a large number of thermodynamic quantities, and therefore plays an important computational role in quantifying phenomena in many types of interacting spin systems. The partition function can be written as
\begin{equation}
    Z = \sum_{\{s_i\}} e^{-\beta H(s_i)},
    \label{equation:partition_function_canonical}
\end{equation}
where $\{s_i\}$ defines the set of all possible configurations of the system. For this study, we will be focusing on Ising models that have $N$ spins, which means the set $\{s_i\}$ is all $2^N$ spin configurations, and $s_i$ are spins which take values of $\pm 1$. $H(s_i)$ denotes the energy, or expectation value, of a classical Hamiltonian (e.g., Ising model) for a specific spin configuration $s_i$. In this way, the partition function is a Hamiltonian dependent property. $\beta$ is the inverse temperature $\frac{1}{k_B T}$ where $k_B$ is the Boltzmann constant and $T$ is the thermodynamic temperature in Kelvin. To make numerical computations, and data visualization, simpler we will use the natural units of the Boltzmann constant by setting $k_B=1$. 

Partition function estimation is an important computational capability because although for many applications having the exact partition function would be valuable, for many Hamiltonians of interest this is quite impractical because it requires enumerating over all spin configurations, making it intractable. There are a limited number of cases of analytic formulas that have been derived for the partition function of specific types of Hamiltonians, typically low-dimensional spin systems~\cite{baxter1985exactly, PhysRev.65.117}. But in general, and especially for very large Hamiltonians, we must instead use approximations of the partition function in order to make realistic computations with these models feasible -- namely, Markov chain Monte Carlo (MCMC)~\cite{metropolis1953equation}. And more generally, a considerable amount of algorithmic work has gone into removing the need to compute the partition function value directly, and instead to make use of ratios of partition functions, thus removing the need to explicitly compute the value of the partition function~\cite{edwards1975theory}, for example using free energy differences for molecular dynamics simulations~\cite{torrie1977nonphysical, zwanzig1954high, valleau1972monte, PhysRevLett.78.2690, bennett1976efficient}. 

\begin{figure*}[t]
    \centering
    \includegraphics[width=0.49\textwidth]{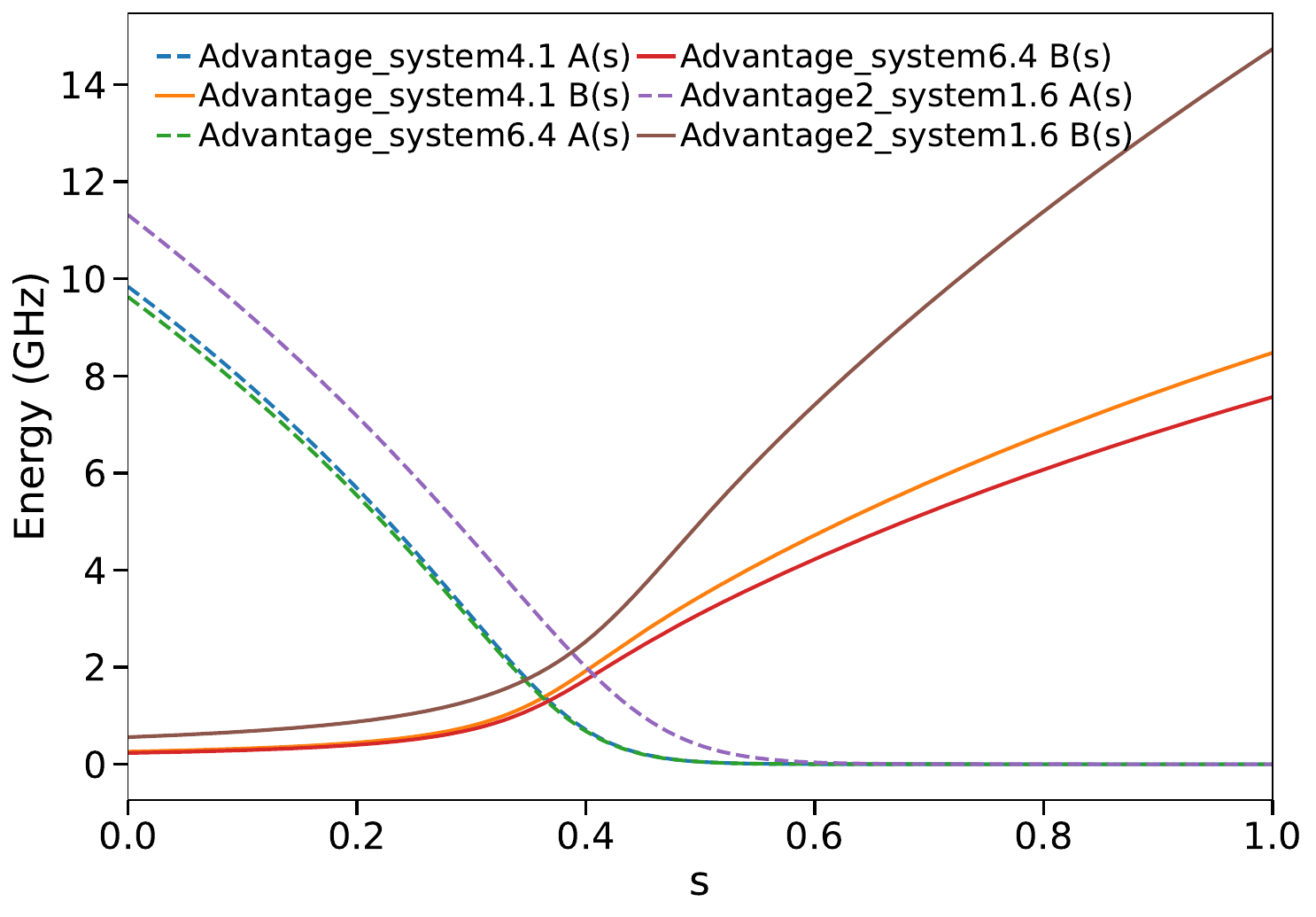}
    \includegraphics[width=0.49\textwidth]{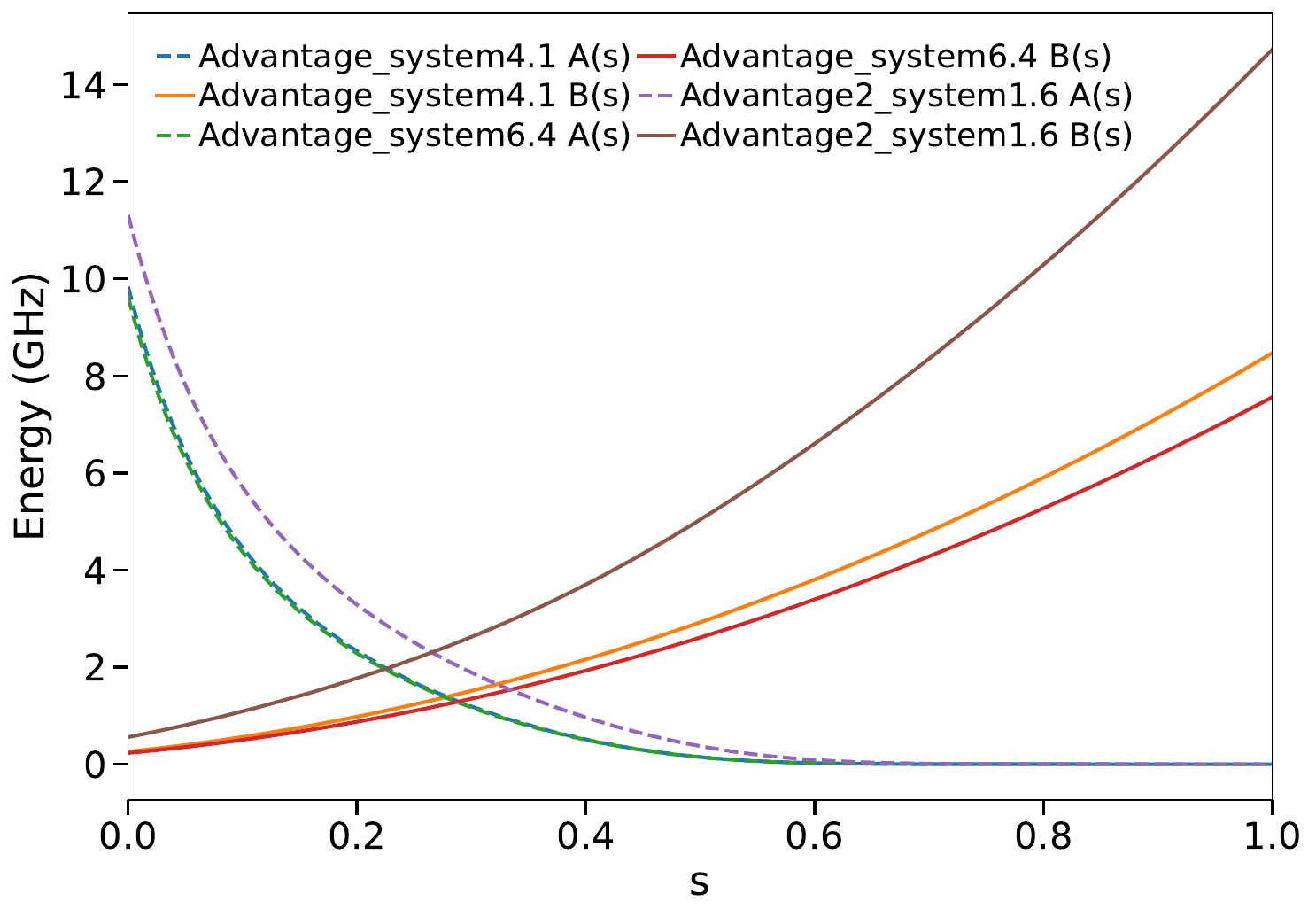}
    \caption{The time-dependent functions $A(s)$ and $B(s)$ used for: (a) the standard forward annealing linear-ramp sampling protocol (b) reverse quantum annealing ``Monte Carlo'' chains, for the three different D-Wave QPUs.}
    \label{fig:anneal-schedule}
\end{figure*}

Quantum annealing (QA) is a type of analog quantum computation, which is based on the adiabatic theorem~\cite{born1928beweis}, and typically the goal of this model of computation is to find a low-energy configuration of an Ising model (which can be equivalently considered a classical discrete combinatorial optimization problem)~\cite{RevModPhys.80.1061, Morita_2008, farhi2000quantumcomputationadiabaticevolution, Santoro_2002, Kadowaki_1998, santoro2006optimization}. More recently, quantum annealers have been increasingly used to probe analog magnetic system simulations, such as of time dynamics of exotic and frustrated types of quantum magnets~\cite{king2025beyond, harris2018phase, king2023quantum, King_2022}. In practice, the commercial company D-Wave has manufactured quantum annealers based on superconducting flux qubits~\cite{johnson2011quantum, Bunyk_2014, PhysRevB.82.024511, Johnson_2010, dickson2013thermally, Lanting_2014}, and these devices can be utilized to probe near term, noisy, analog quantum computation capabilities. This study probes the possibility of using current D-Wave quantum annealers to estimate the canonical partition function\footnote{Note that although we are using quantum annealers, a type of analog quantum computer, we are not attempting to estimate the quantum partition function (e.g., using D-Wave hardware as a quantum thermal sampler~\cite{Izquierdo_2021}), but rather the standard canonical partition function from statistical mechanics} of random $\pm J$ Ising models, defined on the hardware connectivity graphs of the D-Wave processors. The motivation for this study is twofold. First, there are prior studies which have shown evidence that, at least for relatively small Ising models, analog D-Wave quantum annealers can be good samplers of thermal distributions defined by an Ising model (e.g., a classical Hamiltonian)~\cite{PRXQuantum.3.020317, PhysRevApplied.17.044046, Raymond_2016, sandt2023efficient, nelson2021single, PhysRevApplied.8.064025, buffoni2020thermodynamics, PhysRevApplied.11.044083, sathe2025classicalcriticalityquantumannealing, Raymond_2016, Chancellor_2016, PhysRevResearch.6.043050, pelofske2025boltzmannsamplingfrustratedj1}. This suggests that, in particular because the effective sampling temperature can be modified by the total annealing time, D-Wave quantum annealers could be used, in a similar manner to Metropolis-Hastings samplers, could be used to approximate the partition function. The second is that the total time required to generate individual samples on D-Wave quantum annealers is quite fast, suggesting that there is potentially a regime in which D-Wave hardware sampling may be able to compete with classical samplers. Moreover, in general, as analog samplers of classical Hamiltonians, this is a natural question to ask of whether these devices could be potentially good at partition function estimation. Using sampling based noisy analog quantum computation to perform thermodynamic sampling, in particular partition function estimation, we propose as a potential type of approximate heuristic computational tool, similar to MCMC. 

Adiabatic quantum computation (quantum annealing) dynamics are governed by a time-dependent Hamiltonian,
\begin{equation}
    H(t) = A(t)H_{initial} + B(t)H_{diag}.
    \label{equation:QA_high_level}
\end{equation}
where $A(t)$ and $B(t)$ define the ``annealing schedule''. The goal of this study is to use analog quantum computing hardware, quantum annealers, to sample from a specific Ising model, in this case we use a test-case $\pm J$ disordered hardware-compatible $25$-spin Ising model, in such a way that we can reliably and accurately extract an estimate of the true partition function of that Ising model. The most direct way for the analog hardware to work as a partition function estimator is to act as a sampler from the Gibbs distribution defined by that Ising model\footnote{This type of sampling may be known as either thermal sampling, Gibbs sampling, or Boltzmann sampling depending on the context - in this case we will typically refer to this as Gibbs sampling. Each of these processes can refer to more general sampling problems from complex distributions, but they can also refer to the task of sampling from a distribution defined by an Ising model (e.g., classical Hamiltonian), and that is described by $P(s_i) = \frac{e^{- \beta H(s_i)}}{Z}$, where $s_i$ are spin configurations and $Z$ is the partition function. }, for which there are empirical findings that D-Wave hardware is good at this task~\cite{PRXQuantum.3.020317, PhysRevApplied.17.044046, nelson2021single, pelofske2025boltzmannsamplingfrustratedj1}, as well as supporting numerical simulation evidence that quantum annealing can be tuned to be a good thermal sampler~\cite{gyhm2025boltzmannsamplingdiabaticquantum}. In particular this type of sampling approach on D-Wave hardware would require changing the effective sampling temperature, so as to sample a density of states histogram which is representative of the underlying Ising model energy spectrum, which in turn allows estimation of the partition function of a classical Ising model. The control parameters that results in the effective temperature being varied are both the total annealing time and the coupling energy scale programmed on the hardware~\cite{PRXQuantum.3.020317}. 

Prior studies have proposed digital quantum algorithms that can compute the partition function of a Hamiltonian~\cite{giovannetti2025quantumalgorithmestimatingdeterminant}, but in the analog quantum computation paradigm there have been no such algorithmic proposals -- our study is the first to conceptualize analog quantum sampling algorithms for this task. 

\flushbottom

%%%%%%%%%%%%%%%%%%%%%%%%%%%%%%%%%%%%%%%%%%%%%%%%%%%%%%%
%%%%%%%%%%%%%%%%%%%%%%%%%%%%%%%%%%%%%%%%%%%%%%%%%%%%%%%
\section{Methods}\label{section:methods}
%%%%%%%%%%%%%%%%%%%%%%%%%%%%%%%%%%%%%%%%%%%%%%%%%%%%%%%
%%%%%%%%%%%%%%%%%%%%%%%%%%%%%%%%%%%%%%%%%%%%%%%%%%%%%%%

\begin{table*}[ht!]
    \begin{center}
        \begin{tabular}{|l||l|l|l|l|}
            \hline
            D-Wave QPU Chip & Graph name & Qubits & Couplers & Disjoint embedding count \\
            \hline
            \hline
            \texttt{Advantage\_system4.1} & Pegasus $P_{16}$ & 5627 & 40279 & 178 \\
            \hline
            \texttt{Advantage\_system6.4} & Pegasus $P_{16}$ & 5612 & 40088 & 171 \\
            \hline
            \texttt{Advantage2\_system1.9} & Zephyr $Z_{12}$ & 4590 & 41748 & 146 \\
            \hline
        \end{tabular}
    \end{center}
    \caption{D-Wave QPU hardware summary. The maximum programmable annealing time is $2000 \mu$s, and the smallest possible annealing time is $5$ ns, for each of these three devices. A count for the number of disjoint native embeddings of the $25$ spin Ising model is shown in the right column.  }
    \label{table:hardware_summary}
\end{table*}

Our objective is to accurately estimate the partition function of a given Ising model by leveraging the sampling capabilities of a quantum annealer. The general strategy is to estimate the density of states at each energy level of the Ising model, which is then used to calculate the partition function. To sample the necessary energy distributions, we utilize two different quantum annealing protocols: standard linear ramp quantum annealing ($\S$\ref{subsection:methods_standard_QA}) and Monte Carlo chains of reverse quantum anneals ($\S$ \ref{subsection:methods_reverse_quantum_annealing}). In order to compare the D-Wave QPU sampling quality against standard existing methods, we also implement and test the classical algorithms Wang-Landau and Multiple Histogram Reweighting with Markov chain Monte Carlo, which are briefly described in $\S$\ref{subsection:methods_classical_sampling}. When reporting the performance of the various algorithms, the error measure that we will use is logarithmic-scale relative error, defined as
\begin{equation}
    \text{Log Relative Error} = \frac{\abs{\ln{Z^*} - \ln{Z}}}{\ln{Z}}, 
\end{equation}

which is convenient to use in this context because of how large the true partition function can be. $Z^*$ is the current best estimate of the true partition function, and $Z$ is the ground truth partition function, obtained by a complete classical enumeration over all spin configurations. $Z^*$ can be either greater than or less than $Z$, but we assume that the estimate is non-zero and non-negative. We will gather $S$ samples using the various heuristic sampling algorithms, and the key feature for all of these estimation algorithms is that we require $S << 2^N$ in order for the algorithm to be efficient, and offer tractable computations. Efficient partition function sampling algorithms are able to generate samples quickly, but specifically low energy spin configurations which contribute more to the partition function estimate, and moreover are able to sample across the energy spectrum of the Hamiltonian.

\subsection{Classical Partition Function Estimation Algorithms}
\label{subsection:methods_classical_sampling}

As a baseline of the current efficient heuristic sampling methods for numerical computational physics methods, we implement and evaluate Wang-Landau~\cite{wang2001efficient} and Multiple Histogram Reweighting with Markov chain Monte Carlo (MCMC)~\cite{ferrenberg1988new, ferrenberg1989optimized,kumar1992weighted}. In both cases, the Monte Carlo updates we use are random single spin-flips. The Wang-Landau (WL) algorithm works by directly building an estimate of the density of configurations at each discrete energy level bin of the target Hamiltonian, this density of states (DoS) we will denote as $g(E)$. WL performs a type of random walk, but in energy space not configuration space, using the Metropolis-Hastings algorithm \cite{metropolis1953equation} where the acceptance probability is given by the current DoS estimate, making it a \emph{non-Markovian random walk} at least during the learning stage. WL convergence is determined by the histogram flatness criterion, which we set to $0.90$. WL is very useful because once a good $g(E)$ is computed, quantities such as the partition function can be easily computed for any temperature: this does not require re-executing a numerical simulation for every temperature of the model. This is because the partition function can be written as 

\begin{equation}
    Z = \sum_Eg(E)e^{-E/k_BT},
    \label{equation:Z_estimate_density_of_states_form}
\end{equation}
which notably gives a substantial reduction in the combinatorial explosion that results from computing the standard partition function because now we only need densities for each energy level instead of all configurations and typically classical Hamiltonians, especially the types of $\pm J$ spin-glass-models that we consider in this study, have a comparatively small number of energy levels.

MHR on the other hand uses converged MCMC simulations over a range of temperatures to then estimate thermodynamic quantities, in this case the partition function, at an arbitrary temperature within the range of temperatures that were run initially with MCMC. MHR is powerful because it allows in principle many estimates at many different temperatures from an initial fixed upfront computation at a series of temperatures. Here, our reference implementation targets a temperature of $T=4$, and the MCMC runs are performed at $T=[3.5, 3.75, 4.0, 4.25, 4.5]$. The reweighting step is done by iteratively solving for the free energies $\{f_i\}$ and estimating the partition function at the target temperature using the sampled energies. Importantly, the MCMC convergence at lower temperatures can already become computationally intensive even for the $25$ spin model we evaluate. More details of the implementations are given in Appendix~\ref{appendix:classical_algorithms_and_pseudo_code}. 

\subsection{D-Wave Quantum Annealing Processor Hardware Implementation} \label{section:methods_quantum_annealer_details}

The physical system of interest is a Transverse Field Ising Model (TFIM) implemented by the D-Wave analog quantum computers; the Hamiltonian is given by:
\begin{align}
    {\mathcal H} =& - \frac{A(s)}{2}  \sum_i \hat\sigma_x^{i}
    + \frac{B(s)} {2} \left( \sum_i h_i \hat\sigma_z^{i} + \sum_{\langle ij \rangle} J_{i j} \hat\sigma_z^{i} \hat\sigma_z^{j} \right).   
    \label{equation:DWave_QA_Hamiltonian}
\end{align}
where $s\in[0,1]$ parametrizes the anneal schedule, $\hat\sigma_{x,z}^i$ are Pauli operators on spin $i$, $h_i$ and $J_{ij}$ are programmable longitudinal fields and couplings, and $\langle ij\rangle$ denotes interacting pairs of qubits. The schedules $A(s)$ and $B(s)$ interpolate between a driver term dominated by transverse field ($s\!\approx\!0$) and the classical Ising problem Hamiltonian ($s\!\approx\!1$). Fig.~\ref{fig:anneal-schedule} shows the corresponding functions $A(s)$ and $B(s)$ of the D-Wave quantum processing units (QPUs) used in this study. The transverse field is what facilitates state transitions over the course of the anneal because it does not commute with the programmed classical Hamiltonian. For computational-basis states $z=(z_1,\ldots,z_N)$ with $z_i\in\{\pm1\}$, the corresponding classical Ising energy of a single configuration measured at the end of a single anneal is
\[
E(z) \;=\; \sum_i h_i z_i + \sum_{\langle ij\rangle} J_{ij} z_i z_j. 
\]

In this study we used three different superconducting flux qubit D-Wave QPUs, Table~\ref{table:hardware_summary} summarizes the hardware details\footnote{Although textually cumbersome, we will typically directly quote the exact chip id string of the processor for clarity and reproducibility of our results. }. Importantly, the D-Wave quantum annealers are subject to a variety of noise and bias effects, including open quantum system effects due to coupling to the environment~\cite{PhysRevApplied.8.064025, Amin_2015, buffoni2020thermodynamics, PhysRevApplied.19.034053}, analog control errors such as spurious couplings~\cite{tüysüz2025learningresponsefunctionsanalog, Zaborniak_2021, PRXQuantum.3.020317, PhysRevApplied.17.044046}, spin bath polarization~\cite{lanting2020probingenvironmentalspinpolarization}, and noise drift over time~\cite{Pelofske_2023_noise}. All of these factors mean that the D-Wave hardware must be considered a type of noisy sampler, and at sufficiently long anneal times, an open quantum system~\cite{King_2022, king2023quantum, king2025beyond, tindall2025dynamicsdisorderedquantumsystems, PhysRevApplied.19.034053}. The hardware graphs are known as Pegasus~\cite{dattani2019pegasussecondconnectivitygraph, boothby2020nextgenerationtopologydwavequantum} and Zephyr~\cite{zephyr}, each having a slightly different logical graph structure. Due to local hardware defects, the number of couplers and qubits in each hardware graph are typically less than the logical graph description. 

The Ising model we study is a single fixed hardware-graph native $\pm J$ $25$-spin model, which is a subgraph of both the Zephyr and Pegasus hardware graphs. The $\pm J$ coefficients are chosen at random from $\{+1, -1\}$, leading to a small instance of disordered and frustrated Ising model, the exact coefficients are given in Appendix~\ref{appendix:parallel_embeddings_plots}. This relatively small system size is used so that the ground-truth partition function estimate is tractable to compute exactly, using classical enumeration over all spin configurations. The direct spin-to-qubit mapping of the Ising model into the D-Wave hardware mitigates the need for minor-embedding, which can introduce additional sources of error on the hardware and in general cause thermodynamic sampling problems~\cite{PhysRevResearch.2.023020}. Because the D-Wave hardware graphs are much larger than the $25$-spin model, we can embed many instances of the Ising model in parallel on each QPU graph. Appendix~\ref{appendix:parallel_embeddings_plots} shows the disjoint parallel embeddings on each of the D-Wave QPUs. Then, we can execute a single anneal-readout cycle and in the process obtain many independent samples (not considering any potential cross-talk or interaction between the programmed qubits and couplers)~\cite{PhysRevA.91.042314, parallel_QA, Pelofske_2022_boolean}. To this end, we use the Glasgow subgraph isomorphism solver~\cite{mccreesh2020glasgow}, which is part of the \texttt{minorminer} package~\cite{Chern_2023, cai2014practicalheuristicfindinggraph}, in order to iteratively find many disjoint native embeddings of the fixed Ising model -- the counts for these disjoint embeddings are reported in Table~\ref{table:hardware_summary}.

In $\S$\ref{subsection:methods_standard_QA} and $\S$\ref{subsection:methods_reverse_quantum_annealing}, we describe two QA sampling protocols to build a DoS estimation. For both of these methods, the key idea is inspired by the Wang-Landau density of states technique, see Eq.~\eqref{equation:Z_estimate_density_of_states_form}; every measured sample from the D-Wave hardware (for a particular parameter combination) is combined in an energy-histogram, which can then be interpreted as an energy level density estimate. This idea is powerful for the same reason that WL is powerful as an efficient sampling algorithm: estimates of the partition function at different temperatures, even very low temperatures, require no additional compute time.

\subsection{Sampling with Standard Linear Ramp Quantum Annealing}
\label{subsection:methods_standard_QA}

The most straightforward simulation approach is to employ standard quantum annealing with a linear annealing schedule. Such simulations typically sample many low-energy ground states, yielding poor coverage of the full energy spectrum. However, this is only true for anneals that are relatively long, where prior studies have demonstrated a relatively stable linear relationship between annealing time and $\beta$~\cite{nelson2021single}. For significantly shorter annealing times, these simulations can be interpreted more as quenches. In particular, very fast anneals on D-Wave hardware (on the order of $\approx 5$ ns anneals) result in the formation of excited kinks~\cite{king2023quantum, King_2022, bando2020probing, king2025beyond}, which replicate Kibble-Zurek scaling laws~\cite{kibble1976topology, zurek1985cosmological}. As a result, these fast anneals sample higher-energy configurations of the programmed Ising model, and less frequent sampling of ground-states. 

In addition to annealing time, the programmed energy scale also has a significant impact on sampling quality because of the analog nature of the hardware -- weak coupling leads to higher energy configuration sampling. For some classes of Ising models, the transverse field driving Hamiltonian induces a non-uniform probability distribution over configurations with the same degenerate energy level of the classical Hamiltonian (a type of degeneracy lifting), which is particularly notable for the ground states~\cite{PhysRevA.100.030303, matsuda2009quantum, 9605329, PhysRevLett.118.070502, PhysRevE.99.063314}. This sampling bias presents a challenge for accurate Boltzmann sampling, and therefore also potentially for using quantum annealing to sample the partition function. As shown by previous studies~\cite{PRXQuantum.3.020317, PhysRevApplied.17.044046, nelson2021single}, an effective strategy to mitigate this effect is to reduce the programmed energy scale on the D-Wave hardware, thus resulting in less biased degenerate state sampling. Accordingly, the other hardware parameter varied in this study is the energy scale of the $J$ couplers that define the Ising model, although the degree to which this bias affects the sampling is dependent on the degeneracy properties of the Ising model of interest. 

The general intuition of this simulation is that faster anneals result in higher energy configurations being sampled, and longer anneals result in lower energy configurations being sampled. Smaller $J$ coupling results in higher energy configurations being sampled, and larger $J$ coupling energy scale results in lower energy configurations being sampled. $J$ is defined in normalized hardware-specific programmable units, given by Eq.~\eqref{equation:DWave_QA_Hamiltonian} and the energy scales in Fig.~\ref{fig:anneal-schedule}; we use a grid search over many coupling strengths ranging from $0.0001$ (which is likely very near to the hardware precision limit, and is therefore very close to uniform random sampling) up to $1.0$ (the maximum allowed on the hardware). The annealing times we similarly vary in a grid search from $5$ nanoseconds up to $100 \mu s$. In summary, the two analog hardware control parameters are \emph{total annealing time} and \emph{J coupling energy scale}. The third implicit parameter, inherent as a sampling computation, is the total number of samples drawn for each parameter.

\begin{figure}[ht!]
\centering
    \includegraphics[width = \columnwidth]{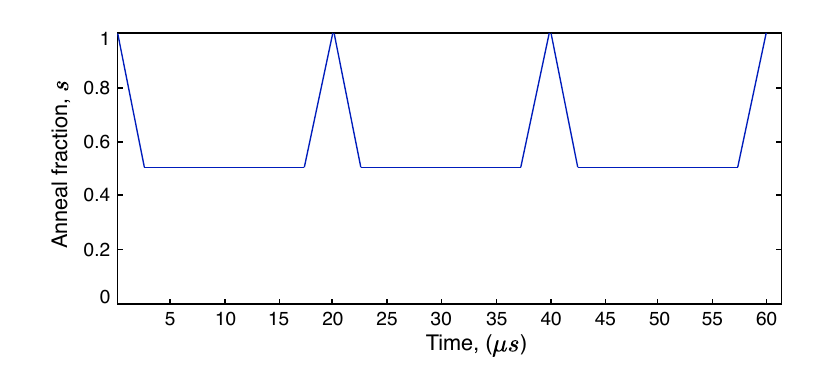}
    \caption{An example representative reverse quantum annealing Monte Carlo chain consisting of three reverse annealing cycles. Each cycle performs a $2.5$ $\mu$s ramp from $s=1$ to $s=0.5$, followed by a $15$ $\mu$s pause at $s=0.5$, and then a ramp back to $s=1$. Each time the hardware reaches $s=1$, the state of all qubits is measured (timescale for the measurement are not shown in this schematic); that exact measured spin configuration is then used to initialize the subsequent simulation. }
    \label{fig:RA-general}
\end{figure}

\begin{figure*}[ht!]
    \centering
    \begin{subfigure}[t]{0.49\textwidth}
        \centering
        \includegraphics[height=2.2in]{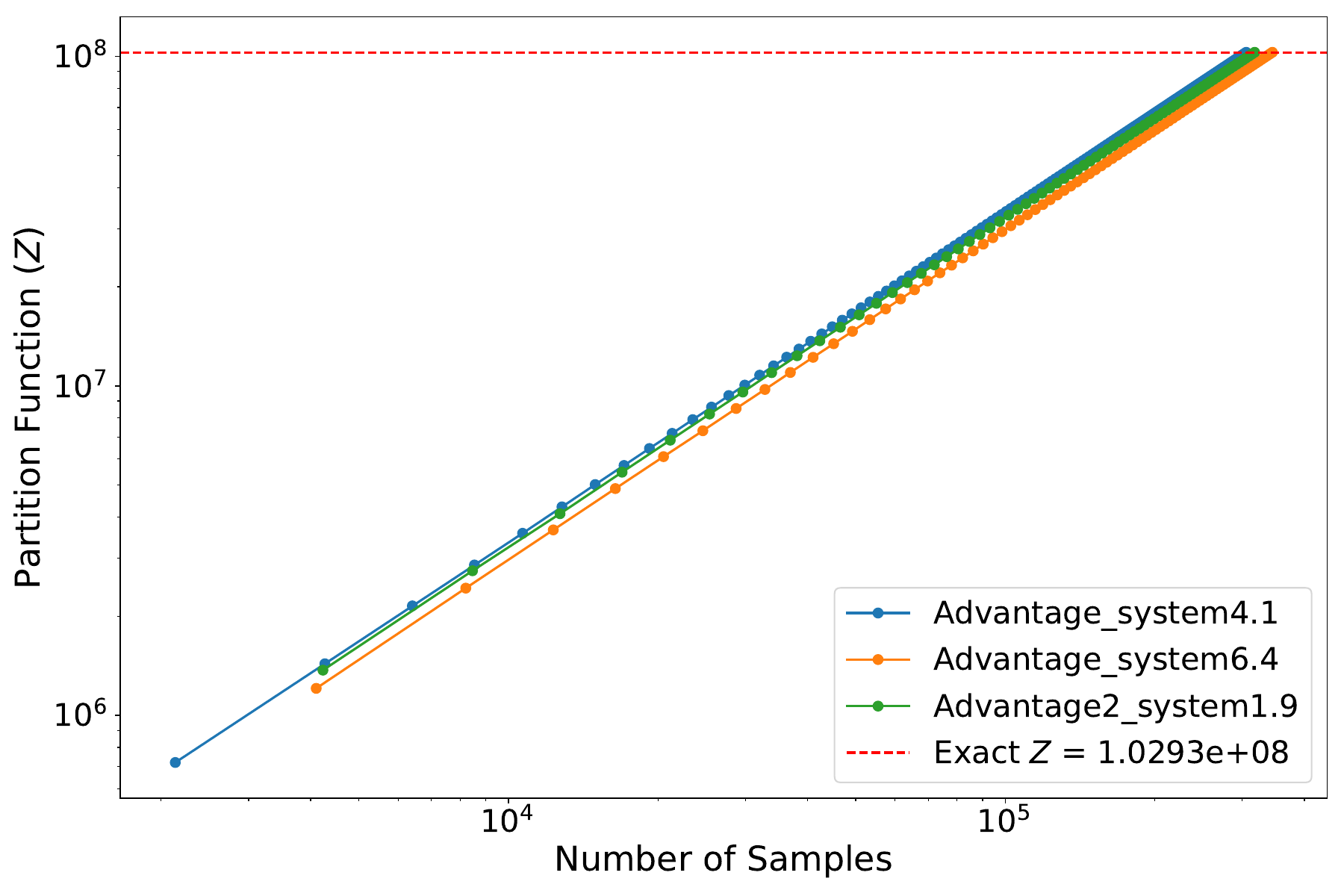}
        \caption{}
    \end{subfigure}%
    ~ 
    \begin{subfigure}[t]{0.49\textwidth}
        \centering
        \includegraphics[height=2.2in]{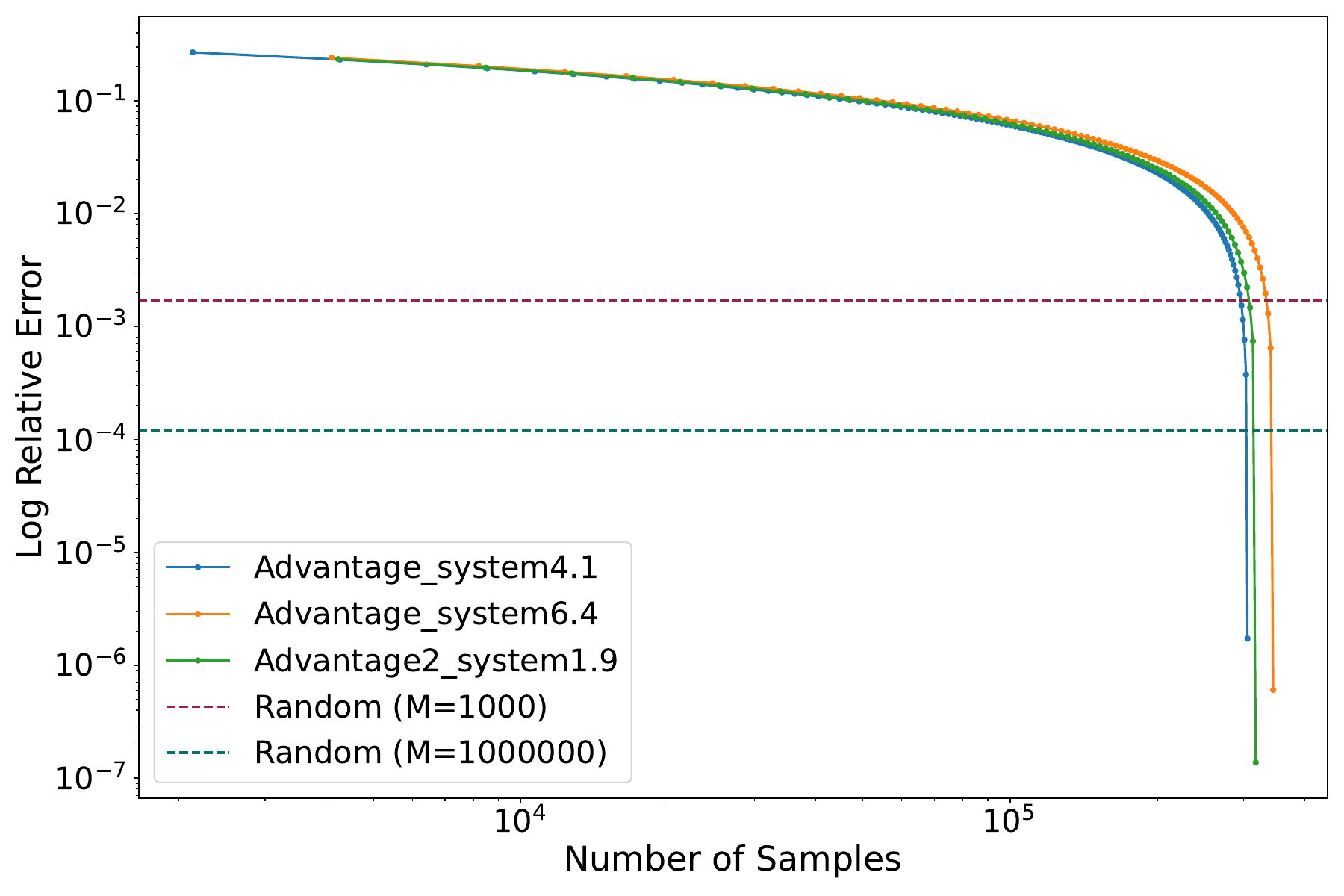}
        \caption{}
    \end{subfigure}
    ~
    \begin{subfigure}[t]{0.49\textwidth}
        \centering
        \includegraphics[height=2.2in]{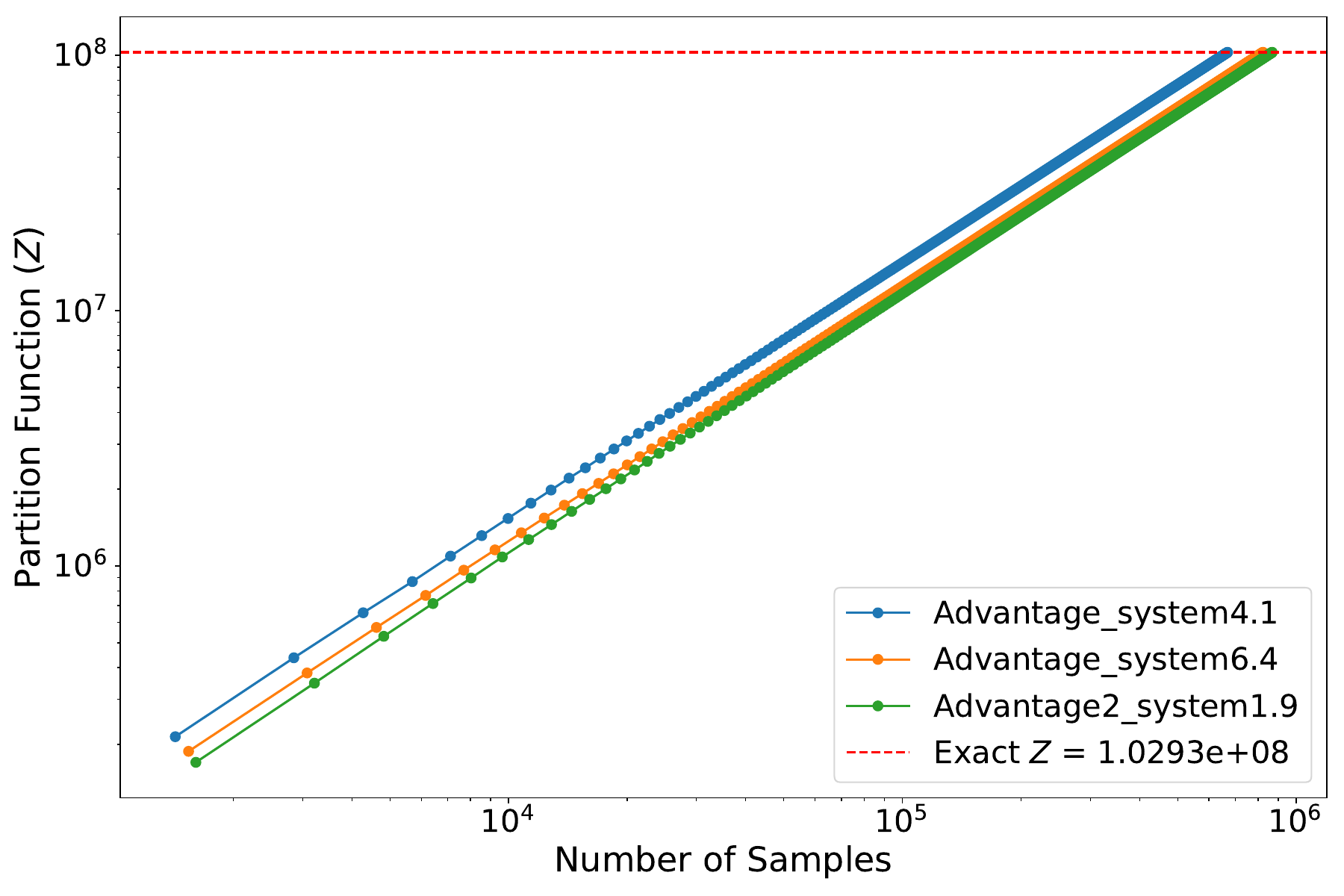}
        \caption{}
    \end{subfigure}%
    ~ 
    \begin{subfigure}[t]{0.49\textwidth}
        \centering
        \includegraphics[height=2.2in]{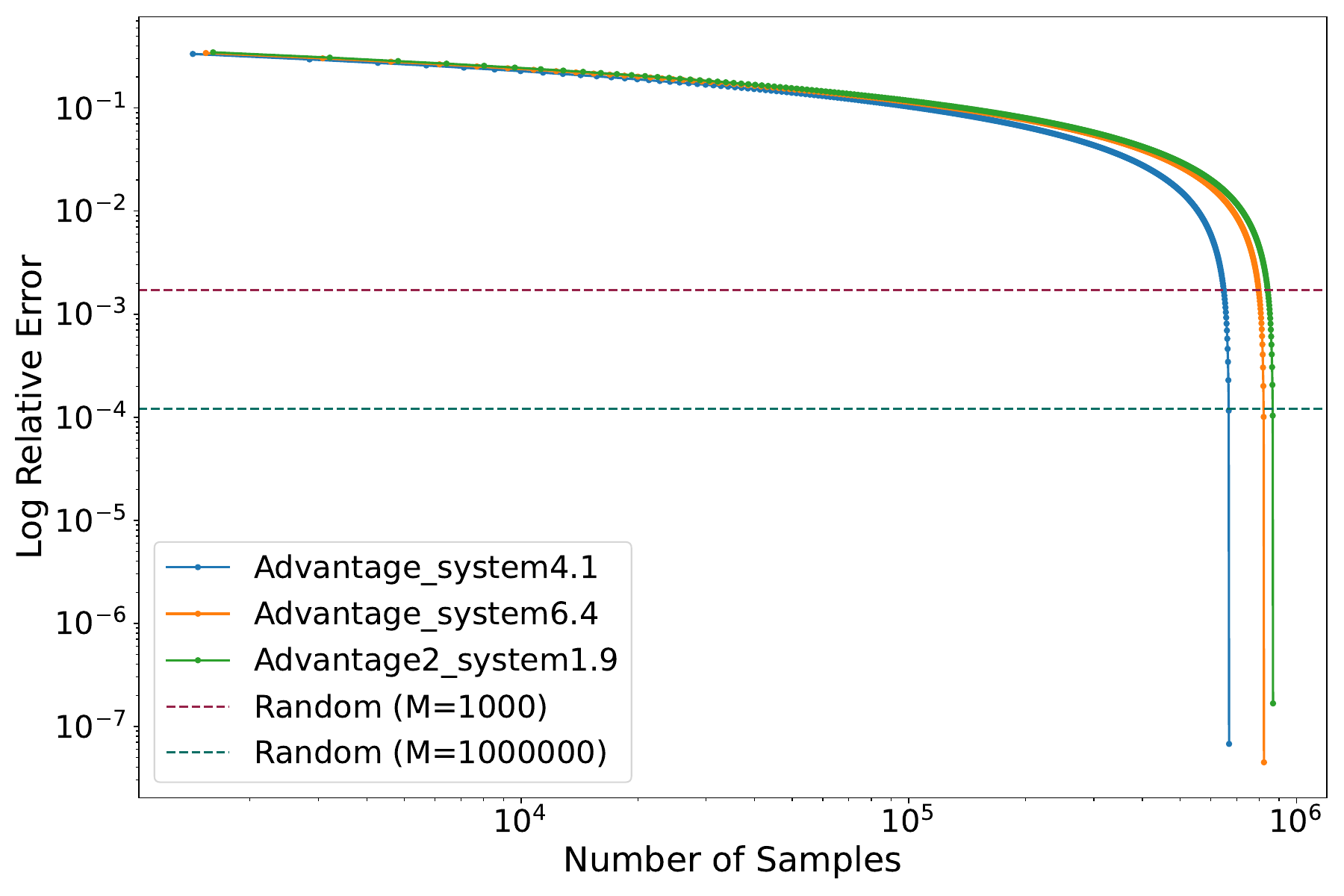}
        \caption{}
    \end{subfigure}
    ~
    \begin{subfigure}[t]{0.49\textwidth}
        \centering
        \includegraphics[height=2.2in]{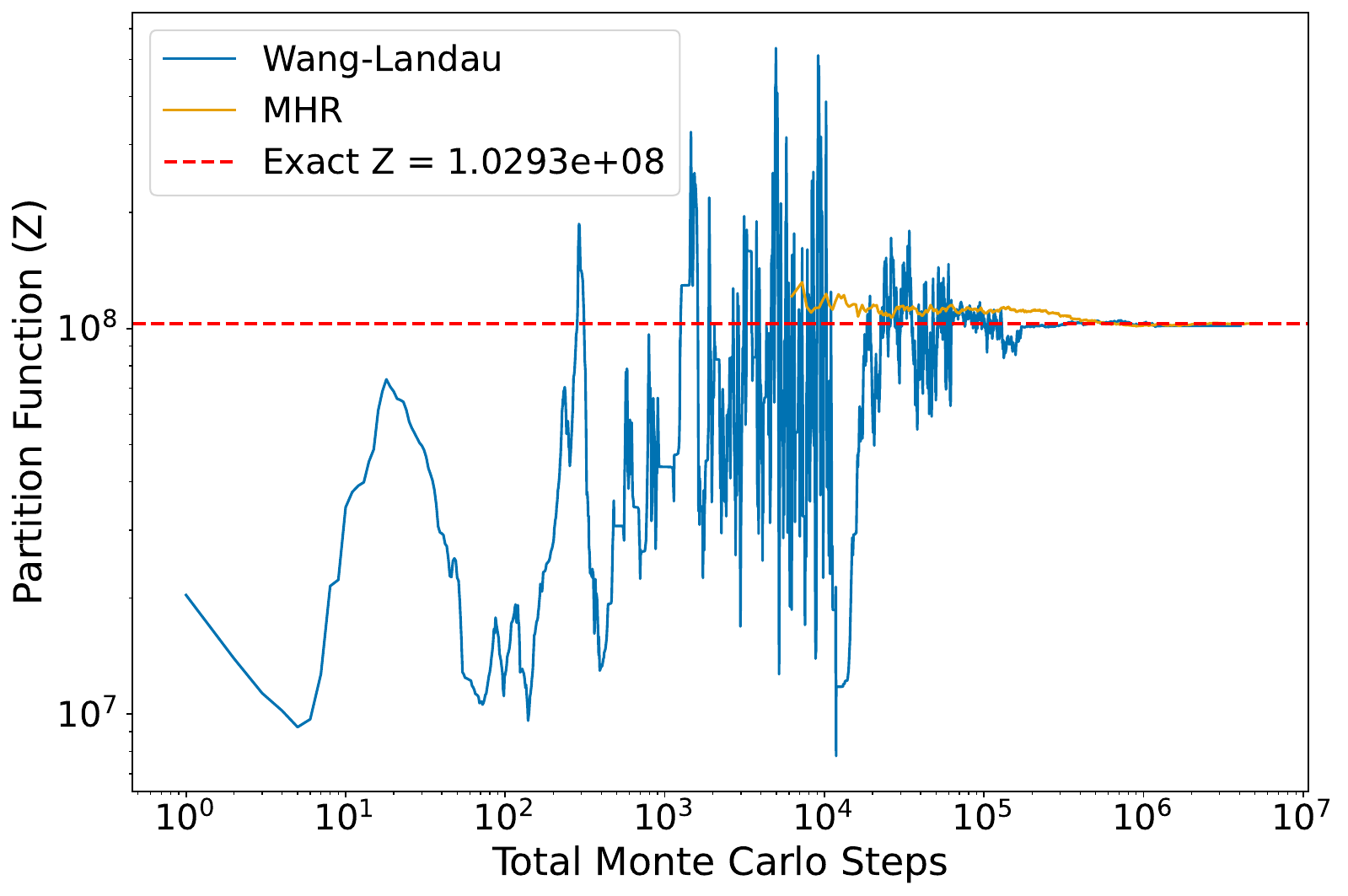}
        \caption{Convergence of the estimated Z towards the exact value (red dashed line) as a function of Monte Carlo steps.}
    \end{subfigure}%
    ~ 
    \begin{subfigure}[t]{0.49\textwidth}
        \centering
        \includegraphics[height=2.22in, keepaspectratio]{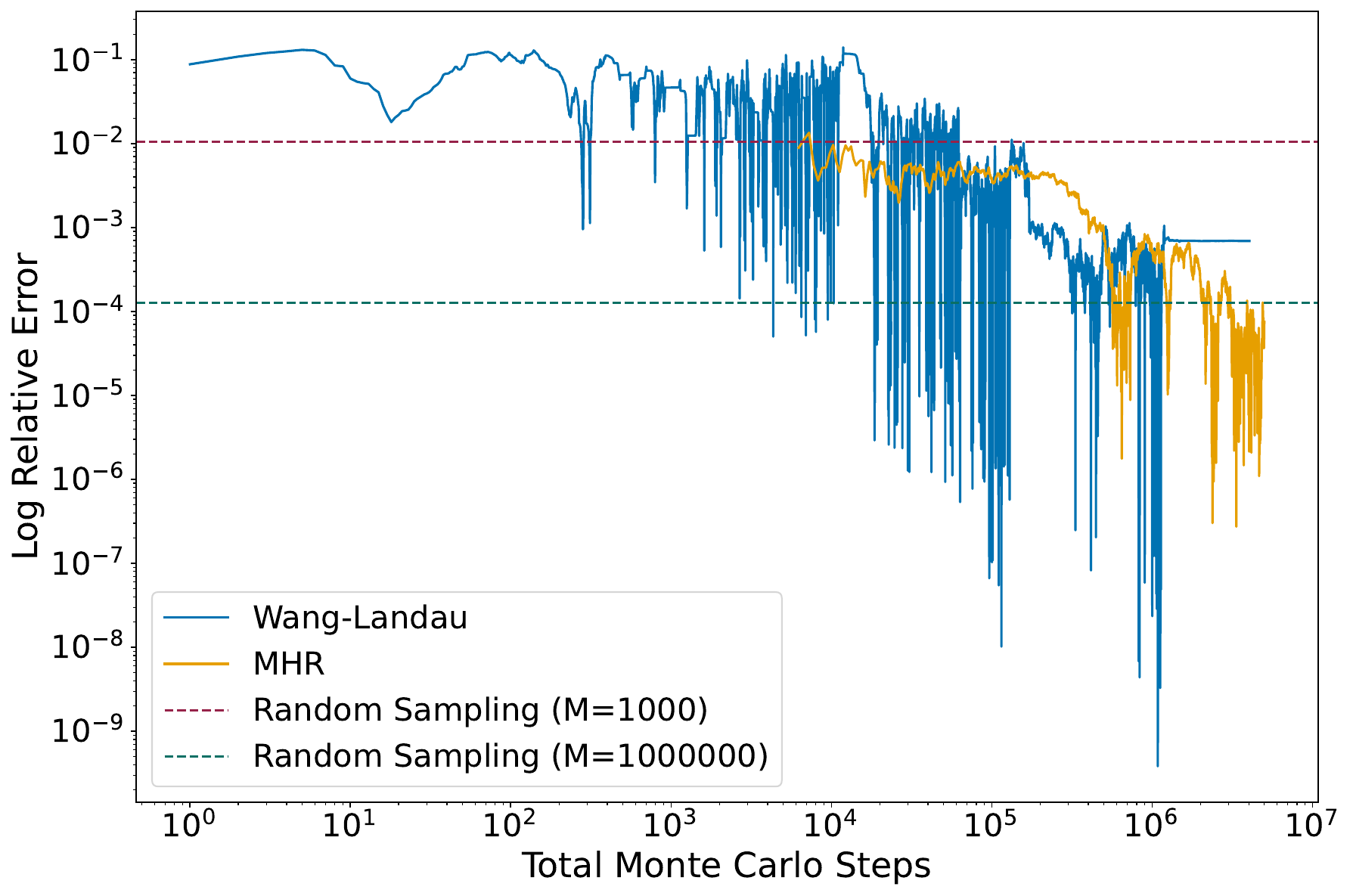}
        \caption{Convergence of the relative logarithmic error for both algorithms, compared against two random sampling baselines.}
    \end{subfigure}
    \caption{A comparison of the quantum annealing sampling techniques, with an optimized set of analog hardware control parameters, (a-d) and the standard classical sampling algorithms Wang-Landau and MHR for estimating the partition function in panels e and f. The x-axis denotes the absolute total sample count that was generated by each respective algorithm, in the case of the classical algorithms this corresponds to the total number of proposed spin updates which is the primary source of compute time and complexity for these algorithms. In the case of QA sampling, the x-axis corresponds to the number of measured samples, which is distributed incrementally from each hardware parameter up to the total final sample count. Random sampling error rates are shown for reference as dashed horizontal dashed lines in the right-column plots.   }
    \label{figure:Fig3_classical_estimation_algorithms_and_best_DWave_comparison}
\end{figure*}

\subsection{Sampling with Monte Carlo Chains of Reverse Quantum Anneals}
\label{subsection:methods_reverse_quantum_annealing}

The second quantum annealing protocol we consider is iterated reverse quantum annealing~\cite{PhysRevApplied.17.054033, PhysRevA.100.052321, PhysRevResearch.3.033006}, also referred to as ``quantum evolution Monte Carlo'' (QEMC)~\cite{king2021scaling, PRXQuantum.2.030317, PRXQuantum.1.020320, King_2018_observation, Pelofske_2023_ra, Henke_2025}\footnote{We will interchangeably use the terms QEMC and iterated reverse annealing. }. The idea of this simulation is illustrated in Fig.~\ref{fig:RA-general}, and it is based on the ``reverse quantum annealing'' algorithm which was originally introduced as a local search heuristic to improve solutions of combinatorial optimization problems~\cite{perdomo2011study, king2019quantumassistedgeneticalgorithm, Golden_2021, Venturelli_2019, Chancellor_2020}. Reverse annealing begins in a classically defined spin configuration, and then introduce some amount of quantum fluctuations by pausing the anneal at a flat and symmetric point in the schedule, denoted as $s$. Having a non-zero $B(s)$ (diagonal Hamiltonian) field at $s$ means that while we have quantum fluctuations, the simulation is biased towards that initial classical configuration we have specified. Then, we repeat this process, where the initial state at each ``reverse anneal'' is defined by the measured state in the last anneal-readout cycle. This sequence forms a sequential chain of reverse anneals, which is analogous to a type of Markov chain Monte Carlo. However, unlike conventional Markov chain Monte Carlo methods, the number of spin flips is not strictly $1$ (or cluster/loop updates), and there is no explicit notion of accept/reject conditions. In this way, this sampling approach is less Monte Carlo like and more like an ``autonomous'' iterative update algorithm. The pause ``location'' in the anneal schedule $s$ must be tuned to promote consistent, monotonic exploration of the energy landscape. The initial classical qubit configuration used to start the QEMC chain is a single random spin configuration with $p=0.5$ between $\{+1, -1\}$ (e.g., an arbitrary high temperature sample). 

As in the standard annealing approach discussed in $\S$~\ref{subsection:methods_standard_QA}, the programmed energy scale serves as a key parameter (which is equivalent to a particular ratio between the transverse field and the lattice coupling which is machine-dependent as seen in Fig.~\ref{fig:anneal-schedule}), controlling the type of sampling the hardware performs. In addition, total reverse annealing times of $2 \mu s$ and $100 \mu s$ are evaluated, and use symmetric anneal-schedule pauses with the quench ramp times being uniformly $0.5 
\mu$s. By combining these controls ($s$, the coupling energy scale $J$, and the annealing time), we generate a sequence of Monte Carlo iterations designed to explore a broad region of the energy spectrum. Shorter annealing times are not possible to utilize for this simulation because of limitations with the ramp rate of anneal-schedule waveforms on the hardware -- if we want a pause of $1 \mu $s, for arbitrary pause $s$ values with rapid quenches the minimum annealing time we can have is $2 
\mu$s. Note these parameters are slightly different from the representative parameters shown in Fig.~\ref{fig:RA-general}. The Monte Carlo iterations define the ``sampling trajectory'' through the energy landscape (coupled with the tuned energy scales and $s$ anneal pause in the anneal schedule). For very long QEMC chains (we use up to $10{,}000$ length chains of reverse anneals), we chain together separate D-Wave device calls to form one uninterrupted chain -- very long device jobs are not allowed on the backend due to a total-compute-time restriction. 

The intuition of the analog control parameters for this simulation are as follows. First, the $J$ coupling energy scale effects follow the same reasoning as with standard quantum annealing ($\S$~\ref{subsection:methods_standard_QA}), as does the total-annealing time, although here we expect the differences between $2 \mu s$ and $100 \mu s$ to be minimal. Next, due to relaxation during the paused reverse anneal, more QEMC iterations will result lower energy configuration sampling, whereas earlier on in the Monte Carlo chain the sampled energy configurations will be higher energy. Lastly, the pause value of $s$ also significantly influences the sampled configuration energy, and also the convergence rate towards the ground-state of the QEMC chain. When $s$ is close to zero, quantum fluctuations dominate and therefore the sampling is closer to random and therefore higher temperature. When $s$ is close to $1$ the state does not change very much from the initial configuration because there are few quantum fluctuations causing state transitions, which means that the temperature of the sampled spin configurations will be close to the temperature of the initial spin configuration in the chain (which, in this case is always a random spin configuration). As described in $\S$~\ref{subsection:methods_standard_QA}, energy scales of $J=0.0001$ to $J=1$ are used. The reverse annealing pauses used are $0.1, 0.2, \ldots, 0.8, 0.9$ (9 in total). 

In summary, the four analog hardware control parameters are \emph{total annealing time}, \emph{J coupling energy scale}, \emph{reverse anneal pause ``location''} (denoted as s) see Eq.~\eqref{equation:DWave_QA_Hamiltonian}, and \emph{total number of Monte Carlo iterations}. There is also an implicit parameter of the total number of samples drawn for each one of these parameters -- however, this is also inherently coupled to the total number of QEMC iterations, at least for a contiguous chain of reverse anneals. Note that because of the parallel embeddings, neglecting any cross-talk error in the hardware, we always have many independent Monte Carlo sequence runs for each parameter.

\begin{figure*}[ht!]
    \centering
    \begin{subfigure}[t]{0.5\textwidth}
        \centering
        \includegraphics[height = 2.3 in, width=\textwidth, keepaspectratio]{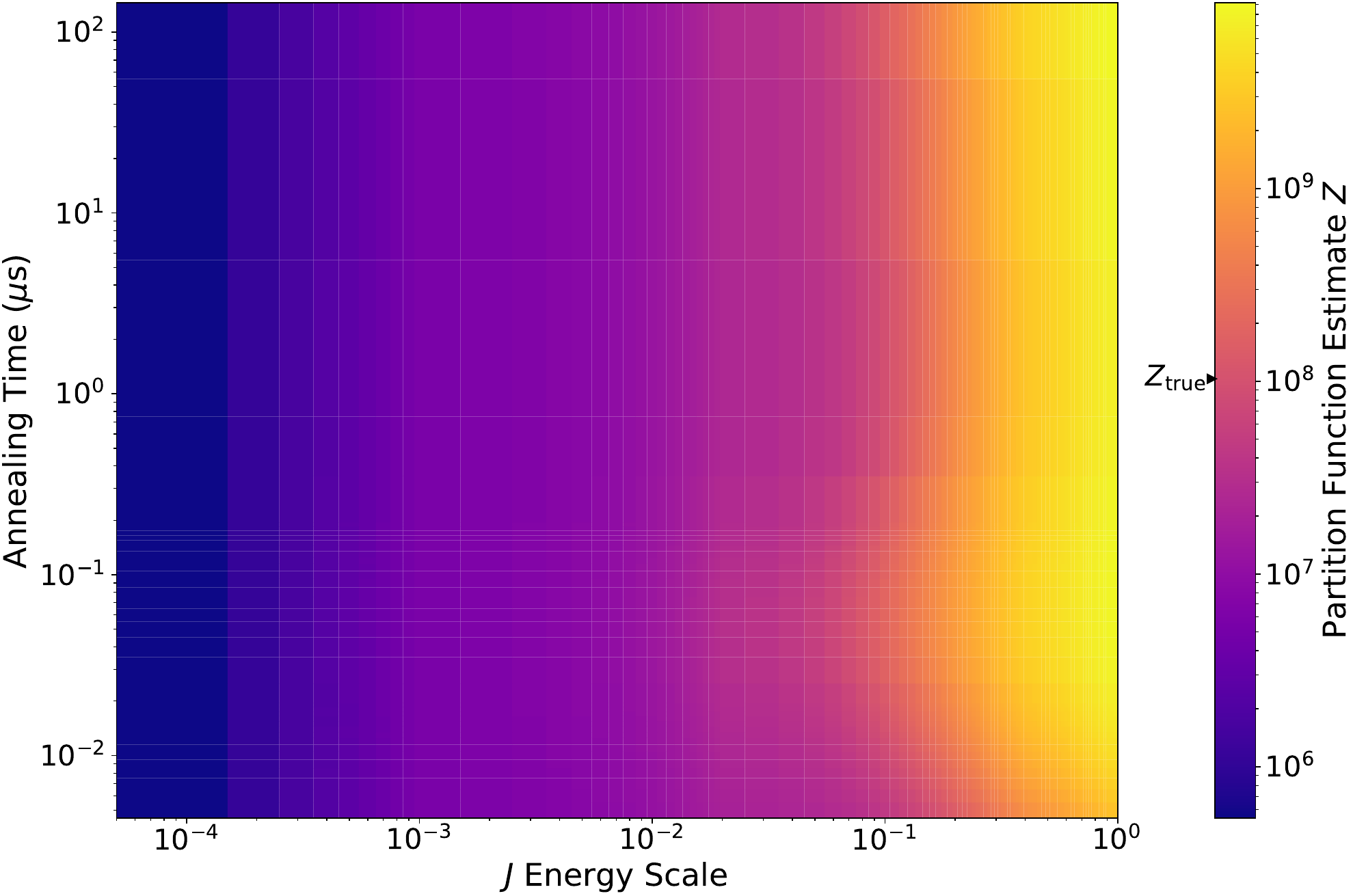}
        \caption{}
    \end{subfigure}%
    ~ 
    \begin{subfigure}[t]{0.5\textwidth}
        \centering
        \includegraphics[height = 2.3 in, width=\textwidth, keepaspectratio]{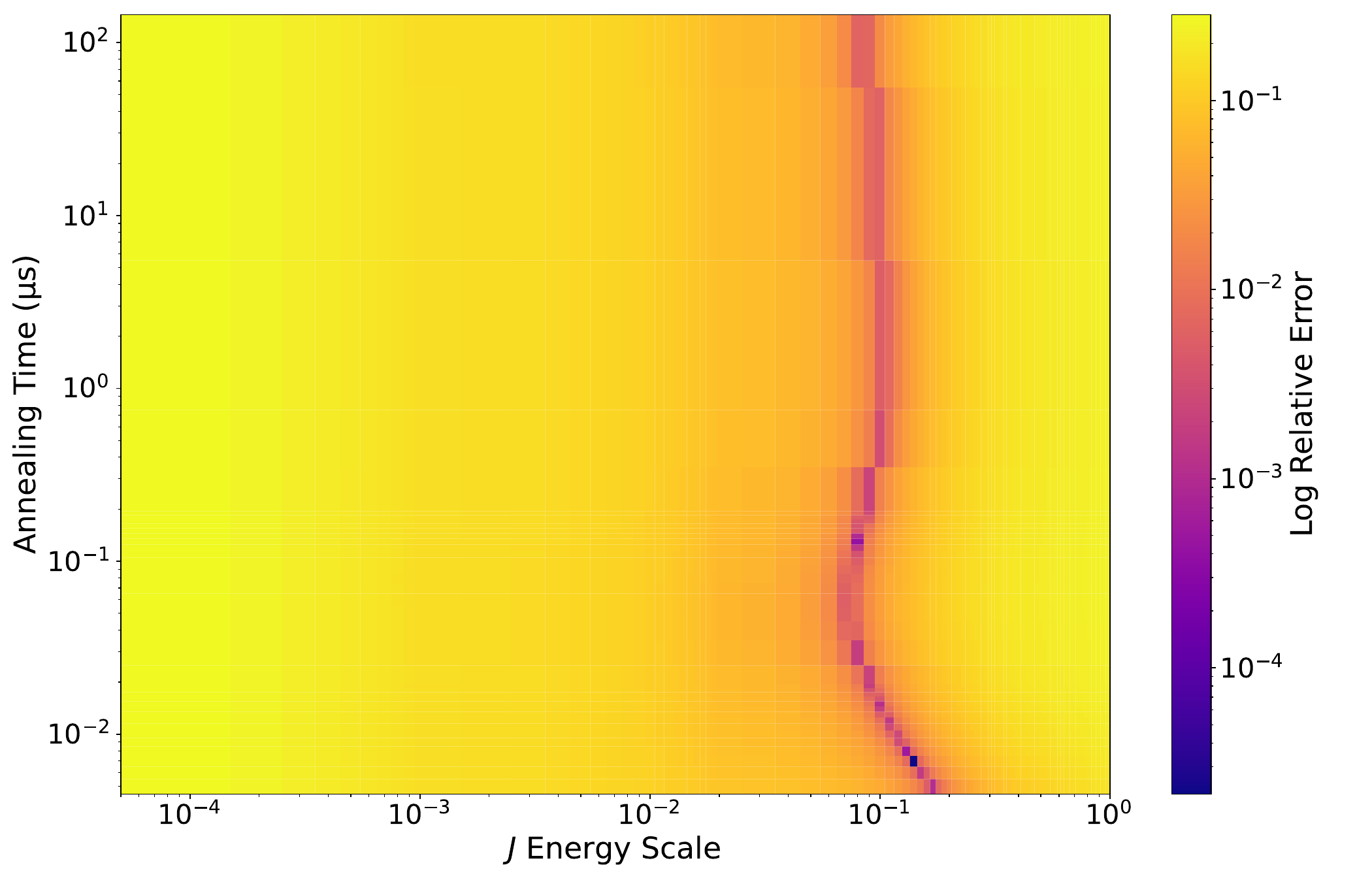}
        \caption{}
    \end{subfigure}
    ~
    \begin{subfigure}[t]{0.5\textwidth}
        \centering
        \includegraphics[height = 2.3 in, width=\textwidth, keepaspectratio]{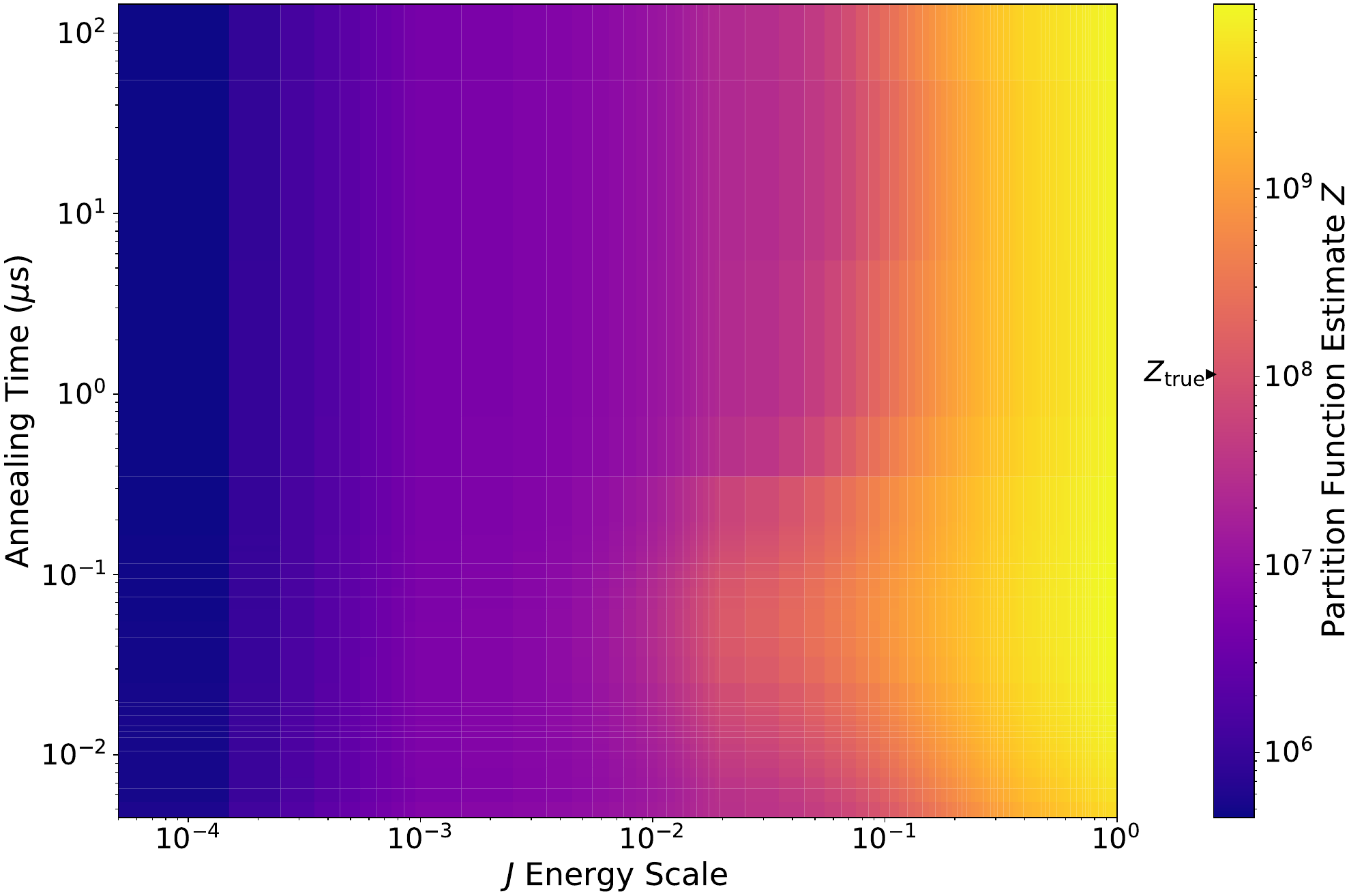}
        \caption{}
    \end{subfigure}%
    ~ 
    \begin{subfigure}[t]{0.5\textwidth}
        \centering
        \includegraphics[height = 2.3 in, width=\textwidth, keepaspectratio]{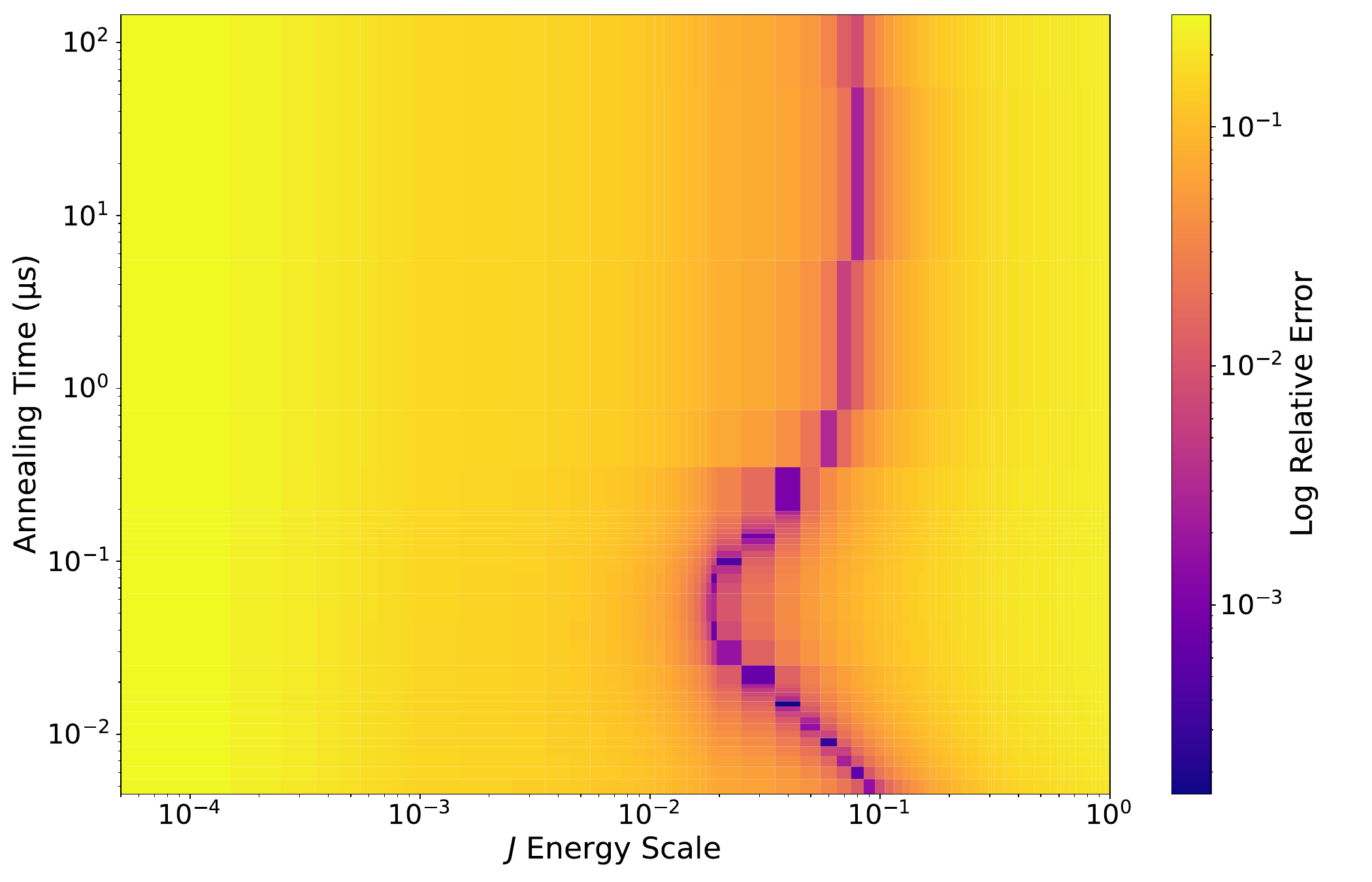}
        \caption{}
    \end{subfigure}
    \caption{Linear-ramp standard quantum annealing convergence with cumulative sampling over larger energy scales. For each fixed annealing time (log scale y-axis), the partition function estimation and log relative error are computed from an energy histogram that accumulates more samples as the $J$ energy scale increases along the x-axis for the \texttt{Advantage\_system4.1} processor (a,b) and \texttt{Advantage2\_system1.9} (c,d). The colormap encodes log relative error (b,d) and the actual partition function value estimate (a,c), where the true value is marked in the legend. The blue line traced out in in panels b,d define the lowest error-rate sampling for this particular protocol, which shows that there is a clear minimum -- any additional higher $J$ energy scales result in high error rates.  }
    \label{figure:Fig4_forward_annealing_cumulative_with_energy_scale}
\end{figure*}

\begin{figure*}[ht!]
    \centering
    \begin{subfigure}[t]{0.5\textwidth}
        \centering
        \includegraphics[height = 2.3 in, width=\textwidth, keepaspectratio]{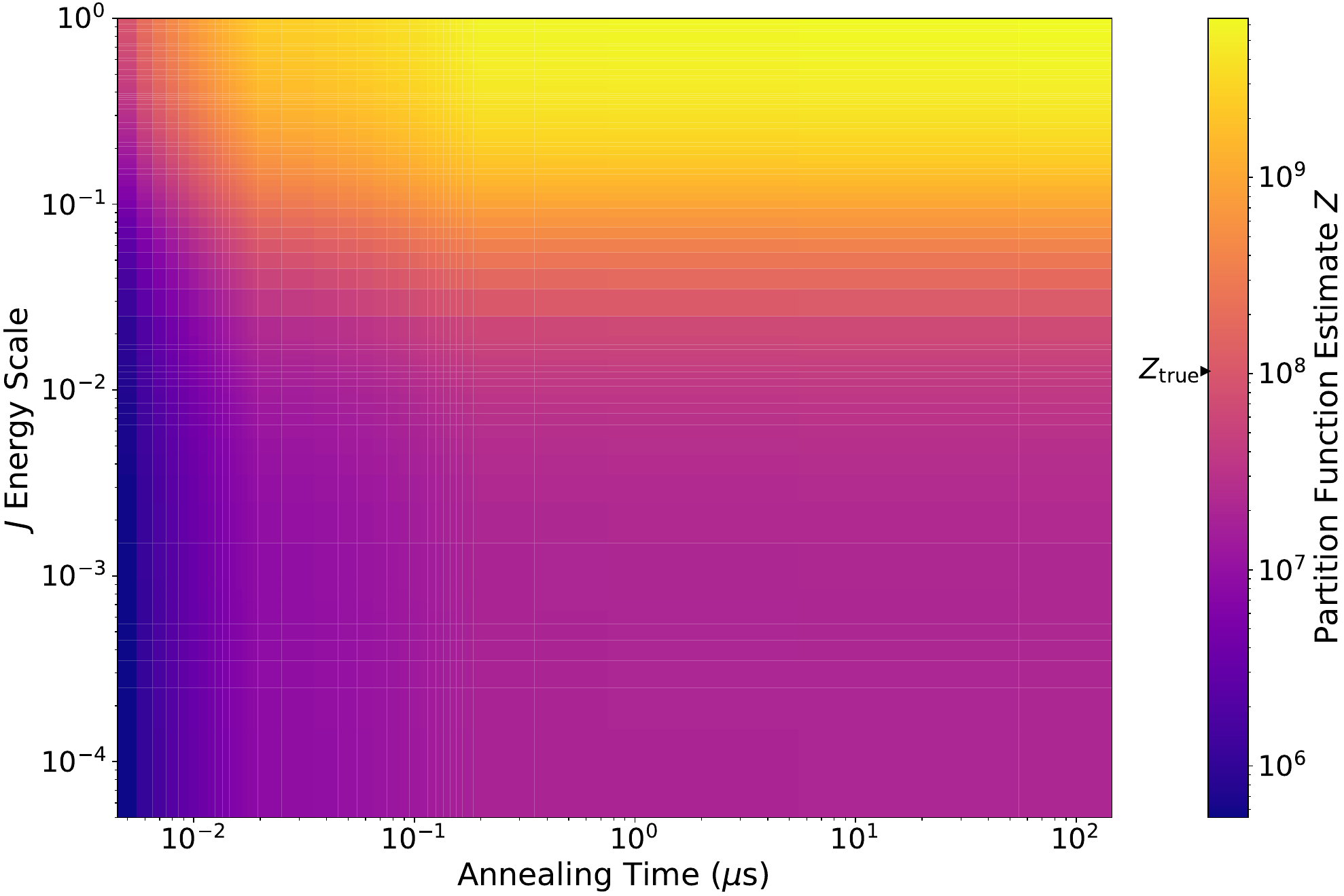}
        \caption{}
    \end{subfigure}%
    ~ 
    \begin{subfigure}[t]{0.5\textwidth}
        \centering
        \includegraphics[height = 2.3 in, width=\textwidth, keepaspectratio]{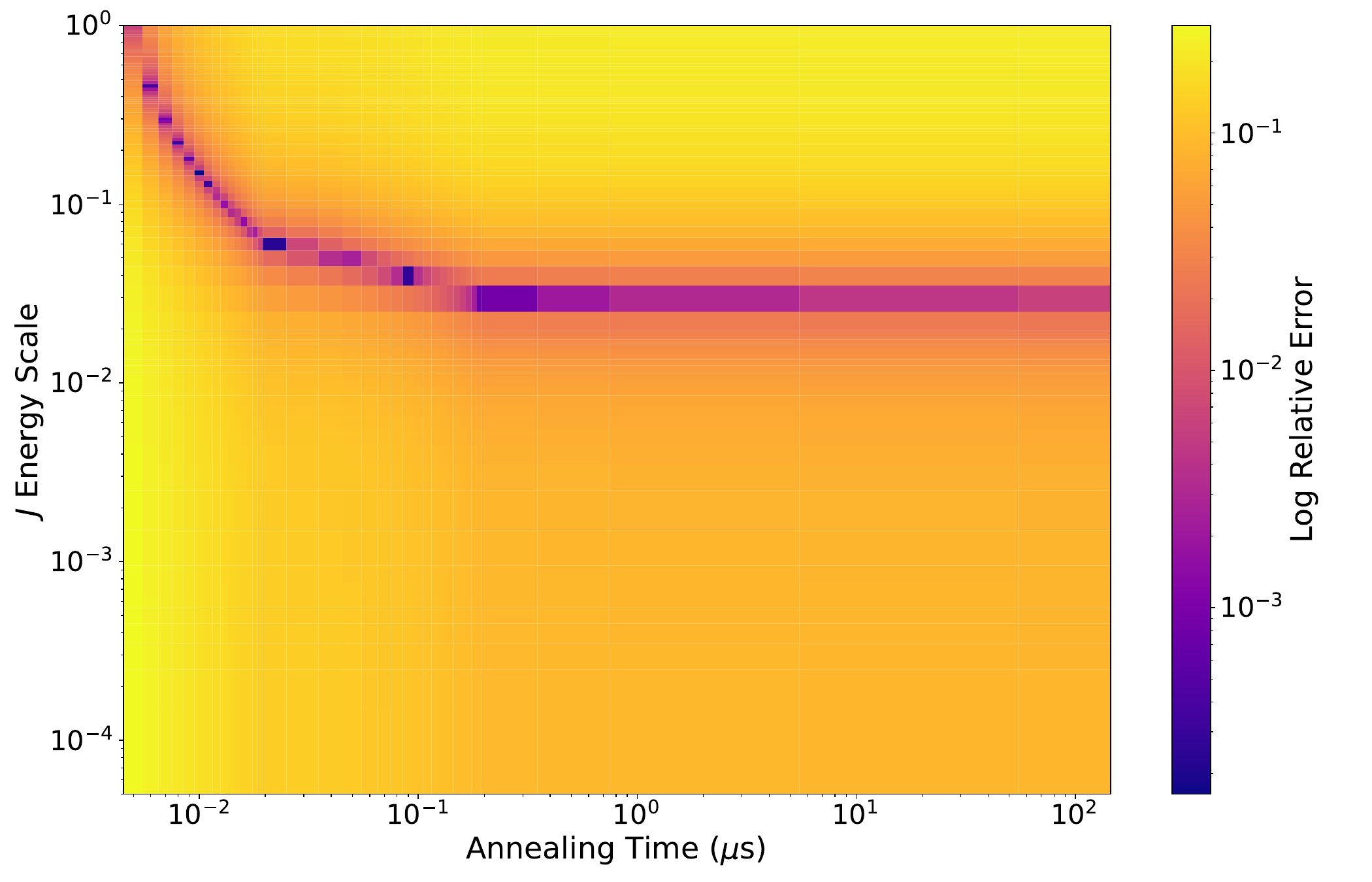}
        \caption{}
    \end{subfigure}
    ~
    \begin{subfigure}[t]{0.5\textwidth}
        \centering
        \includegraphics[height = 2.3 in, width=\textwidth, keepaspectratio]{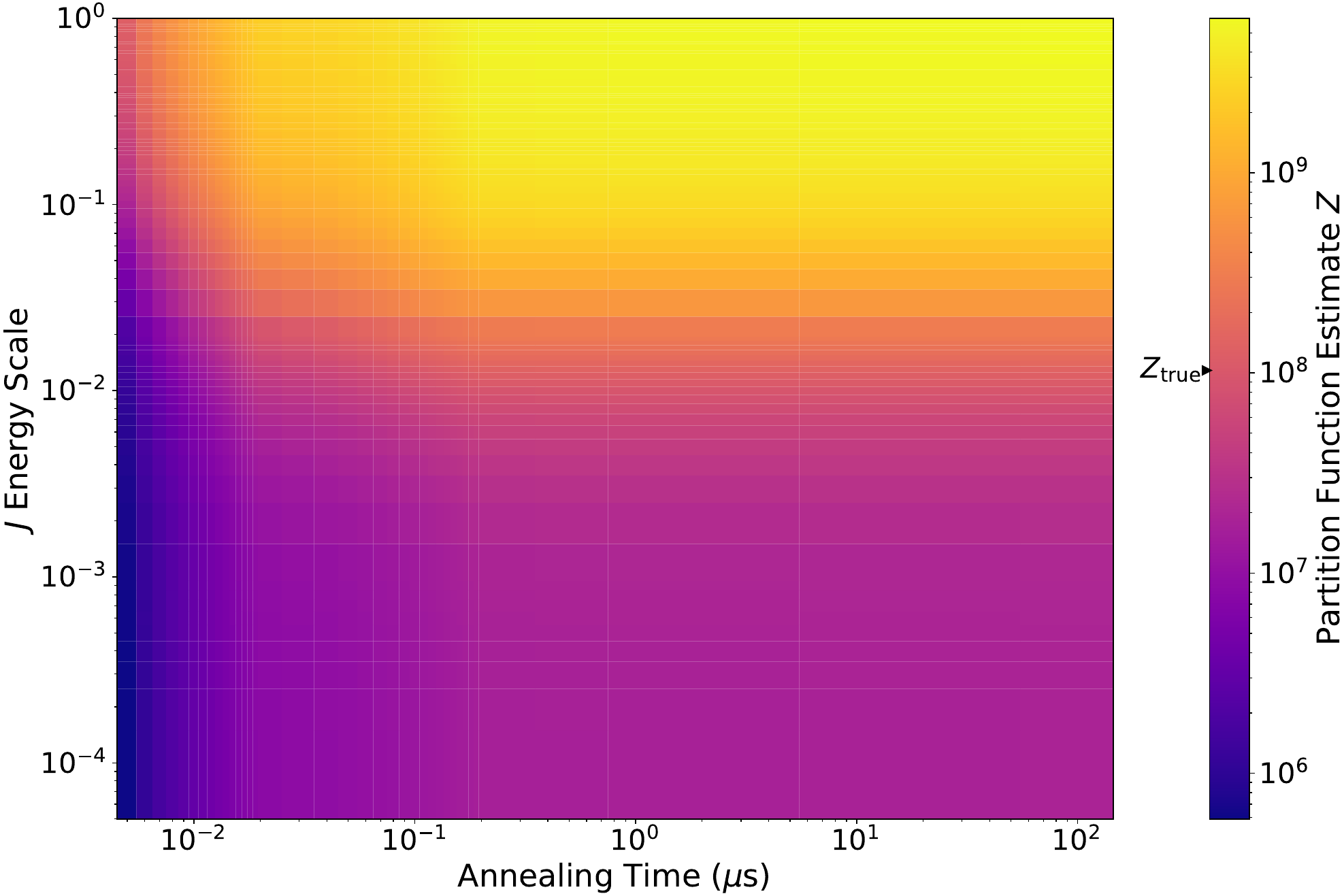}
        \caption{}
    \end{subfigure}%
    ~ 
    \begin{subfigure}[t]{0.5\textwidth}
        \centering
        \includegraphics[height = 2.3 in, width=\textwidth, keepaspectratio]{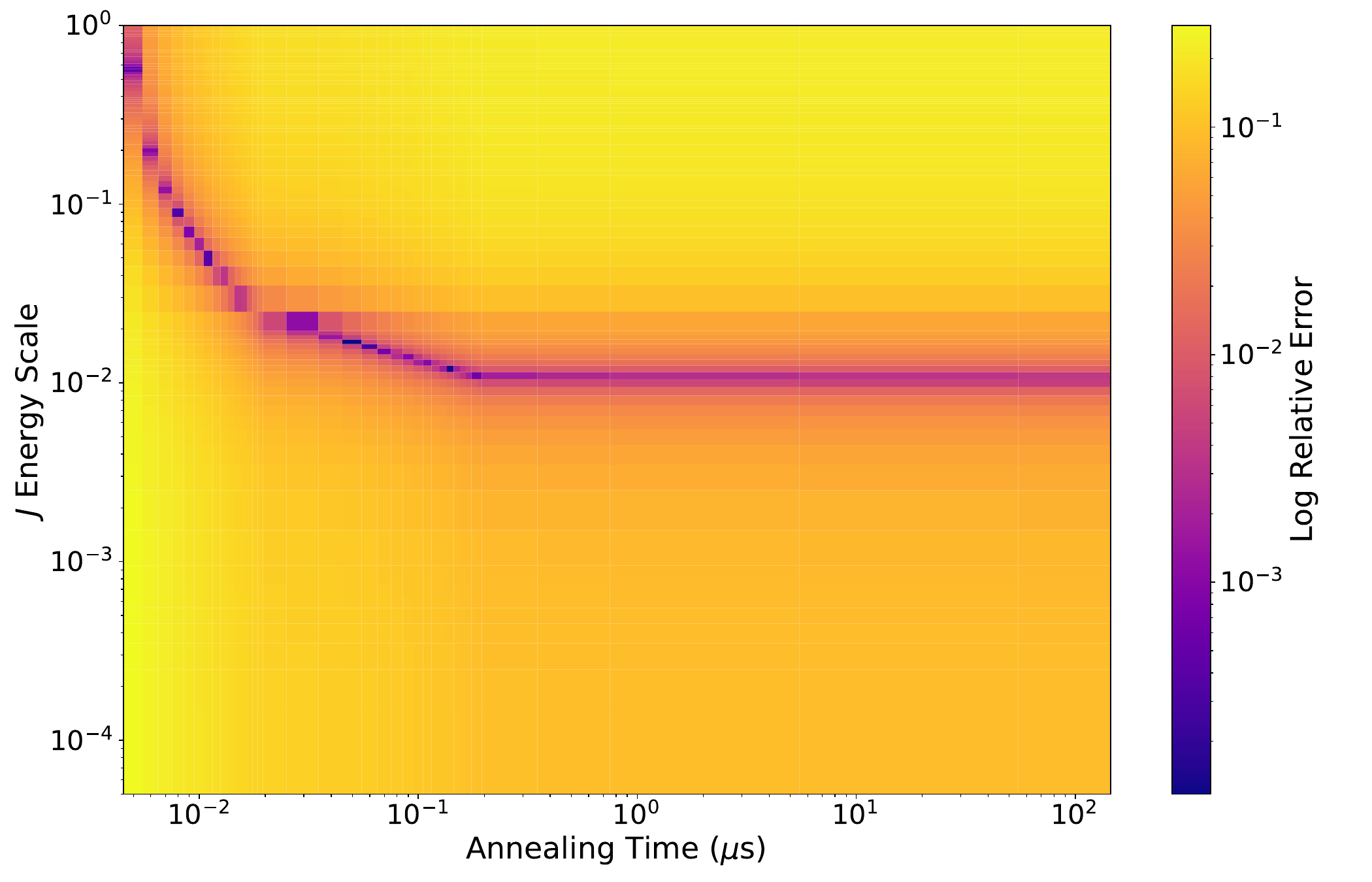}
        \caption{}
    \end{subfigure}
    \caption{Linear-ramp standard quantum annealing convergence with cumulative sampling over increasing annealing times. For each fixed energy scale (log scale y-axis), the partition function estimation and log relative error are computed from an energy histogram that accumulates more samples as the annealing time increases on the x-axis. Results from the \texttt{Advantage\_system4.1} processor (a,b) and \texttt{Advantage2\_system1.9} (c,d). The blue line traced in panels b,d show that the minimum error rate region of the parameter space converges at long annealing times.  }
    \label{figure:Fig5_forward_annealing_cumulative_with_annealing_time}
\end{figure*}

\begin{figure*}[ht!]
    \centering
    \begin{subfigure}[t]{0.5\textwidth}
        \centering
        \includegraphics[height = 2.3 in, width=\textwidth, keepaspectratio]{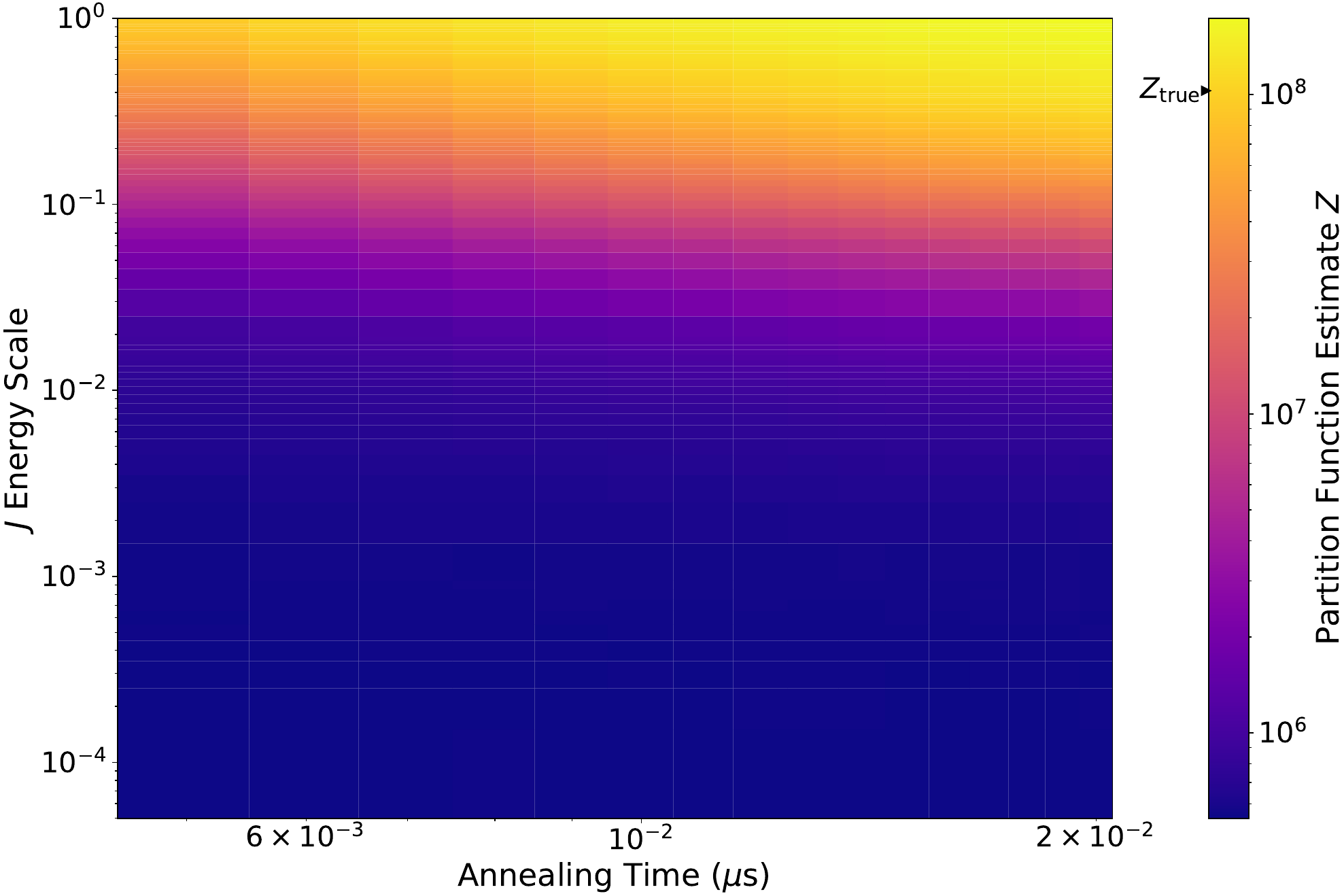}
        \caption{}
    \end{subfigure}%
    ~ 
    \begin{subfigure}[t]{0.5\textwidth}
        \centering
        \includegraphics[height = 2.3 in, width=\textwidth, keepaspectratio]{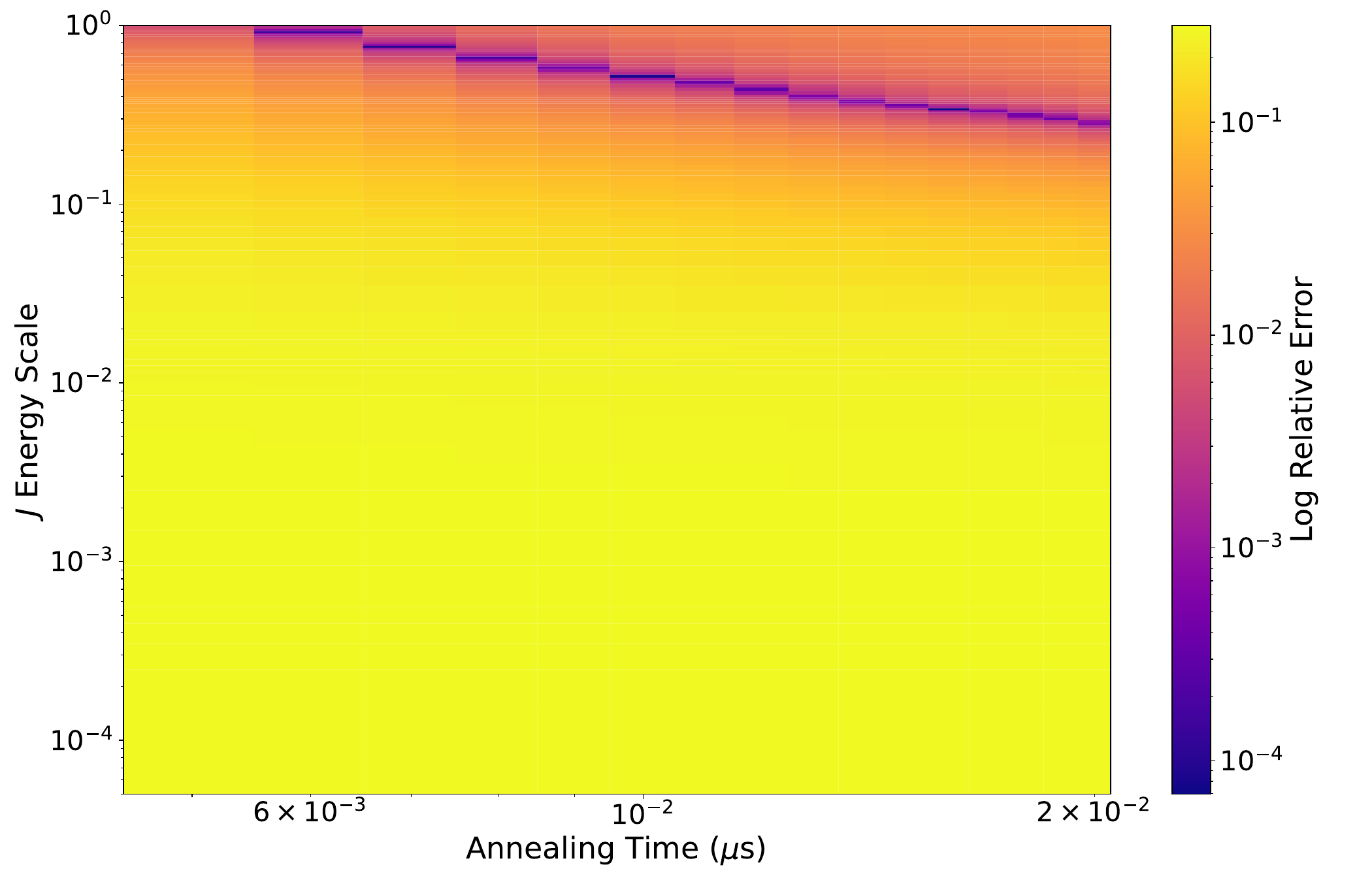}
        \caption{}
    \end{subfigure}
    ~
    \begin{subfigure}[t]{0.5\textwidth}
        \centering
        \includegraphics[height = 2.3 in, width=\textwidth, keepaspectratio]{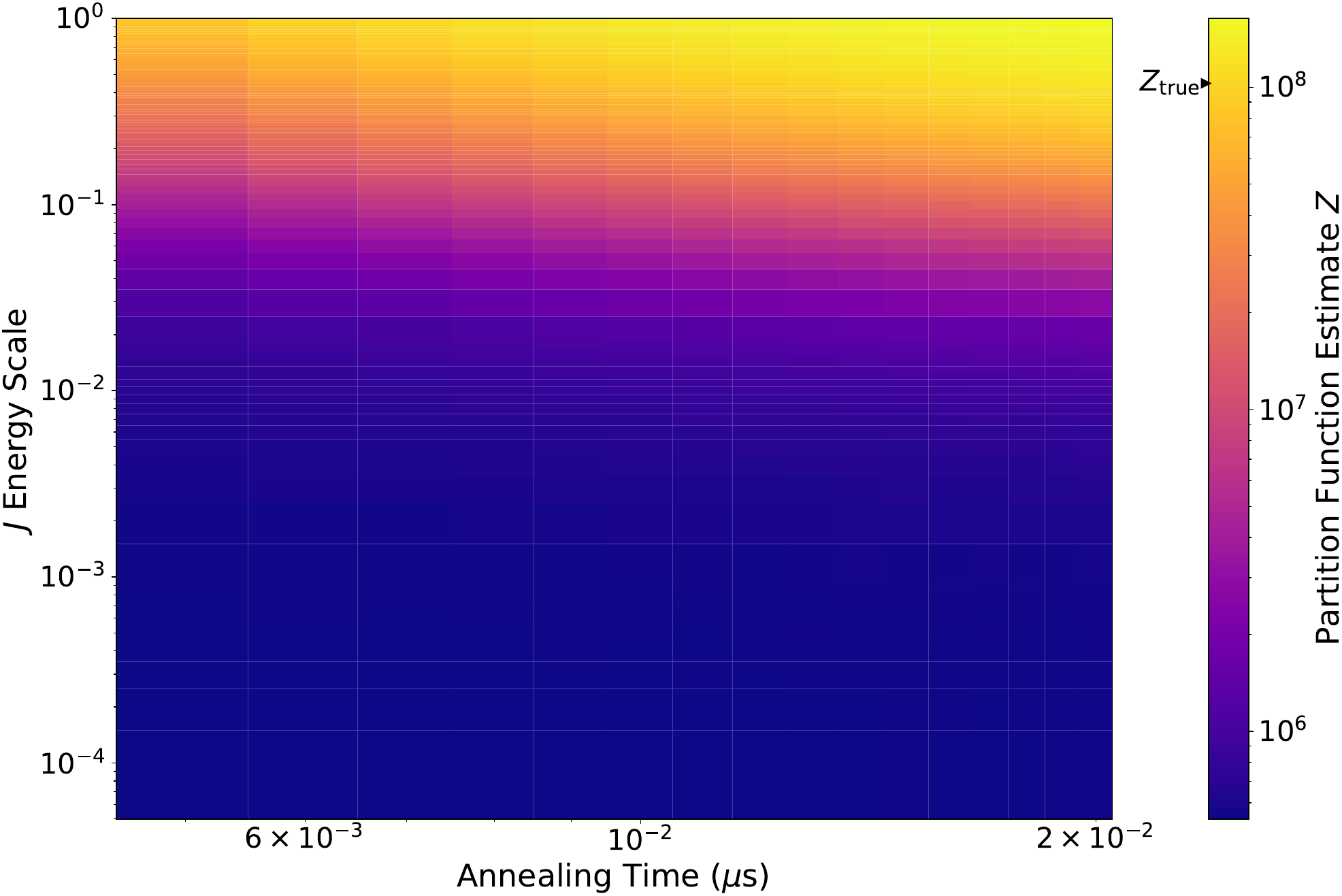}
        \caption{}
    \end{subfigure}%
    ~ 
    \begin{subfigure}[t]{0.5\textwidth}
        \centering
        \includegraphics[height = 2.3 in, width=\textwidth, keepaspectratio]{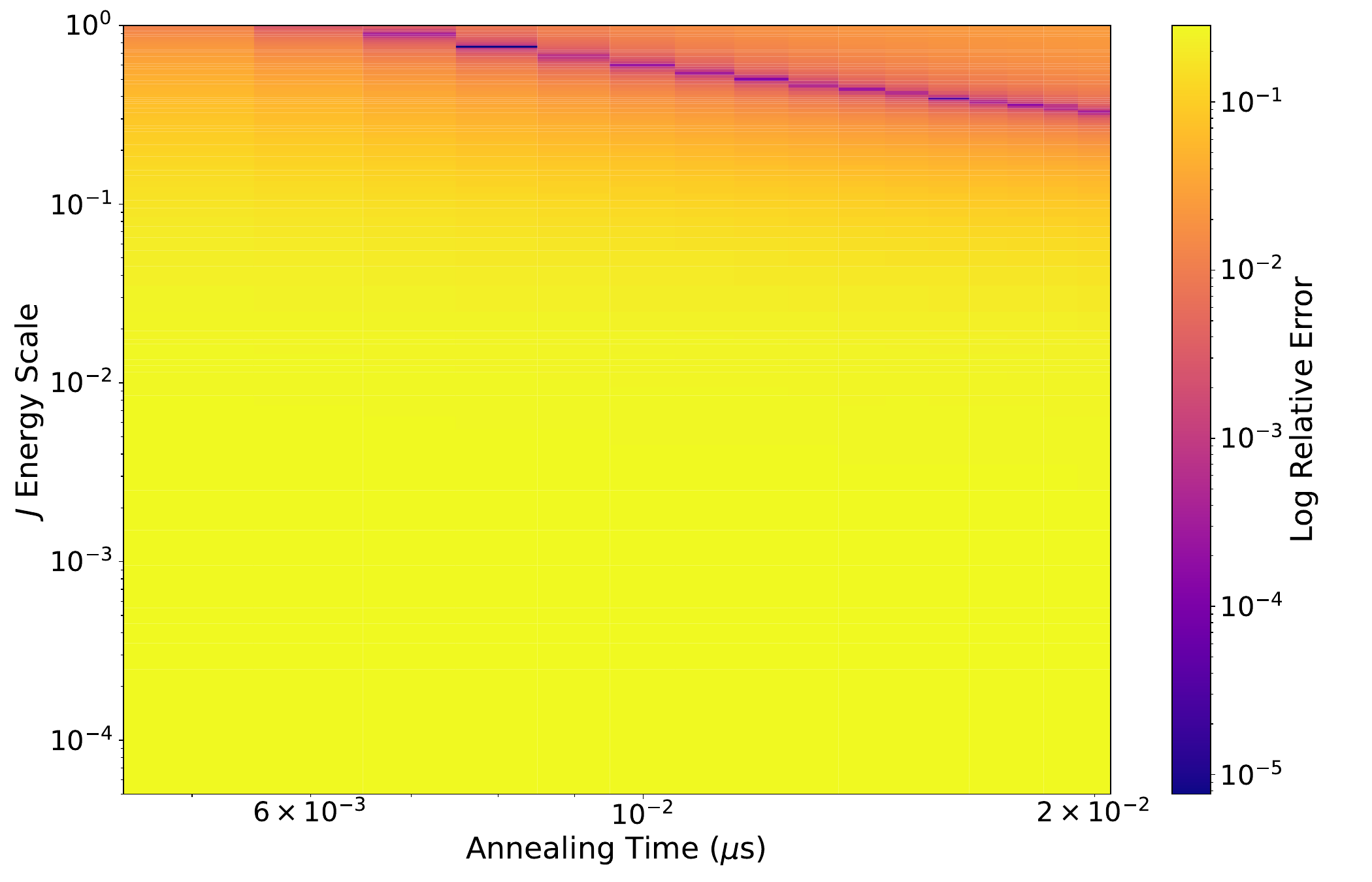}
        \caption{}
    \end{subfigure}
    ~
    \begin{subfigure}[t]{0.5\textwidth}
        \centering
        \includegraphics[height = 2.3 in, width=\textwidth, keepaspectratio]{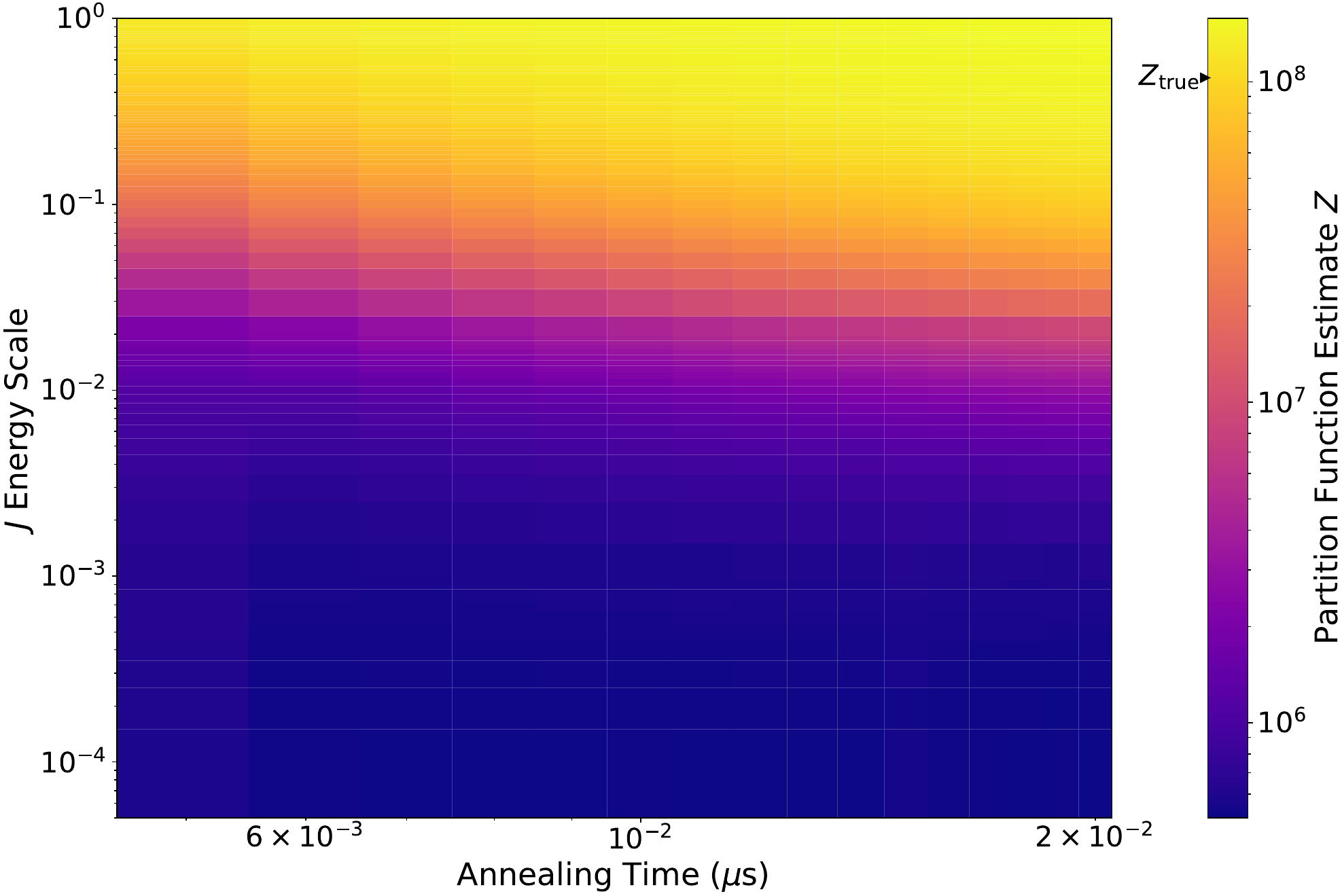}
        \caption{}
    \end{subfigure}%
    ~ 
    \begin{subfigure}[t]{0.5\textwidth}
        \centering
        \includegraphics[height = 2.3 in, width=\textwidth, keepaspectratio]{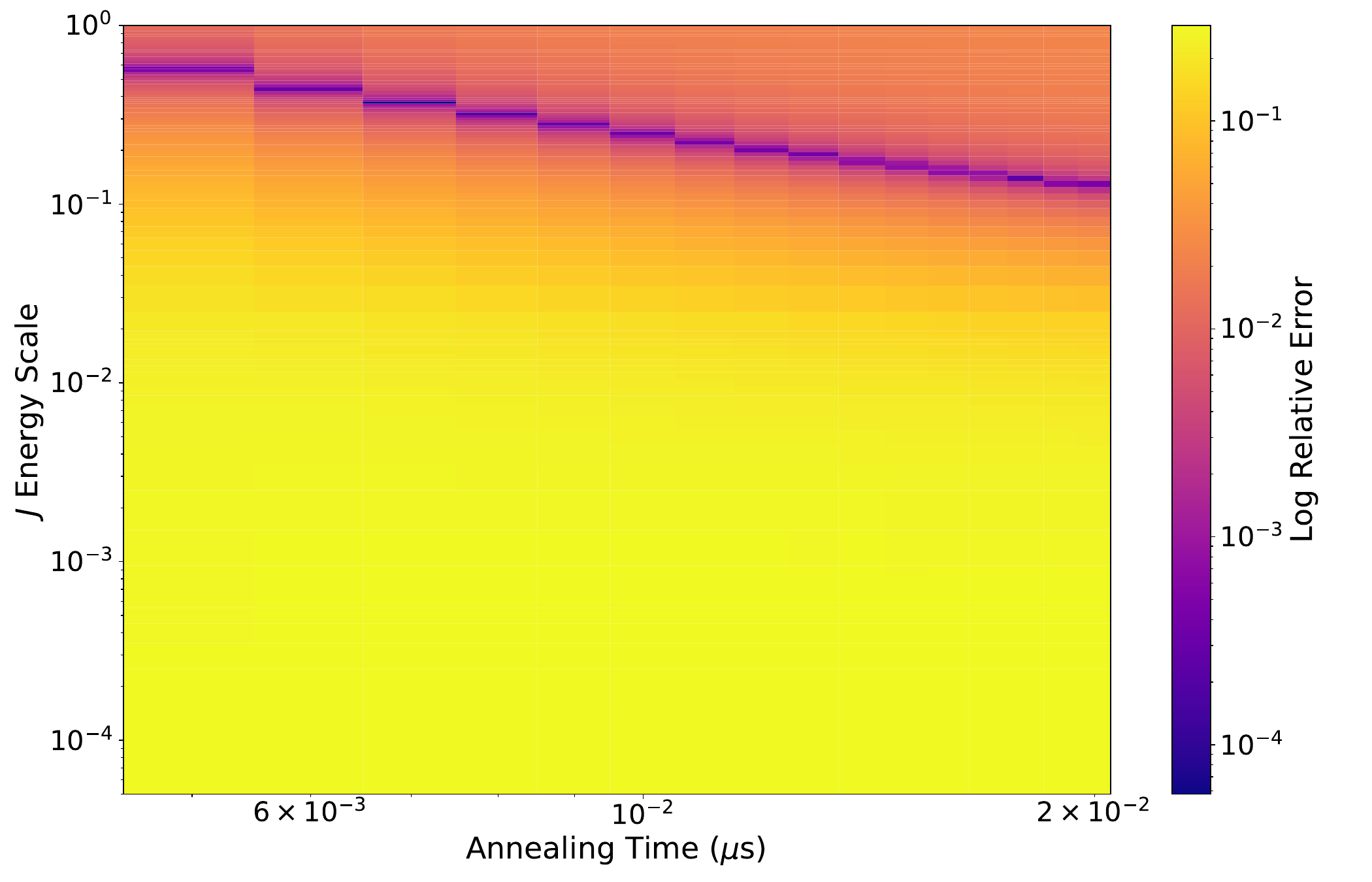}
        \caption{}
    \end{subfigure}
    \caption{Standard linear-ramp quantum annealing partition function estimation (non-cumulative sample accumulation on any axis) with fast annealing times ($5-20~\mathrm{ns}$) a-b: \texttt{Advantage\_system4.1}; c-d \texttt{Advantage\_system6.4}; e-f \texttt{Advantage2\_system1.9}. Each region of the grid corresponds to a total measured spin configuration count of between $178{,}000$ samples and $146{,}000$ samples, depending on the QA processor. }
    \label{figure:fast_annealing_non_cumulative_sampling}
\end{figure*}

\begin{figure*}[ht!]
    \centering
    \begin{subfigure}[t]{0.5\textwidth}
        \centering
        \includegraphics[height = 2.3 in, width=\textwidth, keepaspectratio]{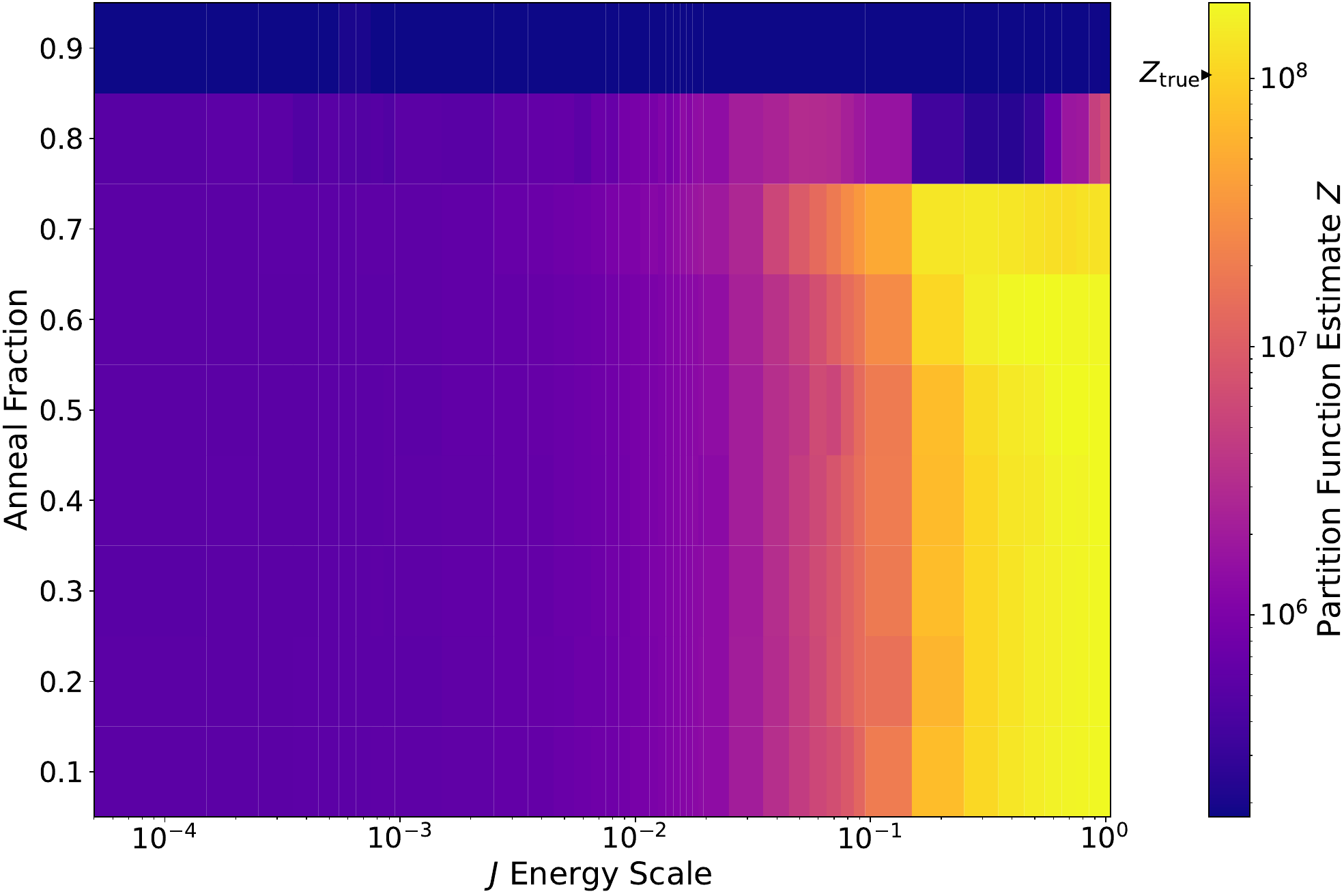}
        \caption{}
    \end{subfigure}%
    ~ 
    \begin{subfigure}[t]{0.5\textwidth}
        \centering
        \includegraphics[height = 2.3 in, width=\textwidth, keepaspectratio]{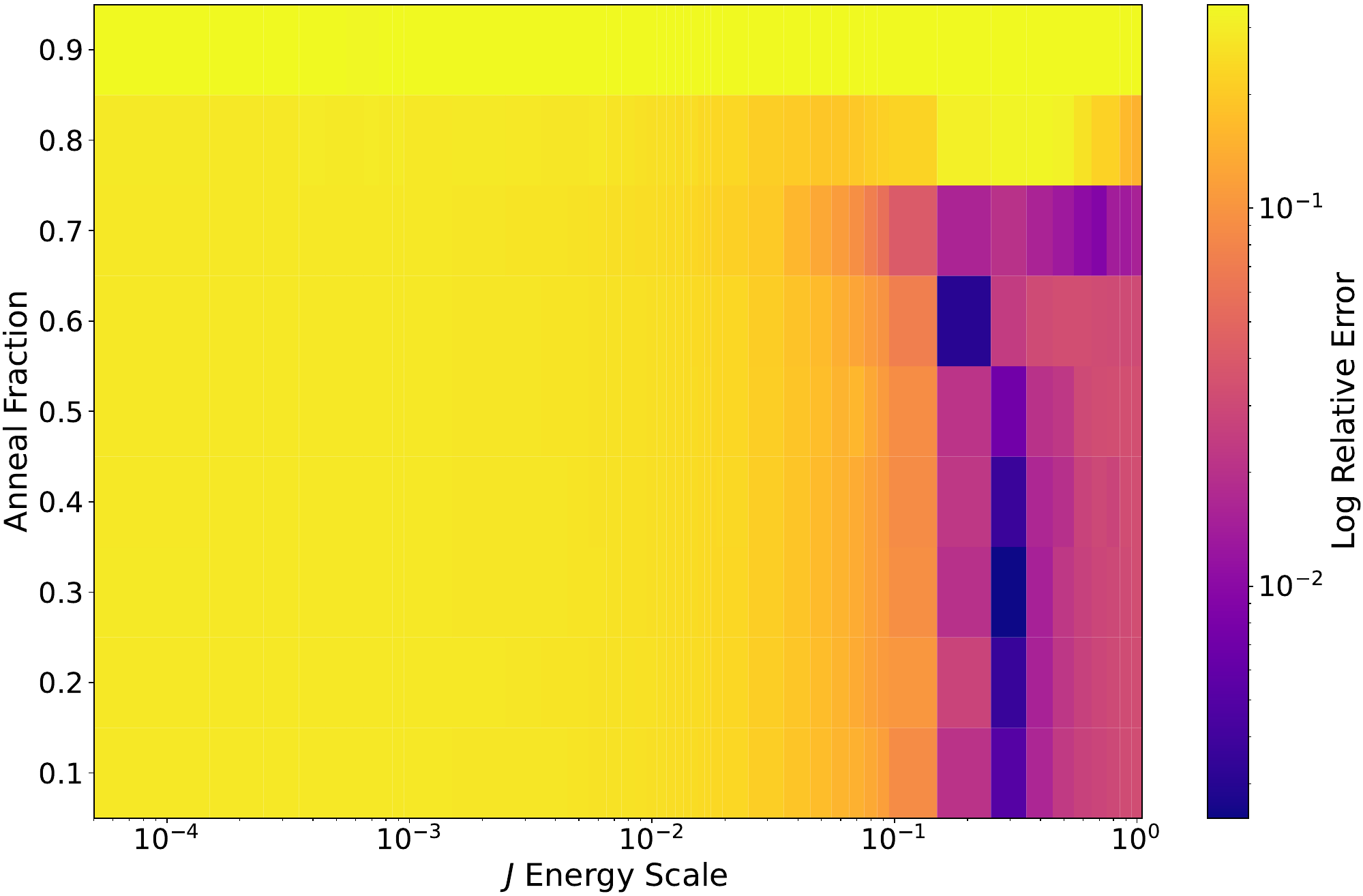}
        \caption{}
    \end{subfigure}
    ~
    \begin{subfigure}[t]{0.5\textwidth}
        \centering
        \includegraphics[height = 2.3 in, width=\textwidth, keepaspectratio]{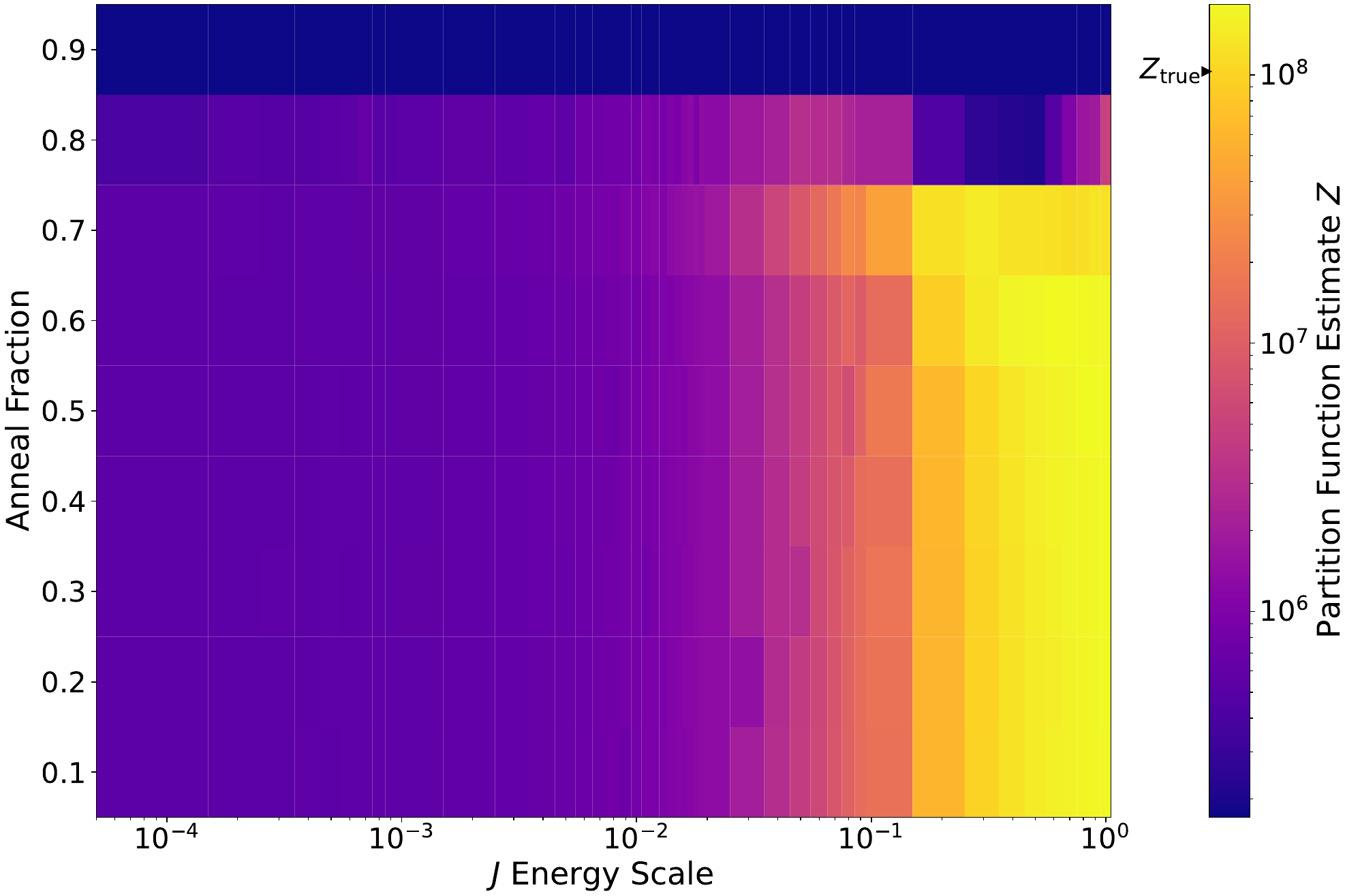}
        \caption{}
    \end{subfigure}%
    ~ 
    \begin{subfigure}[t]{0.5\textwidth}
        \centering
        \includegraphics[height = 2.3 in, width=\textwidth, keepaspectratio]{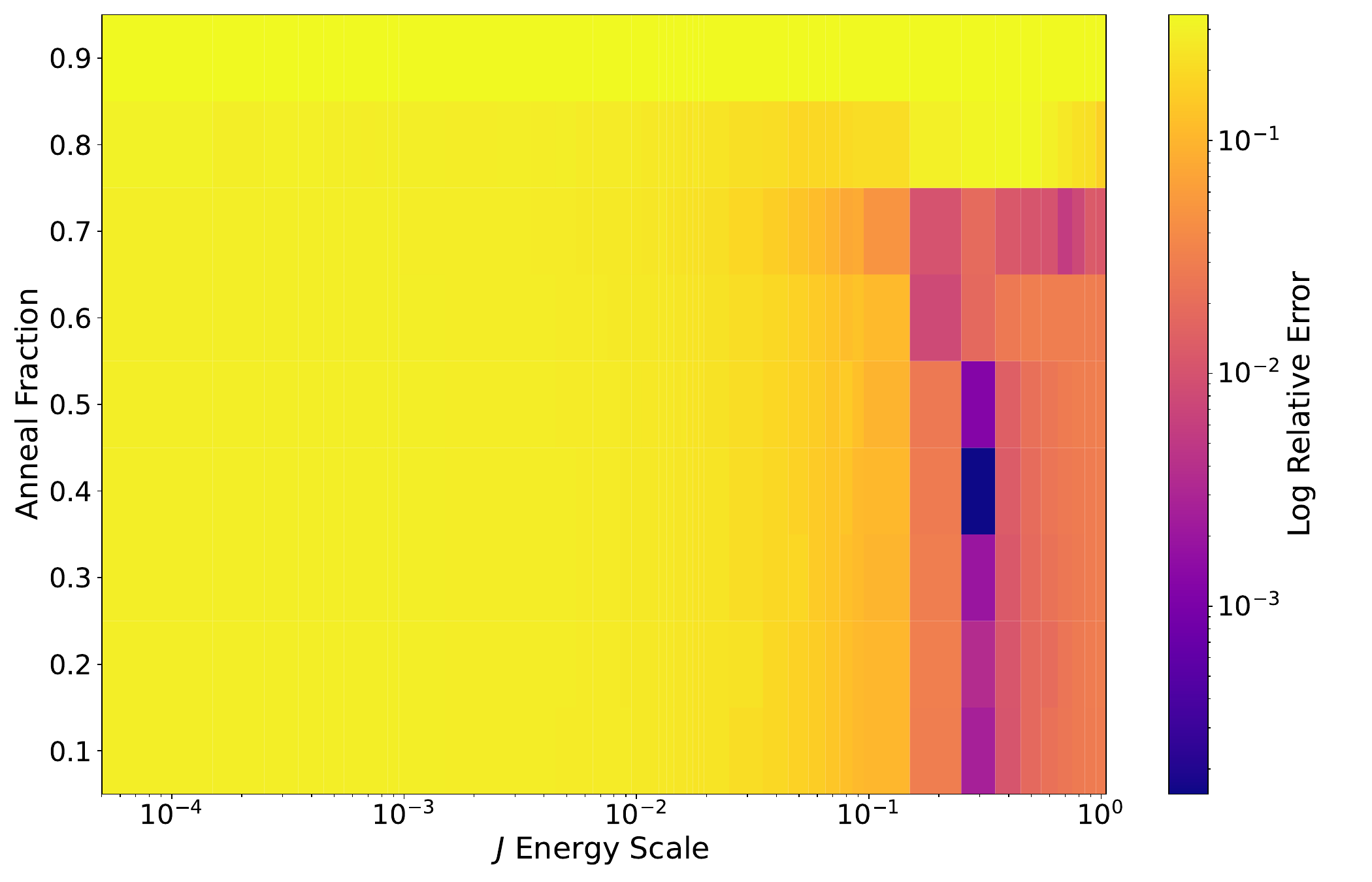}
        \caption{}
    \end{subfigure}
    ~
    \begin{subfigure}[t]{0.5\textwidth}
        \centering
        \includegraphics[height = 2.3 in, width=\textwidth, keepaspectratio]{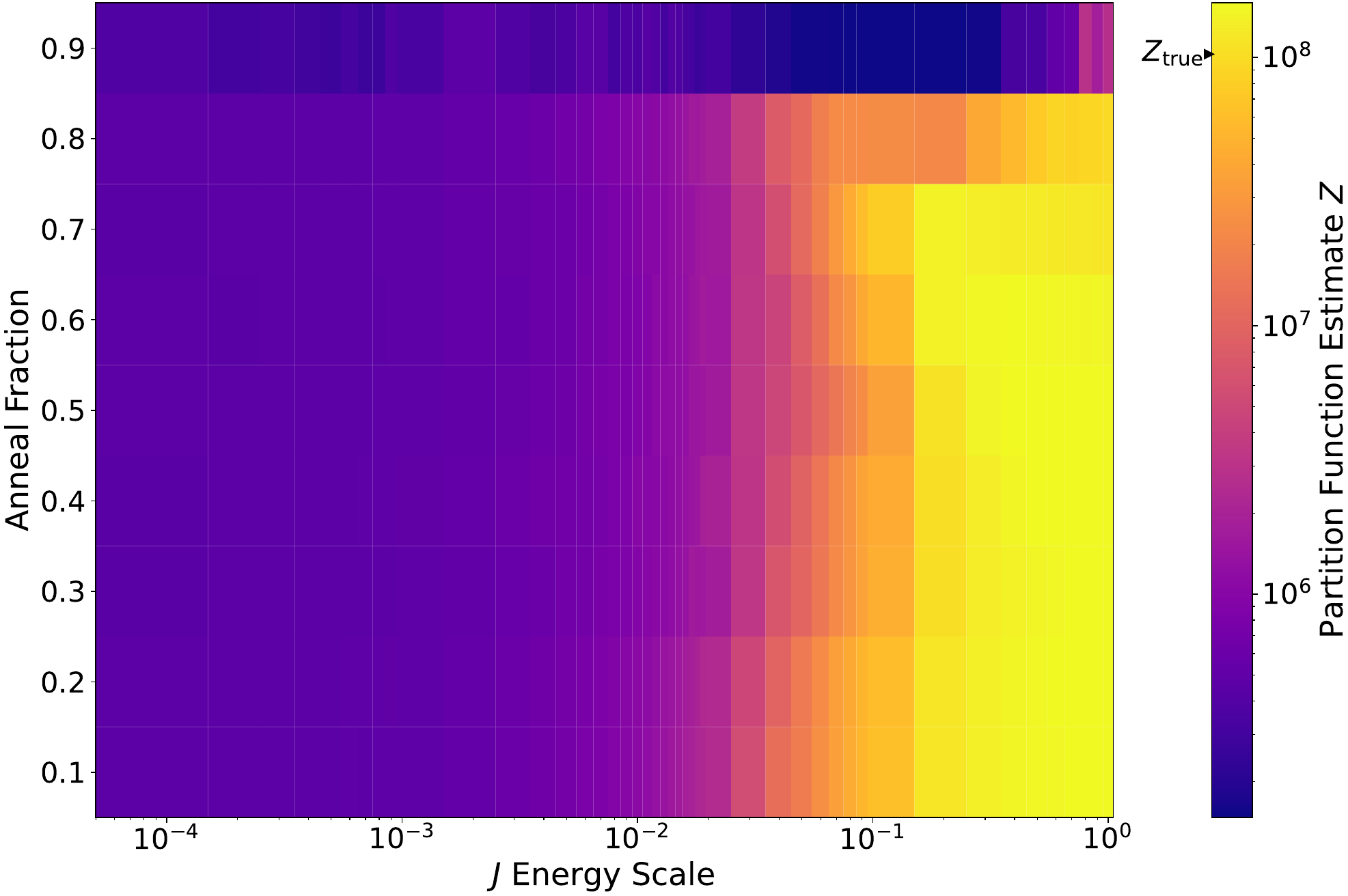}
        \caption{}
    \end{subfigure}%
    ~ 
    \begin{subfigure}[t]{0.5\textwidth}
        \centering
        \includegraphics[height = 2.3 in, width=\textwidth, keepaspectratio]{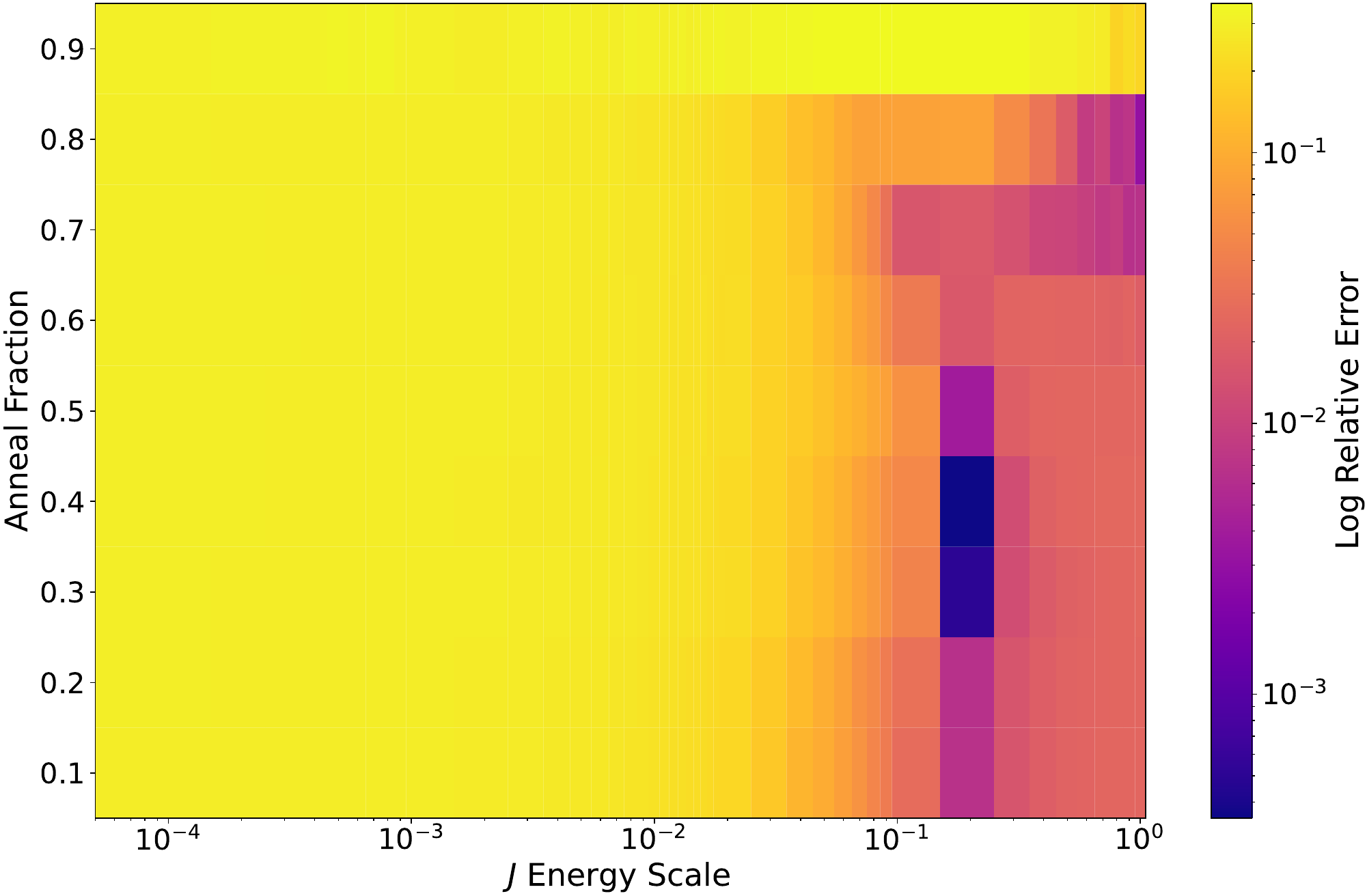}
        \caption{}
    \end{subfigure}
    \caption{Reverse quantum annealing error rate analysis with annealing time of $100 ~\mu\mathrm{s}$ on \texttt{Advantage\_system4.1} (a,b), \texttt{Advantage\_system6.4} (c,d), and \texttt{Advantage2\_system1.9} (e,f). The data represented by each region of the colormap is a distribution defined by the samples drawn from every parallel embedding for reverse annealing iterations up to $1{,}000$; in particular the distributions are not cumulative across the different analog parameters.  }
    \label{figure:reverse_annealing_non_cumulative_error_rates_100ms}
\end{figure*}

%%%%%%%%%%%%%%%%%%%%%%%%%%%%%%%%%%%%%%%%%%%%%%%%%%%%%%%
%%%%%%%%%%%%%%%%%%%%%%%%%%%%%%%%%%%%%%%%%%%%%%%%%%%%%%%
\section{Results}
\label{section:results}
%%%%%%%%%%%%%%%%%%%%%%%%%%%%%%%%%%%%%%%%%%%%%%%%%%%%%%%
%%%%%%%%%%%%%%%%%%%%%%%%%%%%%%%%%%%%%%%%%%%%%%%%%%%%%%%

The core aim of our study is to develop quantum algorithms, run on current D-Wave quantum annealing processors, for sampling good estimates of the partition function of an Ising model where we know ground-truth. The Ising model we study is a specific disordered hardware-lattice defined model, see Appendix~\ref{appendix:parallel_embeddings_plots}, as a test case where ground-truth is tractably computable. For D-Wave QPU sampling, a priori there is not a clear set of parameters, either from the literature or ground-truth numerical computations of the quantum annealing process, which would work especially well for this task -- broadly, the primary intuitive principle is that thermal sampling at several different temperatures should, combined, produce a representative distribution. Longer annealing times and strong $J$ coupling we expect to produce low-temperature sampling, and therefore either shorter annealing times or weaker $J$ coupling is expected to work better for this application, but there is no clear way a priori to obtain \emph{exact} physical parameters for these experiments on hardware. Therefore, we first turn to large parameter space experiments on the D-Wave QPUs, and empirically determine which regions work well and which do not. All quantitative experimental results we report are of the partition function of the Ising model at $T=4$. 

The first natural question we examine is a black-box search over the physical analog parameters to determine which parameters seem to work well. To this end, we utilize a standard machine learning approach of a genetic algorithm where the tunable parameters are the D-Wave hardware analog parameters. The genetic algorithm is penalized for high sample counts, favoring a sparse set of parameters and a smaller number of samples. Fig.~\ref{figure:Fig3_classical_estimation_algorithms_and_best_DWave_comparison} reports the sampling error rates, and sample counts, for these optimized parameters, which shows that the D-Wave QPU sampling can indeed carry out this task reasonably well. A natural question is how this sampling compares to the standard classical sampling algorithms, namely Wang-Landau with Monte Carlo updates and MCMC based MHR, and this is what Fig.~\ref{figure:Fig3_classical_estimation_algorithms_and_best_DWave_comparison} reports in panels e and f. As a reference baseline, we also compare against uniform random configuration sampling, to compute a DoS estimate, for $1000$ and $1{,}000{,}000$ samples -- these error rates are the absolute worst-case sampling. Interestingly, as adaptive convergence-based algorithms, the two classical sampling methods have strong oscillations when the number of Monte Carlo steps is small, and then at sufficiently large simulation steps the simulations converge and perform better than random sampling. The optimized quantum annealing parameters in Fig.~\ref{figure:Fig3_classical_estimation_algorithms_and_best_DWave_comparison} are detailed comprehensively in Appendix~\ref{appendix:GA_algo_parameters}, but we briefly describe them here. The forward annealing \texttt{Advantage2\_system1.9} parameters are $5$ nanoseconds with several different energy scales ranging from $J=0.0001$ to $J=0.92$, with only $75$ samples for each parameter combination being used -- resulting in a total of $317{,}550$ samples ($45$\% unique sample rate). The \texttt{Advantage\_system6.4} forward annealing results have a similar set of parameters, with a single annealing time of $0.11 \mu s$ and energy scales ranging from $0.0001$ to $0.9$, with $84$ samples per parameter and a total of $344{,}736$ samples ($60$\% unique samples). The \texttt{Advantage\_system4.1} parameters were similar as well, but instead used more annealing times ($0.05, 0.17, 1 \mu s$) and J energy scales of $0.017, 0.09, 0.1, 0.9$, and a total sample count of $305{,}448$ ($57$\% unique samples). These results give a good intuitive picture that by sweeping over a range of parameters, typically several different $J$ energy scales, good partition function estimates can be obtained -- the key mechanism behind this is collecting distributions at very different effective sampling temperatures. And this shows that an initial black-box machine learning approach can find a parameter set that works well on a small ``training'' instance.

A natural subsequent question is to find more general parameter range \emph{sweeps} over the energy spectrum and determine how well these work as partition function estimates: in particular if we start from higher temperatures, and then cumulatively build a distribution (with more samples being added to our $g(E)$ estimate as more parameters explored) going to lower-temperature distributions. This is what Fig.~\ref{figure:Fig4_forward_annealing_cumulative_with_energy_scale} explores by fixing annealing time (y-axis) and then varying the $J$ coupling strength on the (x-axis) -- this means that as the x-axis goes to the right, more samples are being used to generate this density of states estimate. What this shows is that this cumulative effective temperature varying strategy does work; for example the \texttt{Advantage\_system4.1} QPU is able to achieve a lower than $10^{-4}$ logarithmic relative error. The reason that is works is that the partition function estimate begins at smaller energy scale under-estimating the true value, and then eventually begins over-estimating the true value, and therefore there is a crossover that occurs with the true partition function value. Fig.~\ref{figure:Fig5_forward_annealing_cumulative_with_annealing_time} shows a similar trend if we instead fix the $J$ energy scale and then increase the total annealing time duration starting at $5$ nanoseconds. Importantly, Fig.~\ref{figure:Fig4_forward_annealing_cumulative_with_energy_scale} and Fig.~\ref{figure:Fig5_forward_annealing_cumulative_with_annealing_time} succinctly demonstrate that both annealing time and $J$ energy scale work as parameters to tune the effective temperature sampling of the Ising model, and therefore one parameter can be fixed while the other is varied, and there will be a minimum error partition function estimation density of states obtained in certain regimes of that parameter space. The complete results for the third Pegasus-graph processor, over these same parameter-cumulative distributions, are given in Appendix~\ref{appendix:additional_results_varying_params_Pegasus_6.4}. Fig.~\ref{figure:Fig5_forward_annealing_cumulative_with_annealing_time} shows something specifically quite notable, which is that the error rate is quite low in panels b and d, where the sample count is small, and also the annealing times are very fast. Based on this observation, Fig.~\ref{figure:fast_annealing_non_cumulative_sampling} then considers these distributions non-cumulatively across different parameters, and focuses on $20$ nanosecond annealing times or faster -- and these results show even lower error rates, down to $10^{-5}$ logarithmic relative error, with even fewer samples than the parameters ranges we found in Fig.~\ref{figure:Fig3_classical_estimation_algorithms_and_best_DWave_comparison}. This parameter regime is specifically interesting because the QPU time used for each parameter setting is very small. The lowest error rate parameters in Fig.~\ref{figure:fast_annealing_non_cumulative_sampling} are as follows; \texttt{Advantage\_system4.1} with a $16$ nanosecond annealing time and $J$ scale of $0.34$ resulted in $6.96 \times 10^{-5}$ log relative error from $178{,}000$ samples and a QPU time of $0.270$ seconds. \texttt{Advantage\_system6.4} with a $8$ nanosecond annealing time and $J=0.76$ resulted in $7.64 \times 10^{-6}$ log relative error from $171{,}000$ samples and a QPU time of $0.209$ seconds. On the \texttt{Advantage2\_system1.9} processor, an annealing time of $7$ nanoseconds and a J coupling of $0.37$ results in an error rate of $5.78 \times 10^{-5}$ with $0.193$ seconds of QPU time used.

These parameter regimes are the lowest error rate results we have found. However, there is a noteworthy property of these sample distributions which is that although the energy level distributions are quite accurate, this is not because of a high degree of solution diversity. Instead, the proportion of unique samples is quite low. For these parameters, the percentage of unique measured spin configurations is $6.47\%, 5.60\%, 3.31\%$ for \texttt{Advantage\_system4.1}, \texttt{Advantage\_system6.4}, \texttt{Advantage2\_system1.9}, respectively. We hypothesize that degeneracy lifting in the closed-quantum system timescales of these fast anneals is causing this bias, although future study is required to determine whether this is the case. The core tradeoff being exploited in this low-error rate sampling is somewhat subtle, because ultimately the estimated $g(E)$ is not replicating the full energy spectrum, but also the very fast quenches are finding many low energy configurations. The sampling here is generating a density of states which have have many low-energies, but not strictly ground-states, and no energies that are in the middle of the energy spectrum. Appendix~\ref{appendix:density_of_states_histograms} examines this in detail. The performance shown in Fig.~\ref{figure:fast_annealing_non_cumulative_sampling} shows that using a relatively small number of samples to obtain ultimately what is a \emph{biased} estimated region of the energy spectrum, by selectively sampling in between the lowest energies and middle of the energy spectrum. The reason that this type of low-energy-biased distribution works as an estimator for this task is because the lower energy configurations contribute more to the partition function value. This parameter regime working well for partition function estimation is a somewhat counterintuitive empirical finding; it exploits a property of having high sample counts at intermediate low-temperature energies, thus allowing a total sample count to not be as large as for example what Wang-Landau produces, because the weighted contribution of the low-energy states dominates the partition function value. 

Finally, Fig.~\ref{figure:reverse_annealing_non_cumulative_error_rates_100ms} presents a sampling error rate analysis when using the iterated reverse annealing, or QEMC, approach. This, similar to the prior results, shows similar results, where there are some specific parameters (namely, stronger $J$ coupling, an anneal fractions $\approx 0.5-0.3$, can generate error rates that are less than $10^{-3}$ log relative error. Appendix~\ref{appendix:QEMC_convergence} investigates sampling error rate as a function of QEMC iterations, which shows that the iterated procedure converges to a fixed error rate. This shows that QEMC equilibration based methods can also generate reasonably good estimates of the partition function. Using QEMC with a shorter simulation time of $2 \mu$ s resulted in very similar results, which are reported in Appendix~\ref{appendix:RA_t2}. Thus, we have shown several different parameter regime sweeps, e.g., from high temperature to lower temperature sampling, which produces reasonably good partition function estimates, and is a consistent set of algorithm parameters based on a thermodynamic sampling perspective of the D-Wave hardware.

%%%%%%%%%%%%%%%%%%%%%%%%%%%%%%%%%%%%%%%%%%%%%%%%%%%%%%%
%%%%%%%%%%%%%%%%%%%%%%%%%%%%%%%%%%%%%%%%%%%%%%%%%%%%%%%
\section{Discussion and Conclusion}
\label{section:conclusion}
%%%%%%%%%%%%%%%%%%%%%%%%%%%%%%%%%%%%%%%%%%%%%%%%%%%%%%%
%%%%%%%%%%%%%%%%%%%%%%%%%%%%%%%%%%%%%%%%%%%%%%%%%%%%%%%

We have shown that D-Wave quantum annealers can operate as noisy samplers to approximate the partition function of an Ising model. The simulation methods of both direct sampling using forward quantum annealing, as well as iterated reverse annealing, both work for the task of cumulatively sampling a density of states distribution of the underlying Ising model. This approximate density of configurations for each energy level can then be used to estimate the partition function. This approach is very powerful because it allows partition function computation at arbitrary temperatures, even very low-temperatures, once the high-quality density of states energy level histogram is built. In other words, the tradeoff of the quantum annealing sampling approaches we describe, similar to WL and MHR of MCMC samples, is an initial computationally intensive up-front sampling, and then computations for arbitrary temperatures are subsequently very easy to compute. We show that these methods are comparable to the classical sampling heuristics of Wang-Landau and MHR with MCMC sampling. Crucially, this is a proof-of-concept study, where we demonstrate this type of sampling algorithm does work, and we provide several different parameter ranges and simulation techniques which can carry out this task on D-Wave hardware. Future research will examine to what extent this sampling accuracy scales for problems of very different problem sizes -- the ultimate question is whether techniques can be developed to verify if this scaling continues to work for Ising models with thousands of spins, e.g., the scale of the current D-Wave processors. The other relevant question is whether these methods work well for models with significant geometric frustration. Lastly, the fast anneals producing good estimates of the true partition function is an unexpected empirical result, and deserves more in-depth investigation. One characteristic that our results imply is that the very fast annealing-quenches produce classical probability distributions which are not necessarily bad estimates of classical Hamiltonian Gibbs distributions. This result showed that the estimation worked well, despite high levels of configuration measurement collisions (low number of unique samples), likely due to degeneracy lifting. Specifically, quantifying the effect of the transverse field would illuminate the quantum mechanisms of this particular sampling in a coherent analog processor, but this type of study requires Pauli X-basis measurement of the qubits in the hardware; current D-Wave processors only support Z-basis measurement.

These methods that we present, validated by extensive D-Wave hardware computations, are a novel type of equilibrium simulation capability for current analog quantum computers, specifically for statistical mechanics and computational physics tasks. We show parameter regimes for which this sampling is performed with a remarkably small amount of compute time, in large part due to parallel embedding as well as very fast annealing times.

%%%%%%%%%%%%%%%%%%%%%%%%%%%%%%%%%%%%%%%%%%%%%%%%%%%%%%%
%%%%%%%%%%%%%%%%%%%%%%%%%%%%%%%%%%%%%%%%%%%%%%%%%%%%%%%
\section*{Acknowledgments}
\label{sec:acknowledgments}
%%%%%%%%%%%%%%%%%%%%%%%%%%%%%%%%%%%%%%%%%%%%%%%%%%%%%%%
%%%%%%%%%%%%%%%%%%%%%%%%%%%%%%%%%%%%%%%%%%%%%%%%%%%%%%%

We thank Frank Barrows, Cristiano Nisoli, and Kipton Barros for helpful discussions. 
This work was supported by the U.S. Department of Energy through the Los Alamos National Laboratory (LANL). Los Alamos National Laboratory is operated by Triad National Security, LLC, for the National Nuclear Security Administration of U.S. Department of Energy (Contract No. 89233218CNA000001). The research presented in this article was supported by the Laboratory Directed Research and Development program of Los Alamos National Laboratory under project number 20240032DR. The work of T.L. was supported by the U.S. Department of Energy (DOE) through a quantum computing program sponsored by the Los Alamos National Laboratory (LANL) Information Science \& Technology Institute. This research used resources provided by the Los Alamos National Laboratory Institutional Computing Program, which is supported by the U.S. Department of Energy National Nuclear Security Administration under Contract No.~89233218CNA000001. The authors would also like to thank the New Mexico Consortium, under subcontract C2778, the Quantum Cloud Access Project (QCAP), for providing quantum computing resources. LANL report LA-UR-25-30601. 

\section*{Author Contributions}
T.L. collected all experimental data, executed all numerical simulations, created all figures, and wrote all computer code used in this study. E.P. conceived of the project, advised on experimental design, and drafted the initial manuscript. Both authors contributed to the writing of the manuscript, and both authors reviewed the final manuscript.

\section*{Code availability}
All code is available at \url{https://github.com/tvle2/qcss-partition}, and the datasets are available on Zenodo at \url{https://doi.org/10.5281/zenodo.18001652} and \url{https://doi.org/10.5281/zenodo.18000560}.

\section*{Competing Interests}
The authors declare no competing interests.  

\onecolumngrid

\appendix

%%%%%%%%%%%%%%%%%%%%%%%%%%%%%%%%%%%%%%%%%%%%%%%%%%%%%%%
\section{Additional Cumulative Parameter Varying Partition Function Estimation Results}
\label{appendix:additional_results_varying_params_Pegasus_6.4}
%%%%%%%%%%%%%%%%%%%%%%%%%%%%%%%%%%%%%%%%%%%%%%%%%%%%%%%

\begin{figure}[ht!]
    \centering
    \includegraphics[width=0.49\textwidth]{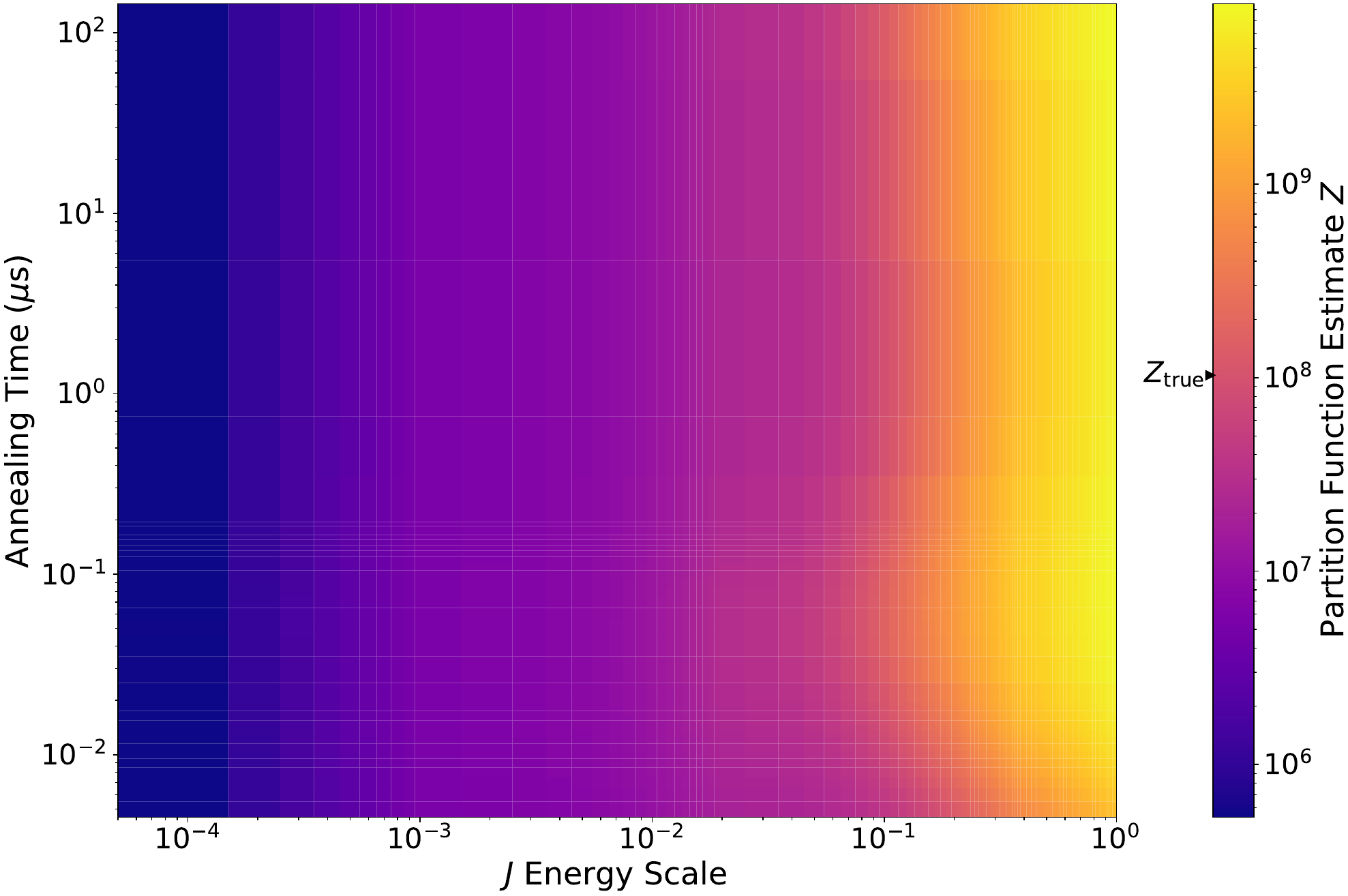}
    \includegraphics[width=0.49\textwidth]{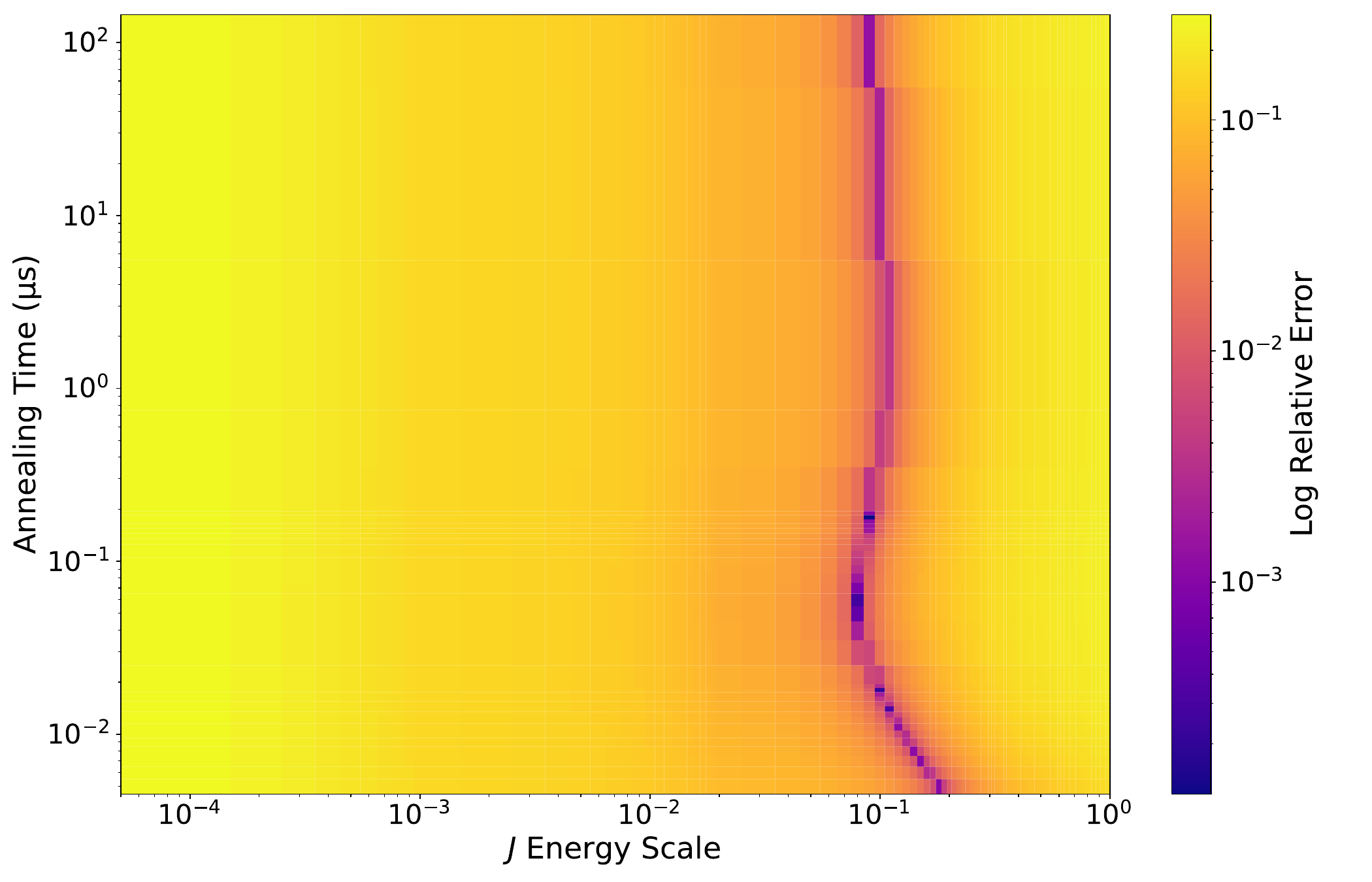}
    \caption{Linear-ramp standard quantum annealing with cumulative sampling over energy scales. For each fixed annealing time, the partition function estimation (left) and log relative error (right) are computed from an energy histogram that accumulates samples as the energy scale increases on \texttt{Advantage\_system6.4}. Continuation of Fig.~\ref{figure:Fig4_forward_annealing_cumulative_with_energy_scale}. }
    \label{figure:appendix_forward_anneal_cumulative_6.4_function_of_J}
\end{figure}

\begin{figure}[ht!]
    \centering
    \includegraphics[width=0.49\textwidth]{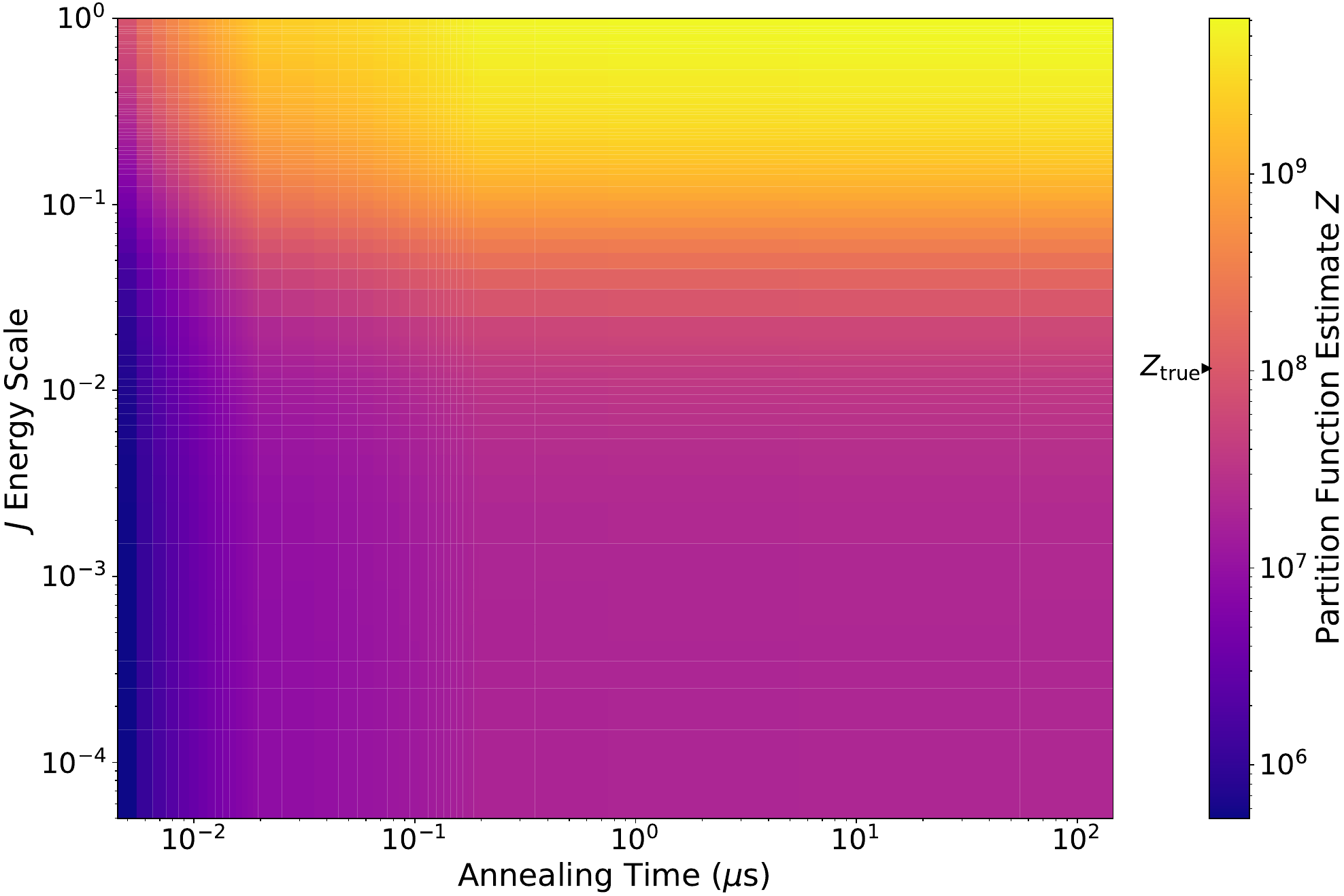}
    \includegraphics[width=0.49\textwidth]{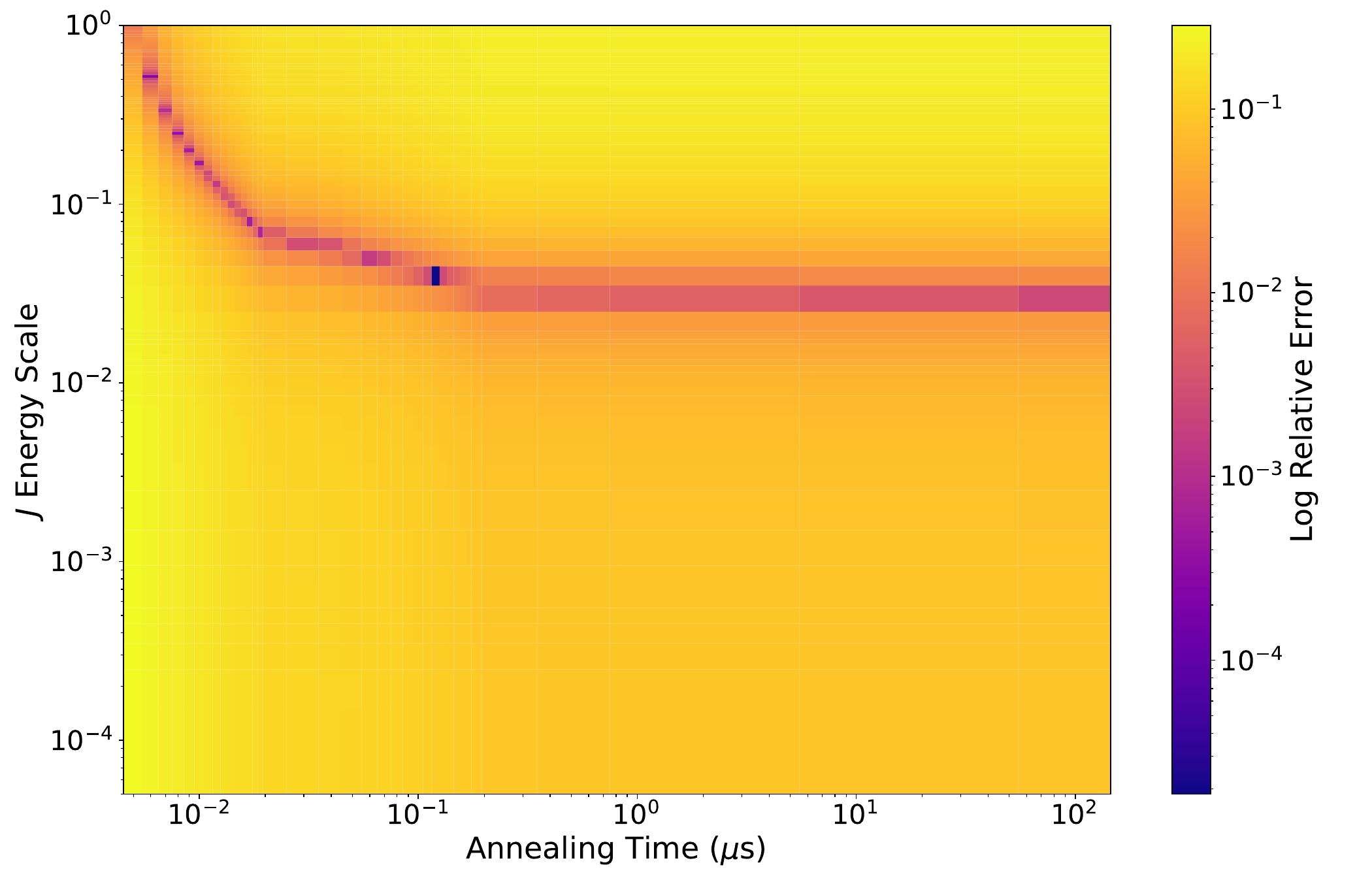}
    \caption{Linear-ramp standard quantum annealing with cumulative sampling over annealing times. For each fixed energy scale, the partition function estimation (left) and log relative error (right) are computed from an energy histogram that accumulates samples as the annealing time increases for \texttt{Advantage\_system6.4}. Continuation of Fig.~\ref{figure:Fig5_forward_annealing_cumulative_with_annealing_time}. }
    \label{figure:appendix_forward_anneal_cumulative_6.4_function_of_annealing_times}
\end{figure}

This section reports additional partition function estimation results, that extend Figures~\ref{figure:Fig4_forward_annealing_cumulative_with_energy_scale} and \ref{figure:Fig5_forward_annealing_cumulative_with_annealing_time} on the \texttt{Advantage\_system6.4} processor, by cumulatively building a distribution at increasingly low temperatures. Fig.~\ref{figure:appendix_forward_anneal_cumulative_6.4_function_of_annealing_times} shows the results when annealing time is varied (from fast to long), and \ref{figure:appendix_forward_anneal_cumulative_6.4_function_of_J} shows the results when $J$ energy scale is varied (from weak to strong coupling).

\begin{figure*}[ht!]
    \centering
    \begin{subfigure}{0.48\linewidth}
        \centering
        \includegraphics[height = 2.5 in, width=\linewidth, keepaspectratio]{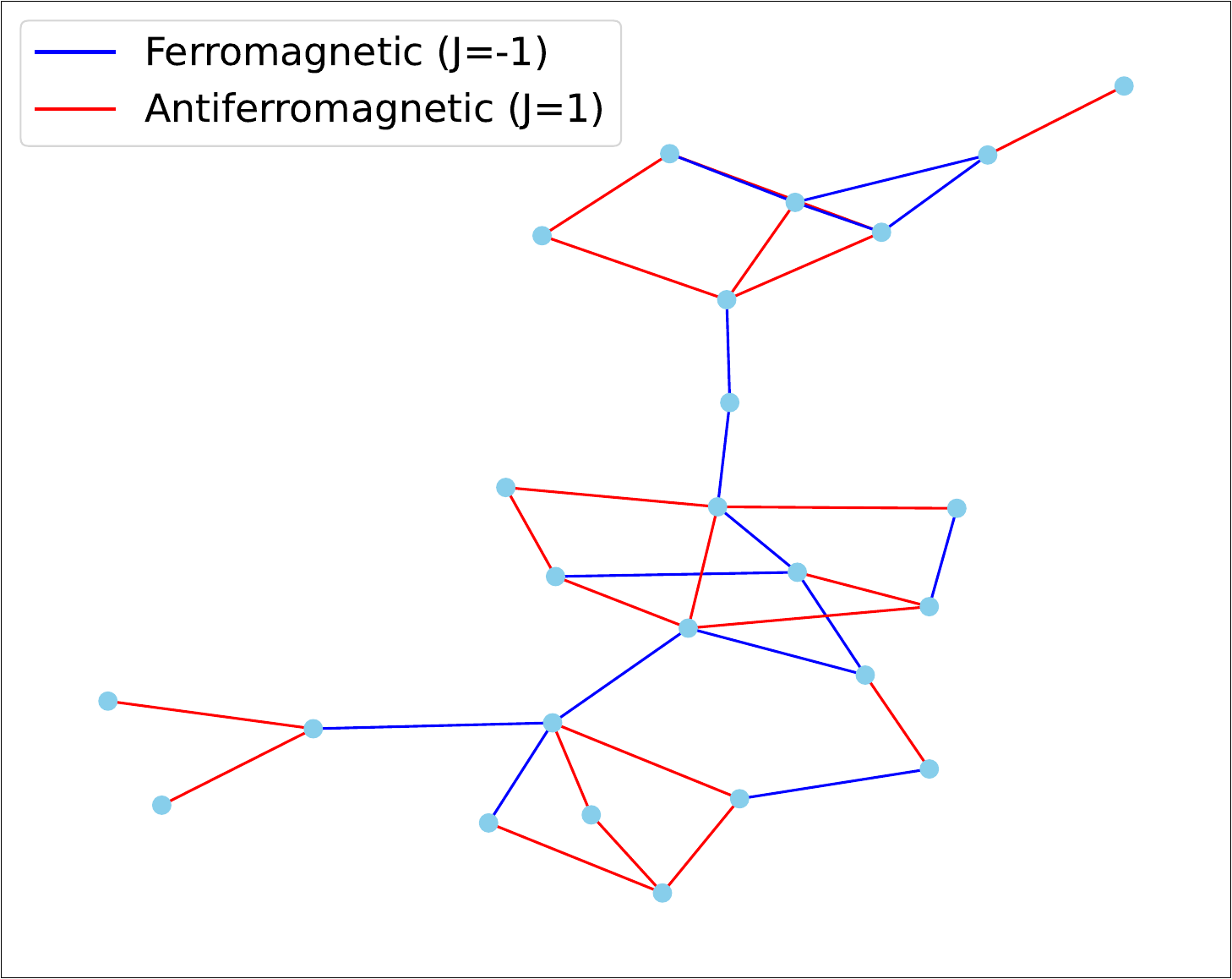}
        \caption{The logical 25-spin Ising problem graph}
    \end{subfigure}\hfill
    \begin{subfigure}{0.48\linewidth}
        \centering
        \includegraphics[height = 2.5 in, width=\linewidth, keepaspectratio]{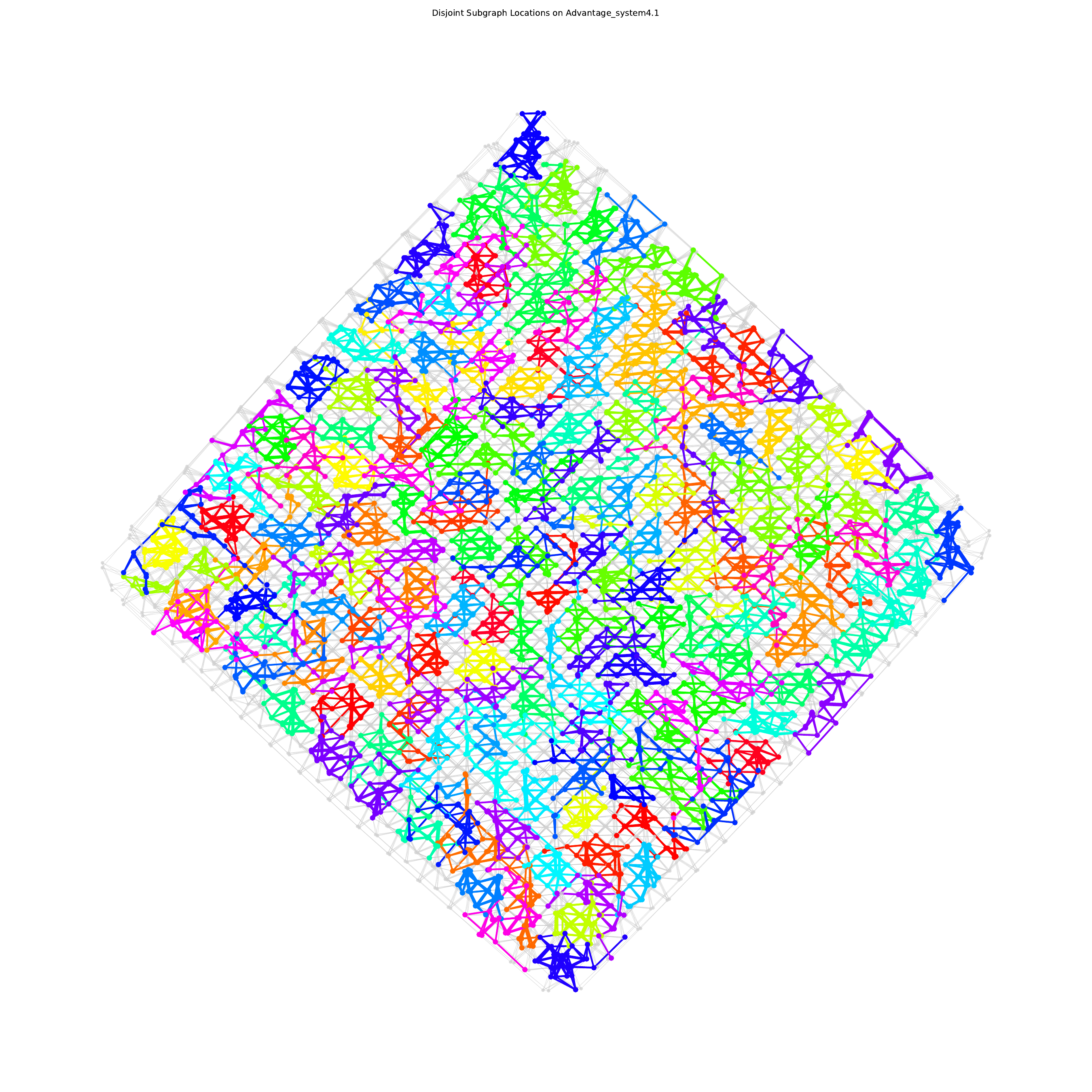}
        \caption{Disjoint embeddings of the problem graph on the D-Wave \texttt{Advantage\_system4.1} QPU}
    \end{subfigure}
    \begin{subfigure}{0.48\linewidth}
        \centering
        \includegraphics[height = 2.5 in, width=\linewidth, keepaspectratio]{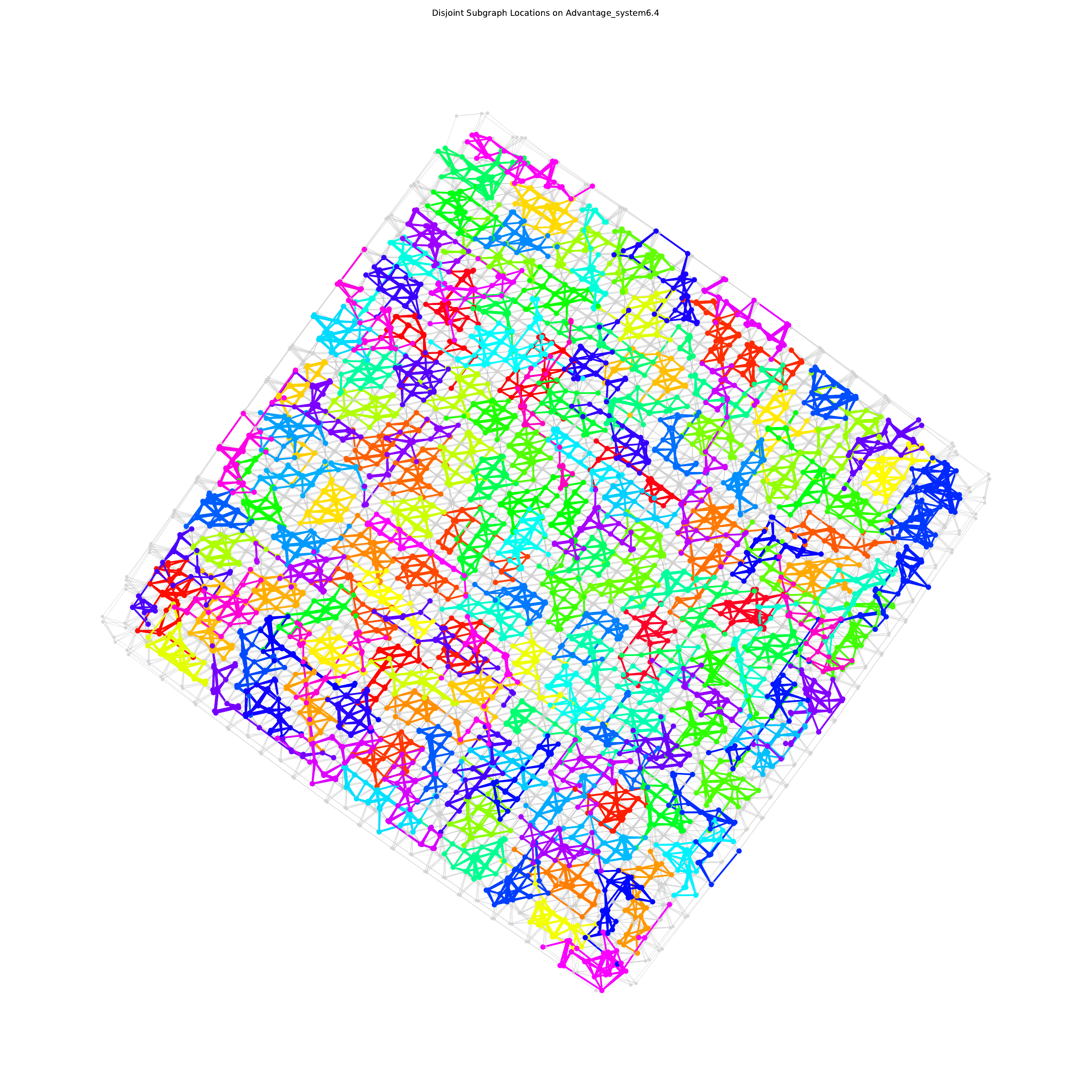}
        \caption{Disjoint embeddings of the problem graph on the D-Wave \texttt{Advantage\_system6.4} QPU}
    \end{subfigure}\hfill
    \begin{subfigure}{0.48\linewidth}
        \centering
        \includegraphics[height = 2.5 in, width=\linewidth, keepaspectratio]{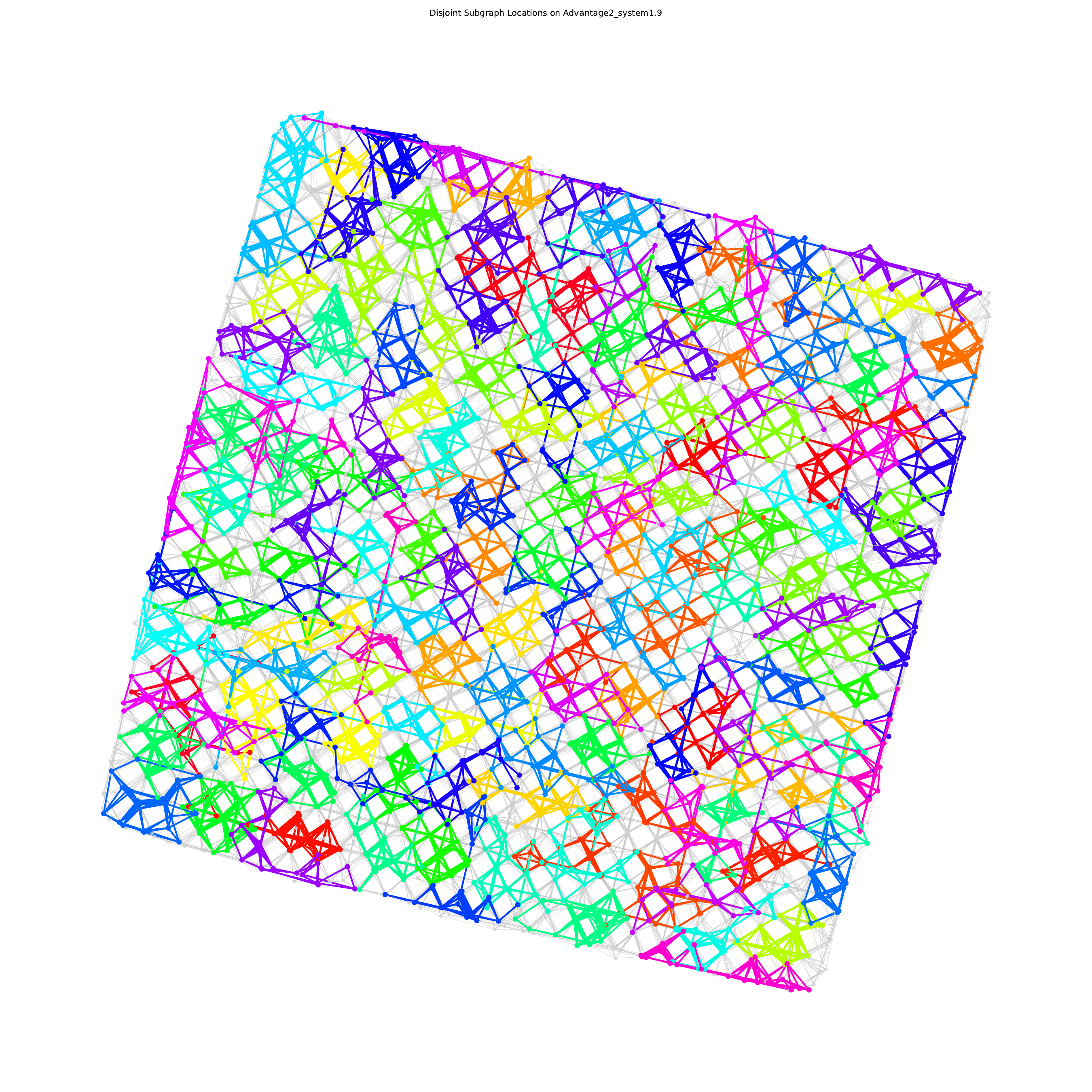}
        \caption{Disjoint embeddings of the problem graph on the D-Wave \texttt{Advantage2\_system1.9} QPU}
    \end{subfigure}
    \caption{Disjoint embeddings of the 25-spin Ising model we sample in this study (a) on three different D-Wave devices. Each colored cluster in (b-d) represents a unique physical embedding on the respective QPU.}
\end{figure*}

%%%%%%%%%%%%%%%%%%%%%%%%%%%%%%%%%%%%%%%%%%%%%%%%%%%%%%%
\section{Parallel Minor Embeddings and Hardware-Defined Ising Model}
\label{appendix:parallel_embeddings_plots}
%%%%%%%%%%%%%%%%%%%%%%%%%%%%%%%%%%%%%%%%%%%%%%%%%%%%%%%

The Ising model we consider in this study is only $25$ spins, but the D-Wave QPUs contain up to many thousands of qubits. Therefore, if we used only a single instance of the $25$ spin model embedded on the hardware, much of the D-Wave hardware would not to be used within each anneal-readout cycle. Therefore, we can make use of disjoint embeddings on the hardware, and execute anneals \emph{in parallel} -- meaning we can obtain an independent sample of the Ising model for each disjoint embedding. Here we show the distribution of disjoint embeddings rendered on the D-Wave hardware graphs.

%%%%%%%%%%%%%%%%%%%%%%%%%%%%%%%%%%%%%%%%%%%%%%%%%%%%%%%
\section{Classical Approximation Monte Carlo Methods Details and Pseudo Code}
\label{appendix:classical_algorithms_and_pseudo_code}
%%%%%%%%%%%%%%%%%%%%%%%%%%%%%%%%%%%%%%%%%%%%%%%%%%%%%%%

Monte Carlo (MC) simulations are widely used for studying statistical mechanics and complex systems. However, obtaining precise thermodynamic quantities and order parameters (e.g., internal energy, specific heat, and magnetization) across a broad range of temperatures typically requires many independent simulation runs. To address this limitation, Wang-Landau algorithm and multiple histogram reweighting method were introduced to predict system behavior at temperatures different from those at which the simulations were performed. In this section, we summarize these two methods, which we employ as standard reference heuristic sampling algorithms for partition function estimation. 

\subsubsection{Wang Landau Algorithm}
\label{subsection:methds_wang_landau}
Proposed by Fugao Wang and David Landau~\cite{wang2001efficient}, the Wang-Landau (WL) algorithm is an adaptive Monte Carlo method to estimate the density of states (DoS) $g(E)$ of a system directly. The partition function can be written in terms of the DoS as follows~\cite{shchur2019properties}:
\begin{equation}
    Z = \sum_Eg(E)e^{-E/k_BT}
\end{equation}
For systems with rugged energy landscapes, accurate DoS estimation is nontrivial. WL addresses this challenge by performing a non-Markovian random walk based on Metropolis-Hastings algorithm \cite{metropolis1953equation}, where the acceptance rule is weighted by the current DoS estimate. A proposal $x \rightarrow x'$ is accepted with probability $\min\{1, g(E(x))/ g(E(x'))\}$, which makes the algorithm non-Markovian. After each visit, the estimate is updated as $g(E) \leftarrow g(E)f$, where $f > 1$ is a modification factor. This update suppresses revisits and drives the energy histogram toward flatness. In practice, the energy range is divided into $N$ bins. For each bin $k$, the algorithm maintains the visit count $H_k$. A histogram is deemed flat when, for a chosen flatness $p \in (0,1)$, the least visited bin has at least a fraction $p$ of the average, i.e. $\min_k H_k \geq p\bar{H}$ with $\bar{H} = N^{-1}\sum_{k=1}^N H_k$. Upon meeting this flatness criterion, the modification factor is reduced, and the histogram is reset. The walk then continues with the updated $f$. Iterating this schedule drives $\ln f$ toward 0 and yields approximately uniform occupancy across energy bins, thereby enabling faster state space traversal than fixed-weight multicanonical sampling \cite{berg1992multicanonical}. The pseudocode of WL is presented in Algorithm \ref{alg:wl_pseudocode}.

\begin{algorithm}[h]
\caption{The Wang-Landau algorithm Pseudocode}
\label{alg:wl_pseudocode}
\begin{algorithmic}[1]
\State Initialize $g(E)\gets 1$ and $H(E)\gets 0$ for all energies $E$ .
\State Choose an initial configuration $x$ and set $E\gets E(x)$.
\State Set modification factor $f \gets f_0>1$ and tolerance $\varepsilon>0$.
\While{$\ln f > \varepsilon$}
    \State Set $H(E)\gets 0$ for all $E$.
    \Repeat
        \State Propose $x'$ and compute $E' \gets E(x')$.
         \State Accept $x'$ with probability
         $
        \alpha(x\to x')=\min\left\{1,\frac{g(E)}{g(E')}\right\}
        $
        \If{accepted}
            \State $x\gets x'$;\;\; $E\gets E'$.
        \EndIf
        \State Update visited energy: $g(E)\gets g(E)\,f$;\;\; $H(E)\gets H(E)+1$.
    \Until{$H$ is flat}
    \State Reduce modification factor $f\gets \sqrt{f}$.
\EndWhile
\end{algorithmic}
\end{algorithm}

\subsubsection{Multiple Histogram Reweighting}
\label{subsection:methods_MCMC_MHR}
Multiple Histogram Reweighting (MHR) \cite{ferrenberg1988new, ferrenberg1989optimized}, also known as the weighted histogram analysis method (WHAM) \cite{kumar1992weighted}, combines samples collected at several inverse temperatures $\{\beta_i\}_{i=1}^R$ to estimate thermodynamic quantities (specifically the partition function in this work) at a target $\beta$. The sampling points $\{\beta_i\}$ are chosen so that the energy distributions of neighboring inverse temperatures overlap, which is essential for stable reweighting. For a system with DoS $g(E)$, the energy histogram at inverse temperature $\beta$ is:
\begin{equation}
    h(E,\beta) = \frac{1}{Z_\beta} g(E)e^{-\beta E}, 
    \quad
    Z(\beta) = \sum_E g(E)e^{-\beta E}.
    \label{eqn:partition_function_def}
\end{equation}
For each simulation $i$ at $\beta_i$, we record a sequence of $N_i$ statistically independent energy measurements $\{E_i^{(k)}\}_{k=1}^{N_i}$ and compute the corresponding count histogram $H_i(E)$, defined so that $H_i(E)$ is the number of samples with energy $E$ and $\sum_E H_i(E) = N_i$. To merge data from different $\beta_i$, MHR introduces the dimensionless free energies:
\begin{equation}
    f_i = -\ln({Z_{\beta_i}}),
\end{equation}
so that $Z(\beta_i)^{-1} = e^{f_i}$. These free energies serve as normalization constants for combining data collected at different $\beta_i$ in the MHR estimator. Given a set of free energies $\{f_i\}$, the MHR estimator for $g(E)$ is \cite{pelissetto_multihistogram}:
\begin{equation}
    g(E) = \frac{\sum_{i=1}^R H_i(E)}{\sum_{j=1}^R N_j e^{-\beta_j E} Z_j^{-1}} = \frac{\sum_{i=1}^R H_i(E)}{\sum_{j=1}^R N_j e^{-\beta_j E + f_j}}
\end{equation}
This expression effectively weights each dataset according to how well it samples energy $E$. The unknown free energies $\{f_i\}$ are calculated by enforcing the defining identity for each sampled inverse temperature:
\begin{equation}
    e^{-f_i} = Z(\beta_i) = \sum_E g(E) e^{-\beta_i E}, \quad \text{with } i = 1,\cdots, R.
    \label{eqn:free_energy}
\end{equation}
which can be solved using iterative methods (e.g., fixed-point iteration, Newton-Raphson). Finally, MHR can be written without explicit histograms by replacing sums over energies weighted by $H_i(E)$ with sums over the recorded energies $E_i^{(k)}$. The MHR estimator of the partition function at a target $\beta$ is:
\begin{equation}
    Z(\beta) = e^{-f(\beta)} = \sum_{i=1}^{R}\sum_{k=1}^{N_i} \frac{e^{-\beta E_i^{(k)}}}{\sum_{j=1}^{R} N_j\, e^{-\beta_j E_i^{(k)} + f_j}}.
\end{equation}
In our implementation, we utilize \textit{dlmontepython} \cite{underwood2021dlmontepython} for the iterative estimations of the MHR free energies, which are then used with energy samples generated by Metropolis MCMC to obtain an estimation of partition function at the target temperature $T=4$. 

\begin{figure*}[ht!]
    \centering
    \includegraphics[width=0.49\linewidth]{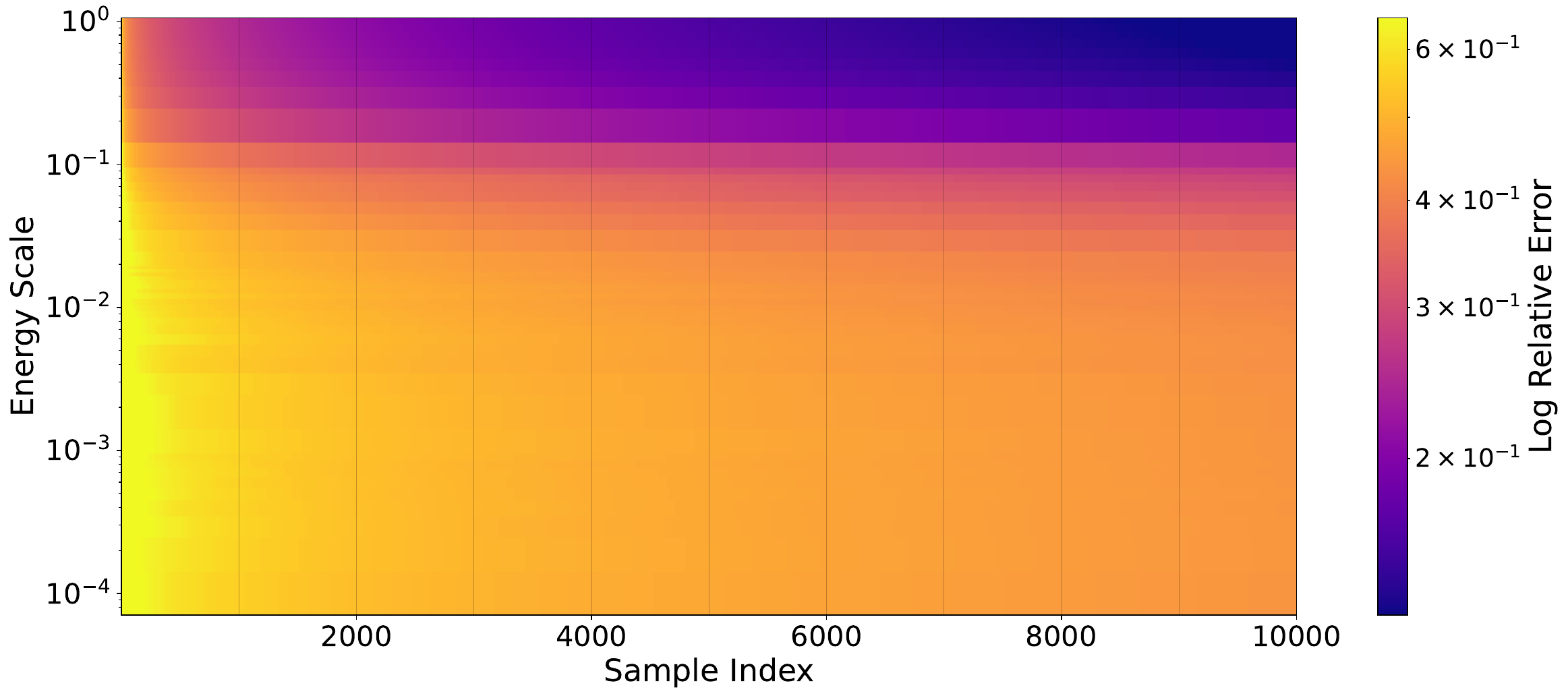}
    \includegraphics[width=0.49\linewidth]{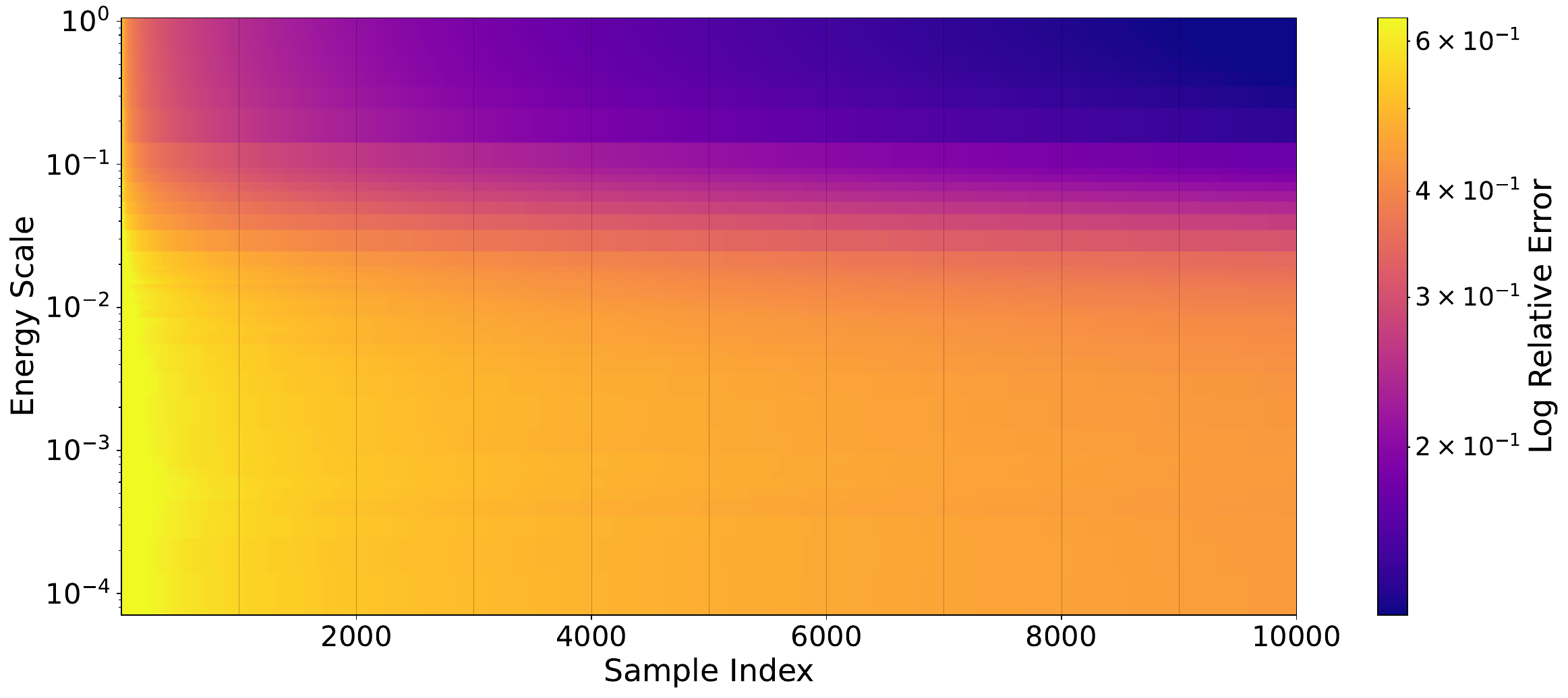}
    \includegraphics[width=0.49\linewidth]{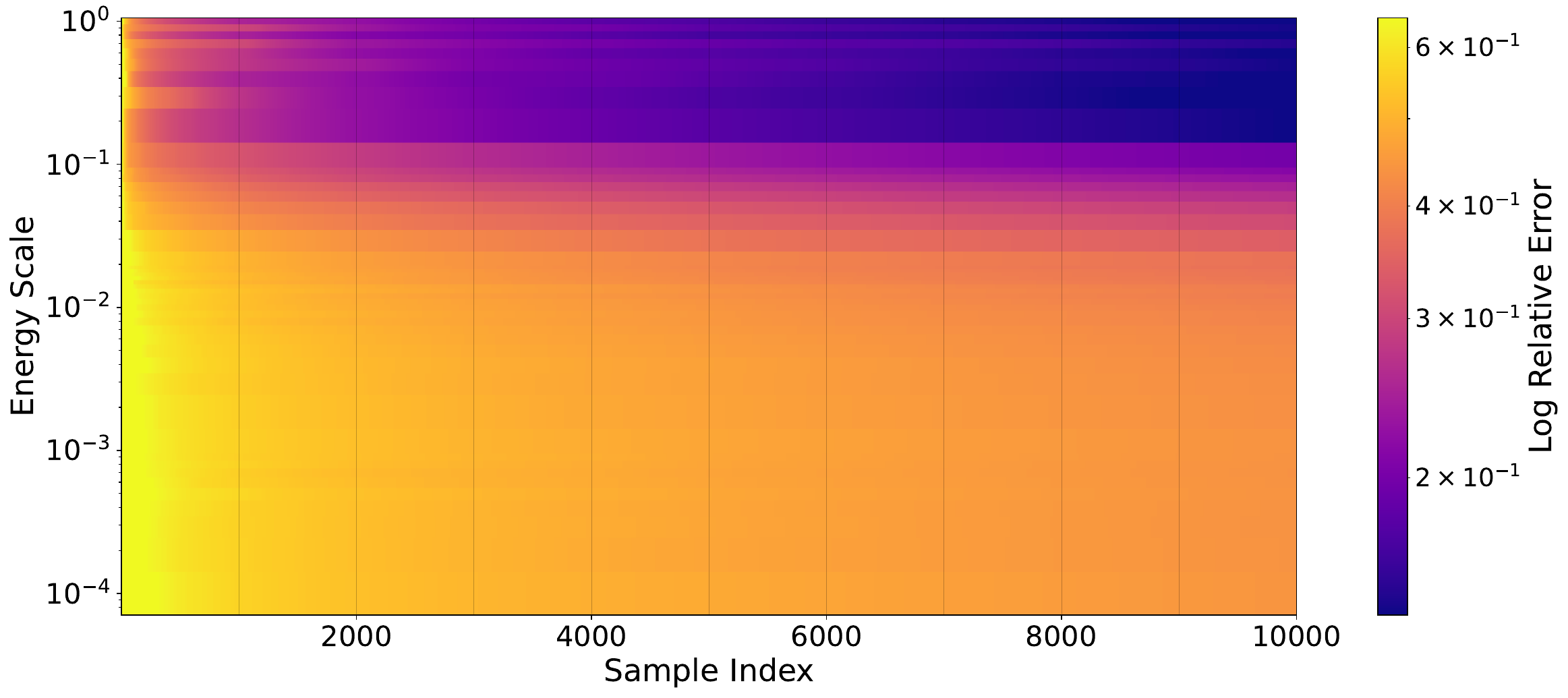}
    \includegraphics[width=0.49\linewidth]{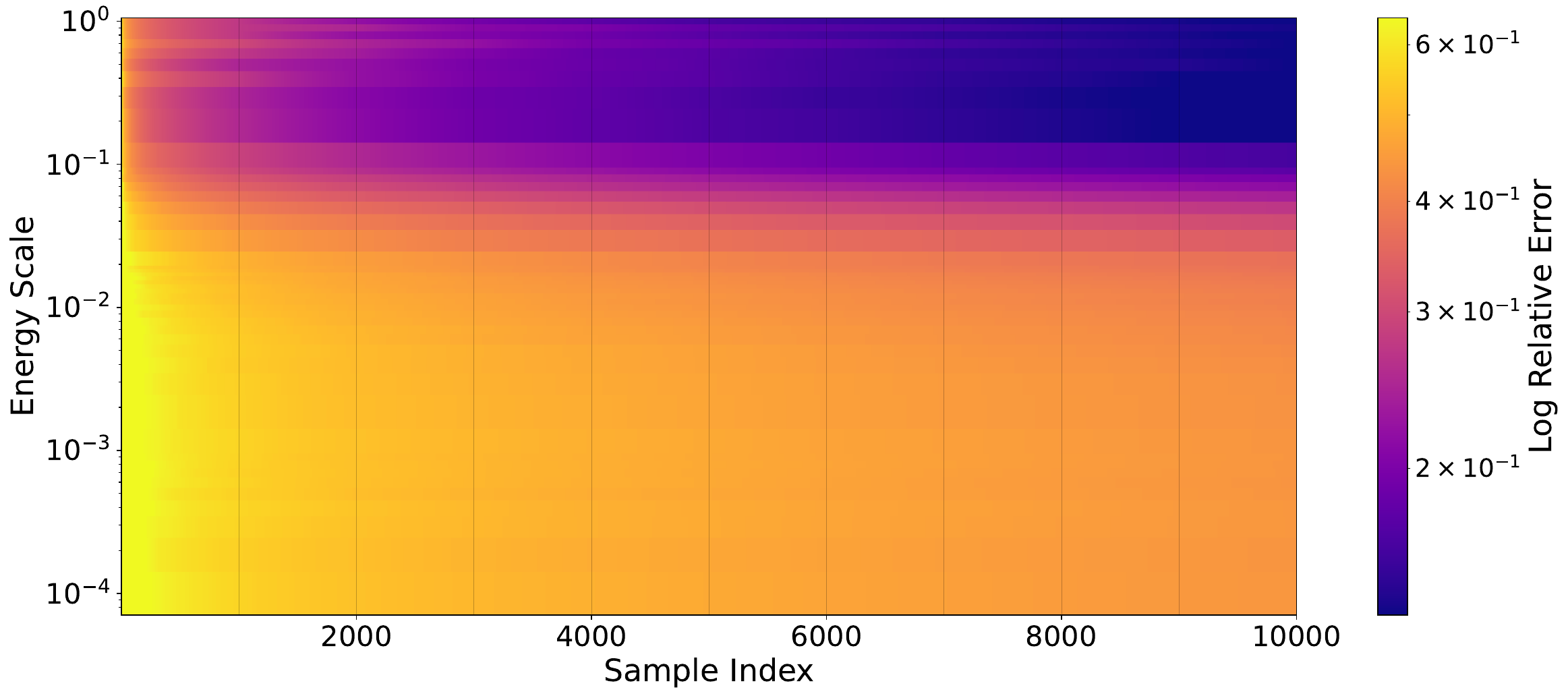}
    \includegraphics[width=0.49\linewidth]{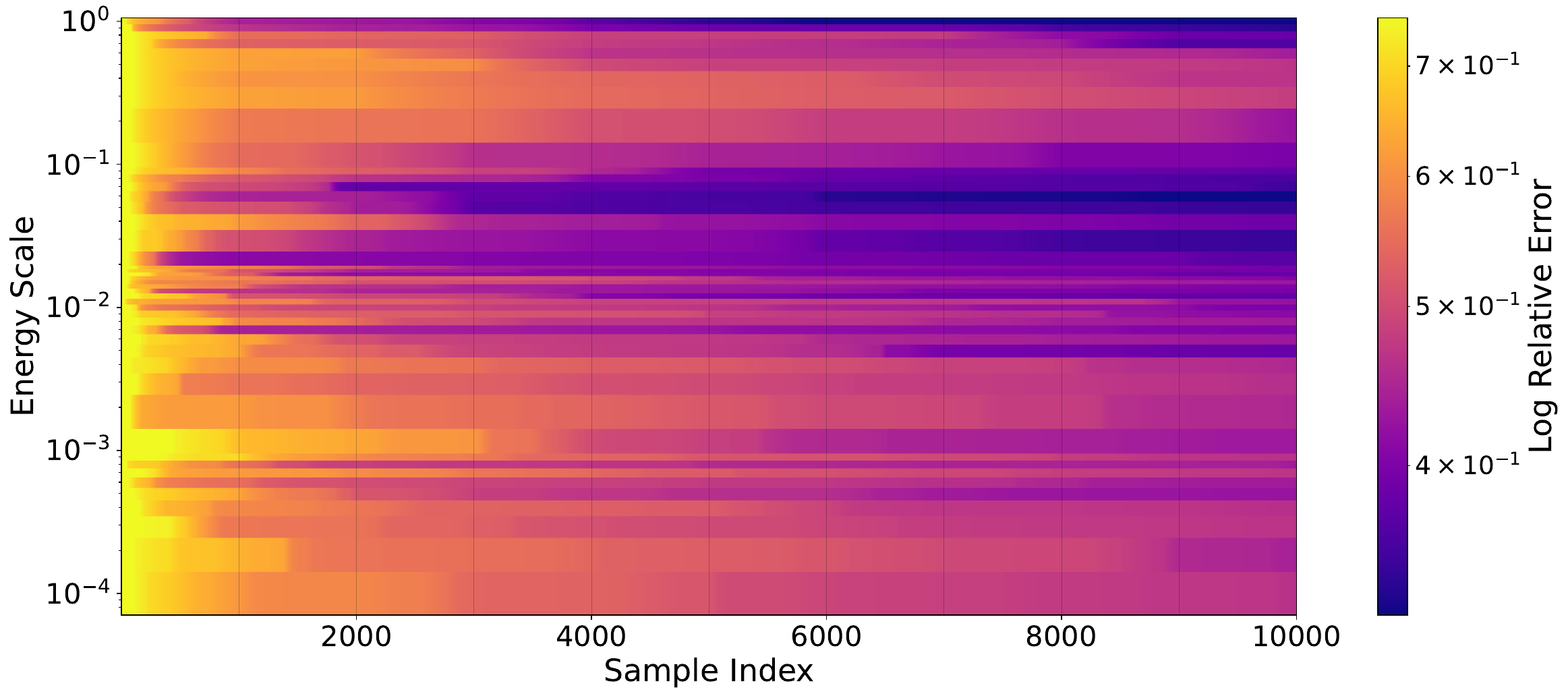}
    \includegraphics[width=0.49\linewidth]{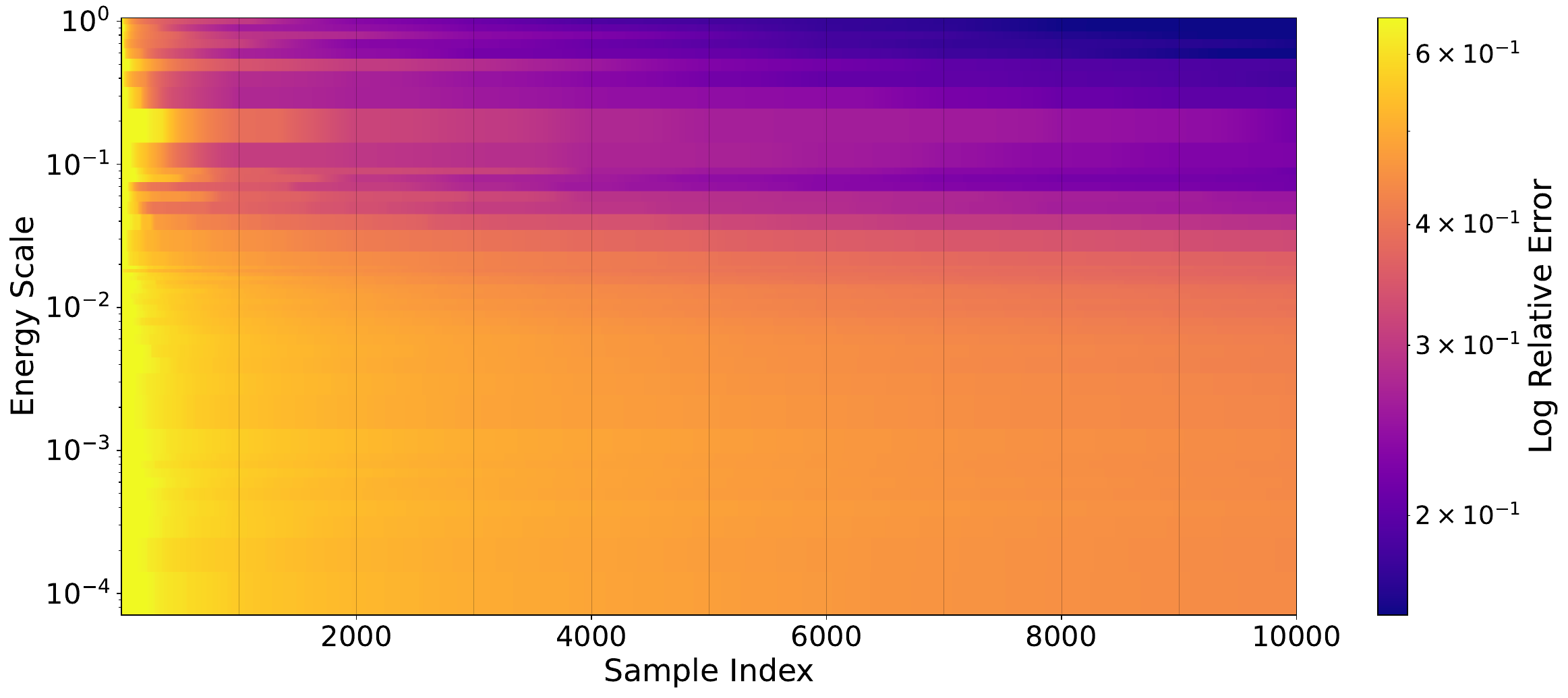}
    \caption{QEMC, or iterated reverse quantum annealing, analysis when using a reverse anneal schedules pause at $s=0.1$ (top), $s=0.7$ (middle), and $s=0.8$ (bottom), where larger $s$ values correspond to weaker transverse fields. The color-scales denote the logarithmic relative error. For each $J$ coupling energy scale (y-axis), as we move right on the x-axis a distribution of samples is accumulated as we progress further along a single QEMC chain -- each x-axis point along each horizontal stripe of data corresponds to a single, additional, measured anneal from one of the parallel-embedded disjoint Ising model instances on the QPU. In all cases, the sample count is not sufficient to reach the true partition function which is why the error rates are fairly high. This shows that at the lowest error rate results come from strong $J$ coupling, along with properly tuned $s$ pauses. The error rates converge in all cases after several thousand iterations. Left column reports data from \texttt{Advantage\_system4.1} and the right column reports data from the \texttt{Advantage2\_system1.9} processor, and all sub-plots used a total simulation time of $100 ~\mu\mathrm{s}$.   }
    \label{figure:QEMC_convergence_analysis}
\end{figure*}

\begin{figure*}[ht!]
    \centering
    \begin{subfigure}[t]{0.5\textwidth}
        \centering
        \includegraphics[height = 2.3 in, width=\textwidth, keepaspectratio]{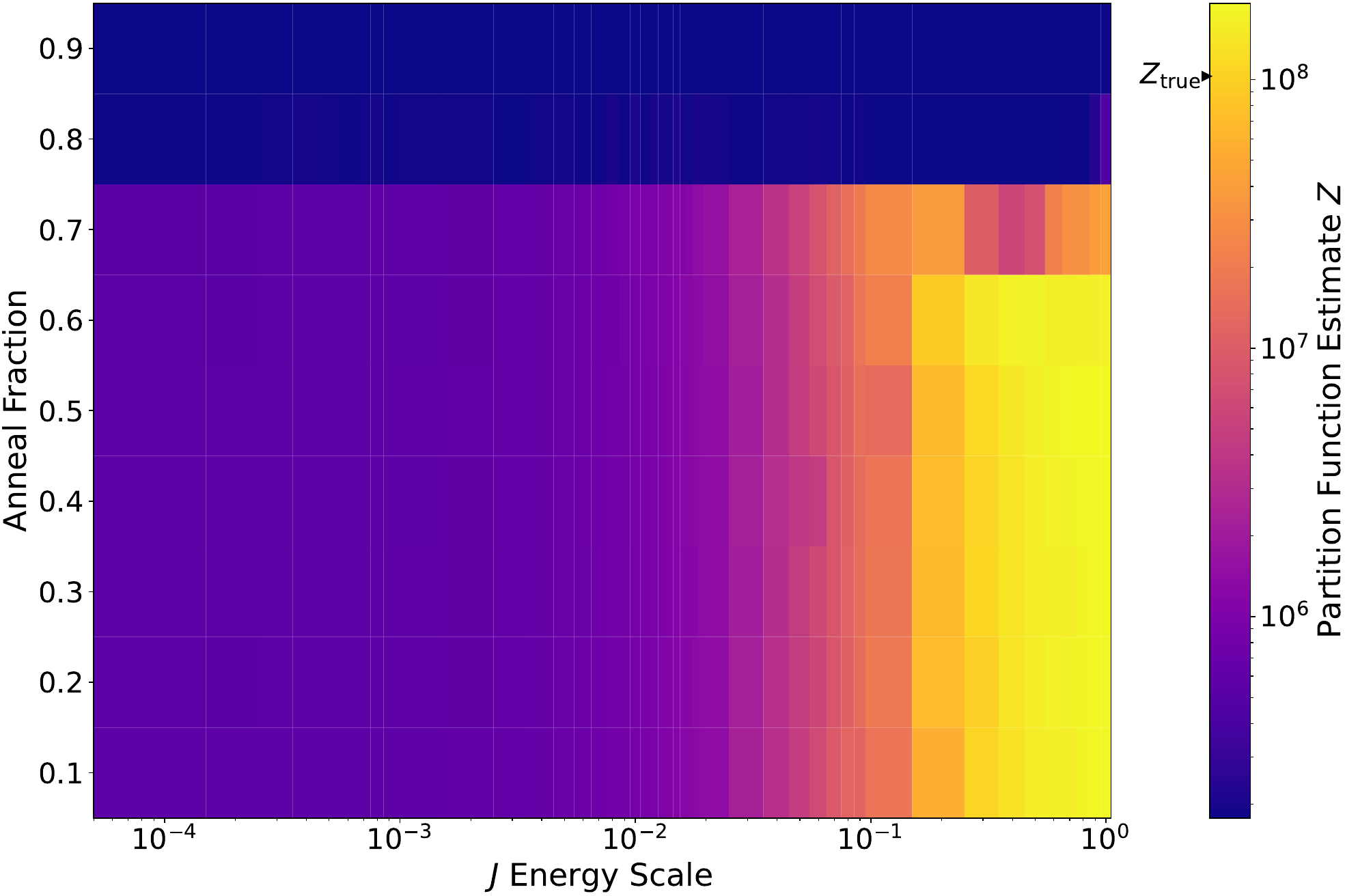}
        \caption{}
    \end{subfigure}%
    ~ 
    \begin{subfigure}[t]{0.5\textwidth}
        \centering
        \includegraphics[height = 2.3 in, width=\textwidth, keepaspectratio]{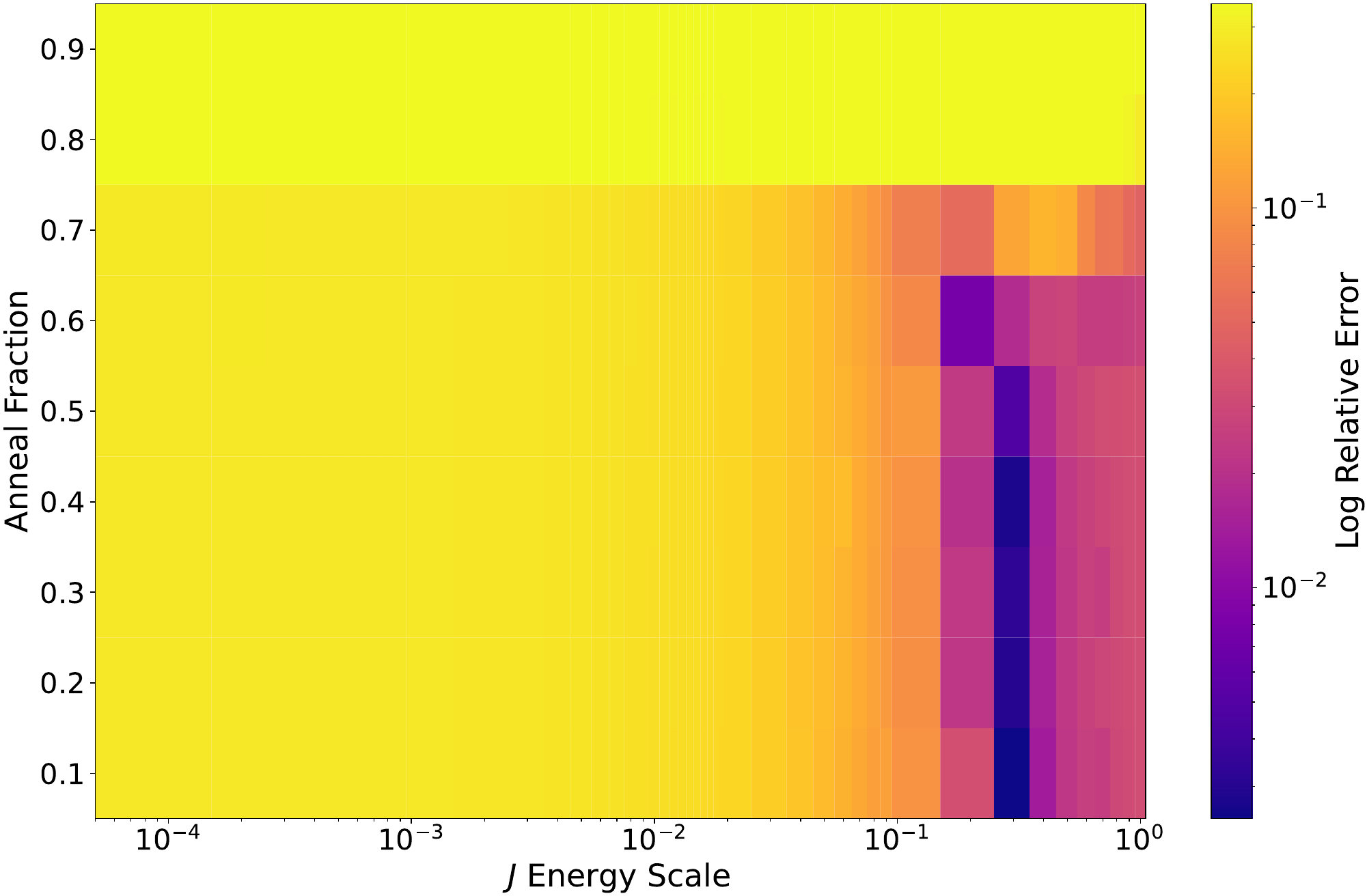}
        \caption{}
    \end{subfigure}
    ~
    \begin{subfigure}[t]{0.5\textwidth}
        \centering
        \includegraphics[height = 2.3 in, width=\textwidth, keepaspectratio]{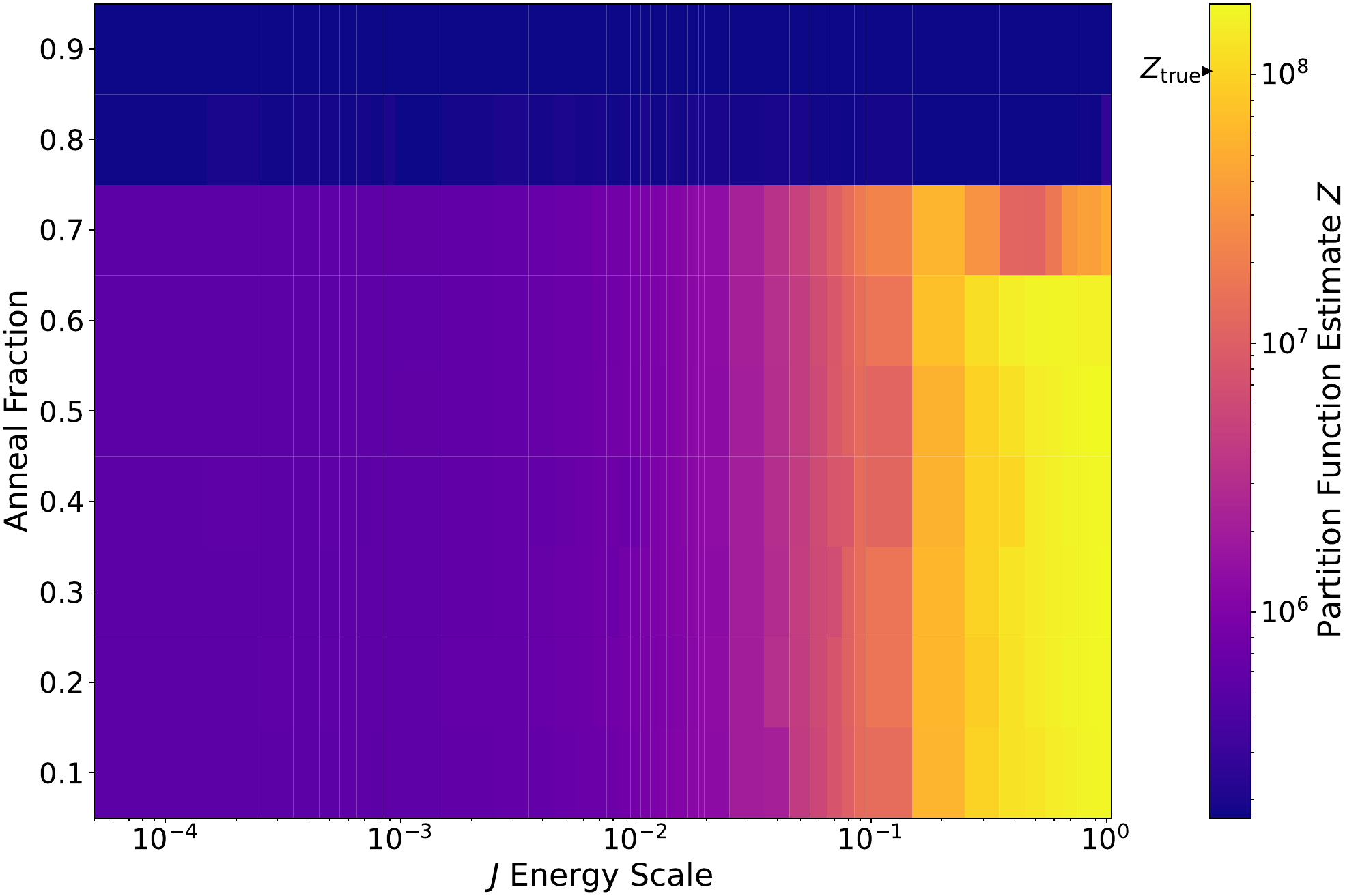}
        \caption{}
    \end{subfigure}%
    ~ 
    \begin{subfigure}[t]{0.5\textwidth}
        \centering
        \includegraphics[height = 2.3 in, width=\textwidth, keepaspectratio]{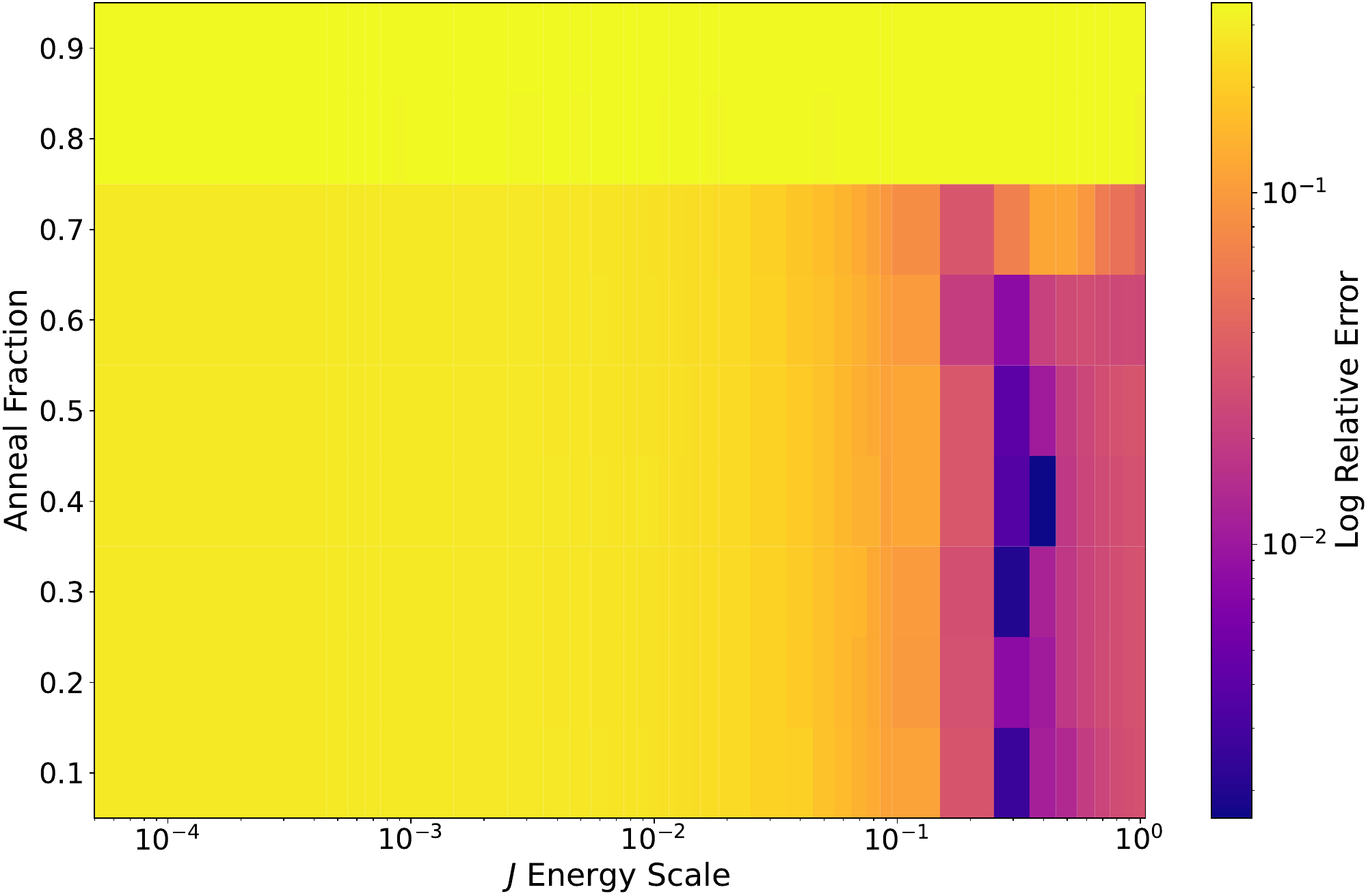}
        \caption{}
    \end{subfigure}
    ~
    \begin{subfigure}[t]{0.5\textwidth}
        \centering
        \includegraphics[height = 2.3 in, width=\textwidth, keepaspectratio]{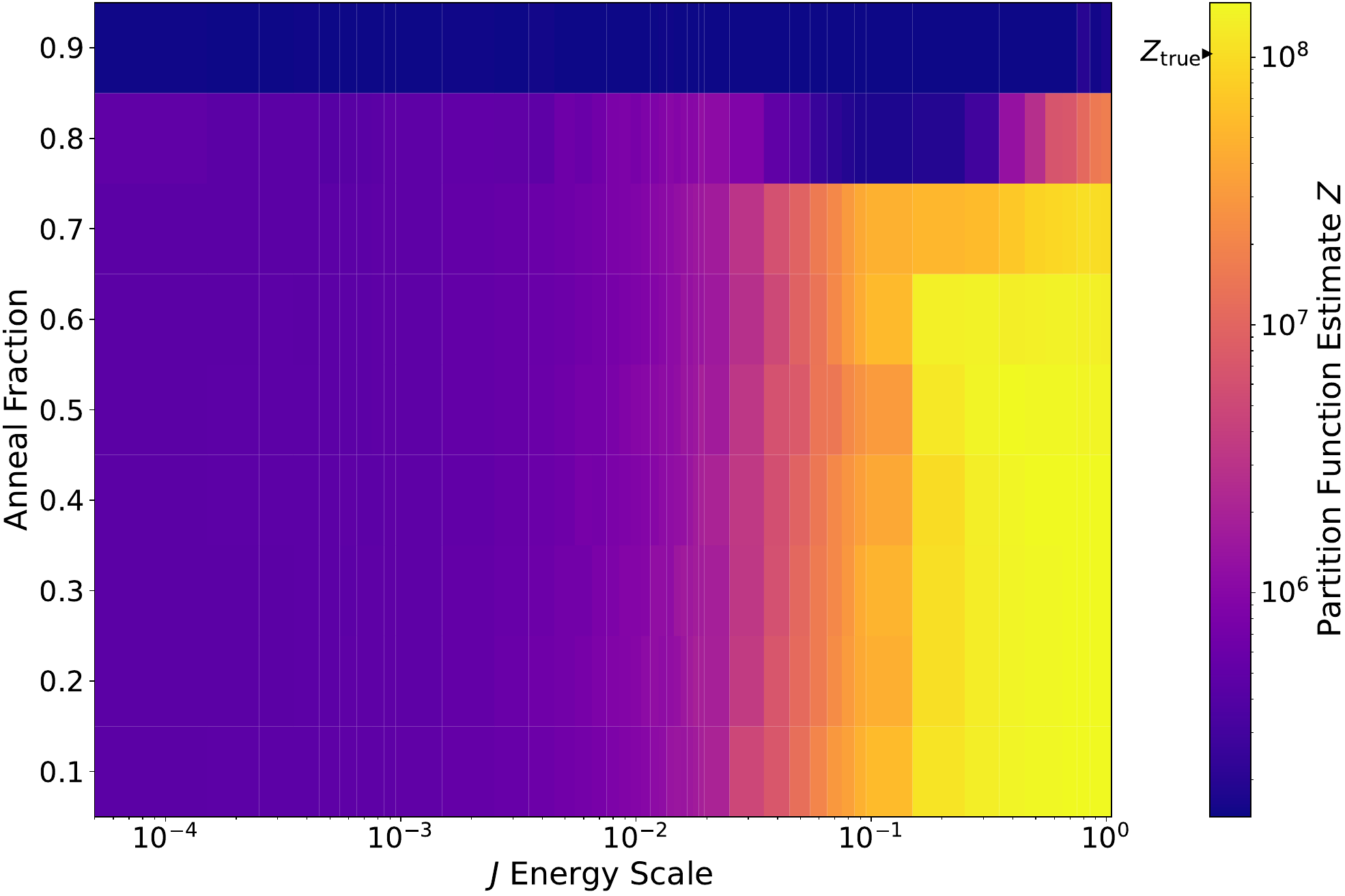}
        \caption{}
    \end{subfigure}%
    ~ 
    \begin{subfigure}[t]{0.5\textwidth}
        \centering
        \includegraphics[height = 2.3 in, width=\textwidth, keepaspectratio]{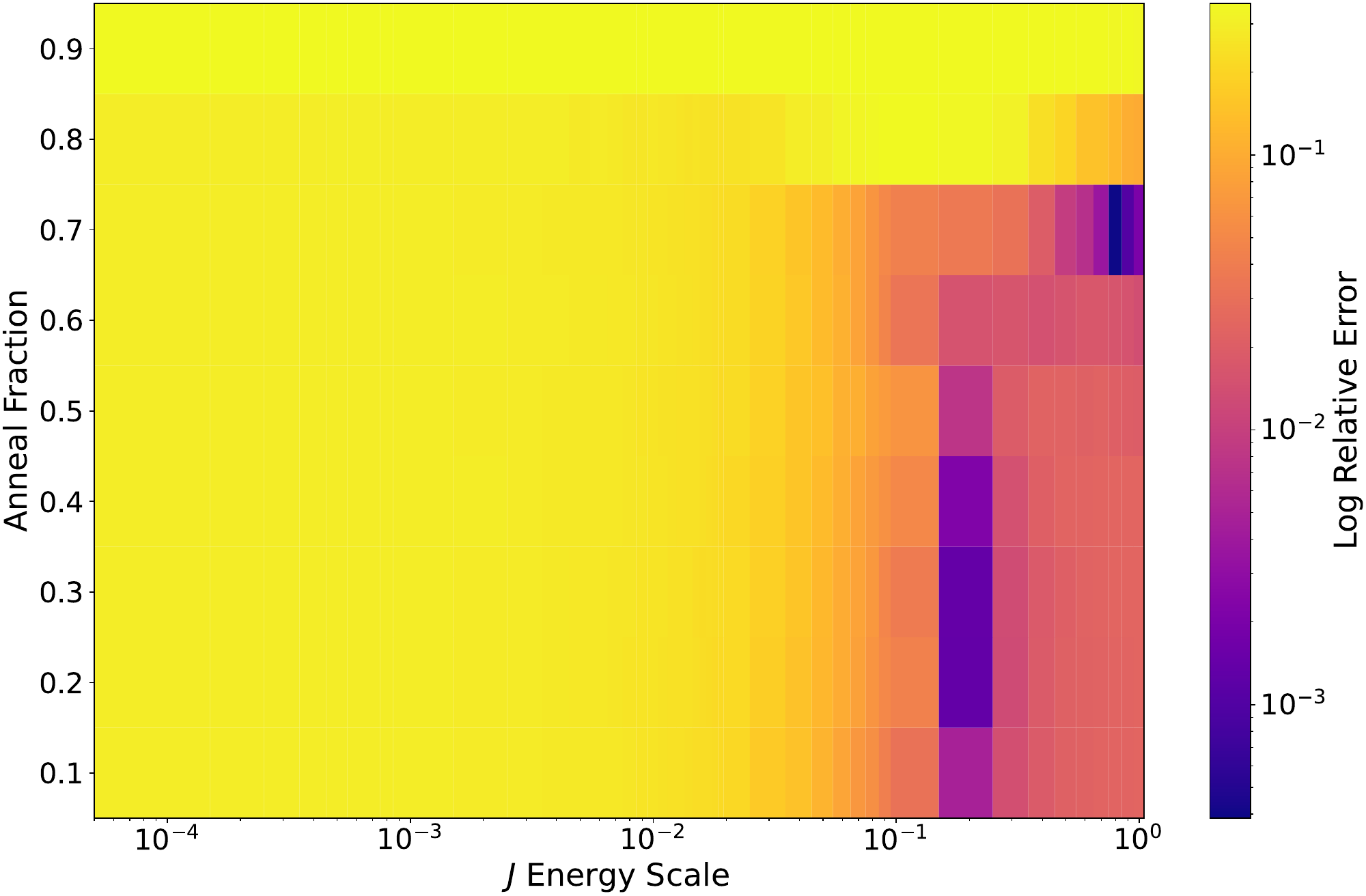}
        \caption{}
    \end{subfigure}
    \caption{Iterated reverse annealing analysis with annealing time of $2 ~\mu\mathrm{s}$ as a function of $J$ energy scale (log x-axis) and reverse-anneal pause (y-axis). Data from the  \texttt{Advantage\_system4.1} processor (a,b), \texttt{Advantage\_system6.4} (c,d),  \texttt{Advantage2\_system1.9} (e,f). The data in these plots are not cumulative on any axis, and the plot style is the same as Fig.~\ref{figure:reverse_annealing_non_cumulative_error_rates_100ms}.  }
    \label{fig:reverse_annealing_non_cumulative_t2_appendix}
\end{figure*}

%%%%%%%%%%%%%%%%%%%%%%%%%%%%%%%%%%%%%%%%%%%%%%%%%%%%%%%
\section{Reverse Quantum Annealing Monte Carlo Chain Sampling Analysis}
\label{appendix:QEMC_convergence}
%%%%%%%%%%%%%%%%%%%%%%%%%%%%%%%%%%%%%%%%%%%%%%%%%%%%%%%

One important parameter in the QEMC simulations is the total length of the Monte Carlo chain -- in other words, how many iterations have been performed (having initialized the chain with a completely random configuration). Fig.~\ref{figure:QEMC_convergence_analysis} analyzes the convergence of QEMC, which demonstrates that the iterated reverse annealing typically converges to a fixed error value, if the pause provides sufficient quantum fluctuations for the state of the system to evolve. Fig.~\ref{figure:QEMC_convergence_analysis} reports sampling produced from only a single one of the disjoint parallel embeddings, for a single full contiguous chain of reverse anneals (for each parameter).

%%%%%%%%%%%%%%%%%%%%%%%%%%%%%%%%%%%%%%%%%%%%%%%%%%%%%%%
\section{2 Microsecond Reverse Quantum Annealing Partition Function Estimation}
\label{appendix:RA_t2}
%%%%%%%%%%%%%%%%%%%%%%%%%%%%%%%%%%%%%%%%%%%%%%%%%%%%%%%

Fig.~\ref{fig:reverse_annealing_non_cumulative_t2_appendix} reports the partition function estimation from using reverse quantum annealing as a function of both the anneal fraction and the $J$ energy scale, with a total simulation time of $2 ~\mu\mathrm{s}$, as compared to Fig.~\ref{figure:reverse_annealing_non_cumulative_error_rates_100ms} where $100 ~\mu\mathrm{s}$ was used. This demonstrates that there is relatively little difference between $100 ~\mu\mathrm{s}$ and $2 ~\mu\mathrm{s}$ simulations.

\begin{figure*}[ht!]
    \centering
    \includegraphics[width=0.48\textwidth]{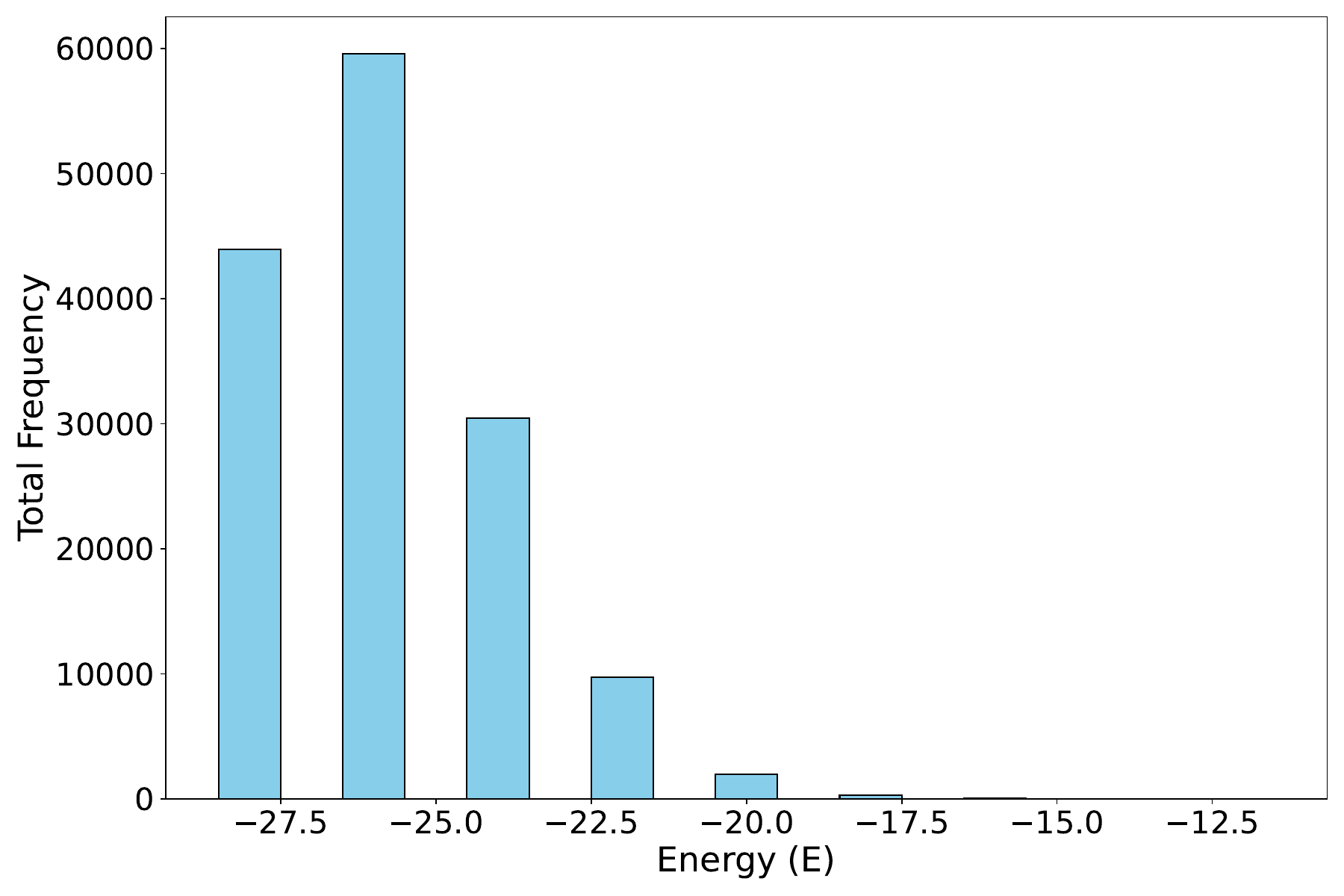}
    \includegraphics[width=0.48\textwidth]{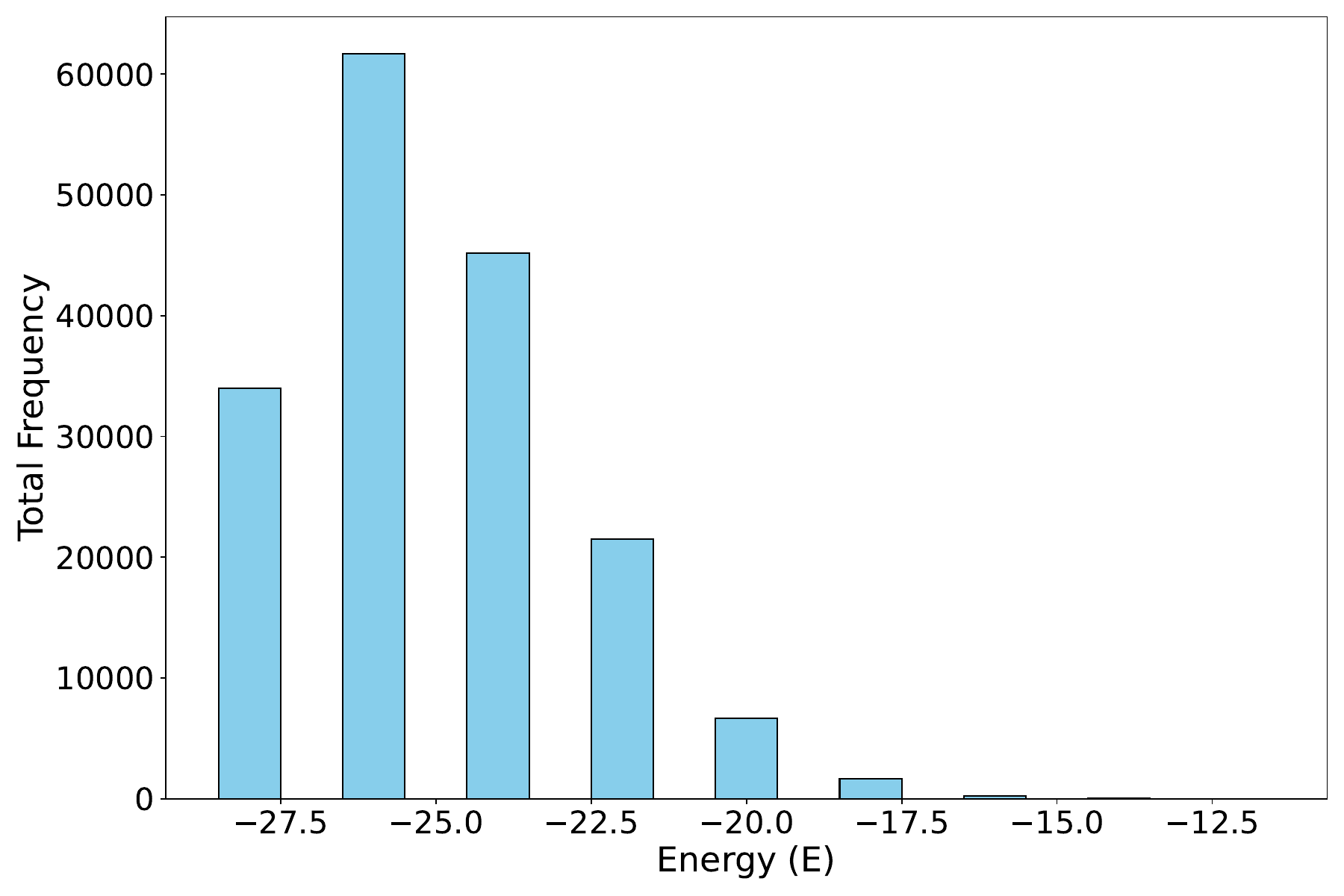}
    \includegraphics[width=0.48\textwidth]{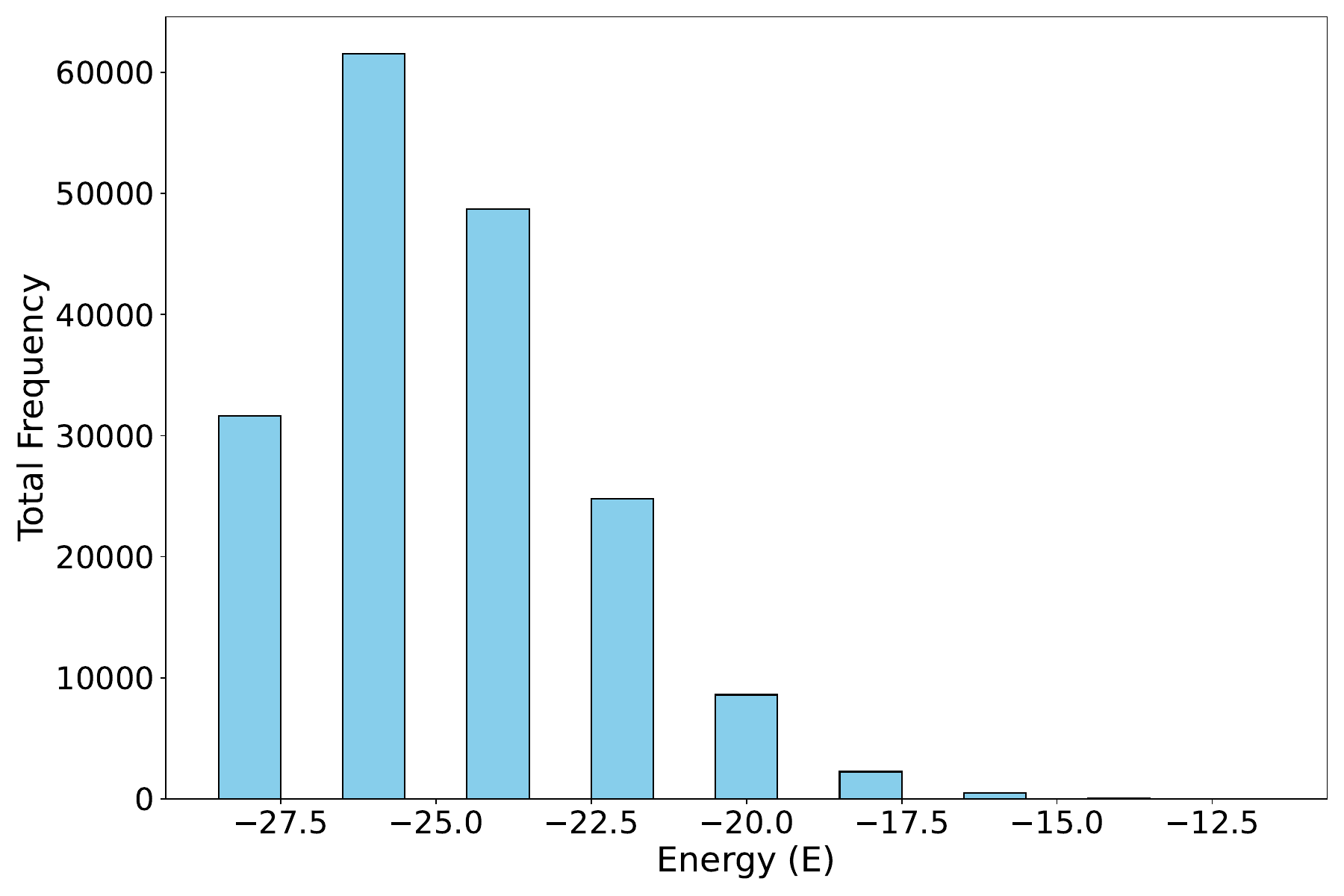}
    \includegraphics[width=0.48\textwidth]{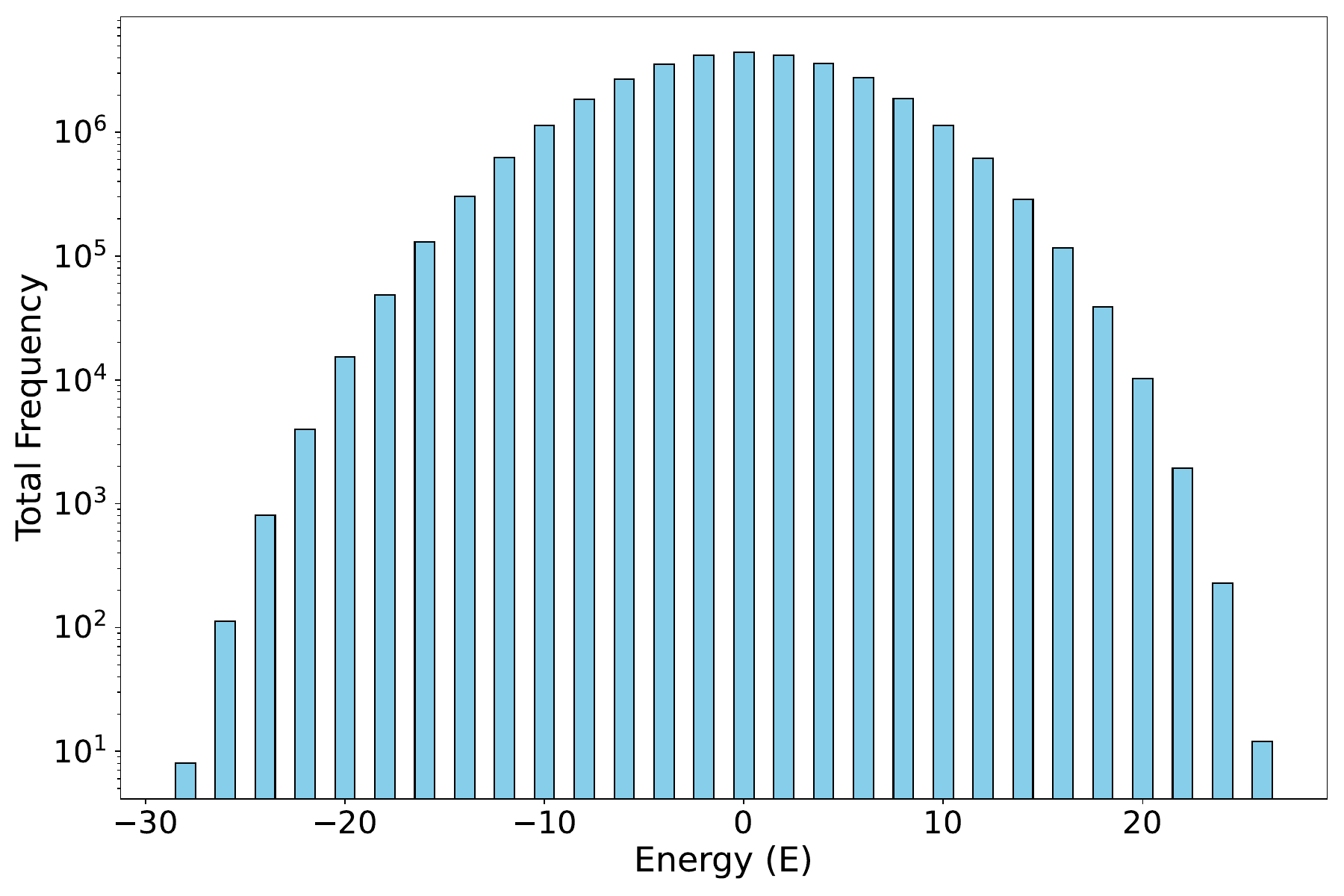}
    \caption{Histograms of the density of states estimates for the lowest error-rate parameters shown in Fig.~\ref{figure:fast_annealing_non_cumulative_sampling} (\texttt{Advantage2\_system1.9} upper-left, \texttt{Advantage\_system6.4} upper-right, and \texttt{Advantage\_system4.1} lower-left) and full $2^N$ configuration energy histogram with logarithmic scale (lower-right), spanning energies from -28 to 26. These experimental energy distributions (all but lower-left plot) come from relatively fast anneal-quenches generated by each QPU, which is why the energy distributions are not primarily ground-state energies, but instead also have sampled many higher-energy configurations.   }
    \label{figure:density_of_states_histograms}
\end{figure*}

%%%%%%%%%%%%%%%%%%%%%%%%%%%%%%%%%%%%%%%%%%%%%%%%%%%%%%%
\section{Tuned Density of States Estimates from the Analog Hardware Sampling}
\label{appendix:density_of_states_histograms}
%%%%%%%%%%%%%%%%%%%%%%%%%%%%%%%%%%%%%%%%%%%%%%%%%%%%%%%

Fig.~\ref{figure:density_of_states_histograms} reports the energy histogram DoS estimates for the  lowest-error rates from the fast-anneal-quenches shown in Fig.~\ref{figure:fast_annealing_non_cumulative_sampling}, along with the full distribution of all $2^N$ energies.

%%%%%%%%%%%%%%%%%%%%%%%%%%%%%%%%%%%%%%%%%%%%%%%%%%%%%%%
\section{Genetic Algorithm Optimized Parameters}
\label{appendix:GA_algo_parameters}
%%%%%%%%%%%%%%%%%%%%%%%%%%%%%%%%%%%%%%%%%%%%%%%%%%%%%%%

In this section, the complete set of genetic-algorithm optimized analog hardware parameters that are used in Fig.~\ref{figure:Fig3_classical_estimation_algorithms_and_best_DWave_comparison} are given. The complete parameter set that was optimized over is a full range of $J$ energy scales and annealing times, spanning the feasible range that these parameters can be programmed on the hardware. The genetic algorithm optimization treated each individual parameter as a ``trait'', and then mutation rate, population size, and crossover proportion were then tuned to give reasonably good convergence, resulting in a black-box generation of a heuristic parameter combination that generates a good DoS estimate. The cost function used was to minimize log relative error, and to penalize high sample count. For optimized parameter set forward anneal sampling on \texttt{Advantage2\_system1.9}, the $J$ energy scales were \emph{0.0001, 0.0002, 0.0003, 0.0009, 0.002, 0.01, 0.018, 0.019, 0.02, 0.07, 0.1, 0.11, 0.13, 0.16, 0.17, 0.23, 0.25, 0.26, 0.29, 0.32, 0.33, 0.35, 0.38, 0.39, 0.52, 0.7, 0.8, 0.9, 0.92}. The annealing time selected was only $5$ nanoseconds. A total of $317550$ samples were used, with each parameter combination contributing $75$ samples. For \texttt{Advantage\_system4.1}, the selected energy scales are \emph{0.017, 0.09, 0.1, 0.9}, and the selected annealing times are \texttt{0.05, 0.17, 1}, with $143$ samples per parameter combination and a total of $305448$ samples. On \texttt{Advantage\_system6.4}, energy scales of \emph{0.0001, 0.0005, 0.0006, 0.0009, 0.001, 0.003, 0.008, 0.009, 0.019, 0.03, 0.04, 0.05, 0.06, 0.1, 0.12, 0.13, 0.15, 0.21, 0.22, 0.26, 0.48, 0.8, 0.84, 0.9} were selected and an annealing time of $0.11 \mu s$. 

For the QEMC simulations on \texttt{Advantage2\_system1.9}, a single pause $s$ of $0.5$ was selected, an annealing time of $2 \mu s$, and J energy scales of \emph{0.0003, 0.0007, 0.002, 0.003, 0.007, 0.013, 0.014, 0.017, 0.05, 0.08, 0.7}, with a QEMC chain length of $541$ (per parameter) and a total sample count of $868846$. For \texttt{Advantage\_system6.4} QEMC simulations, J energy scales of \emph{0.002, 0.003, 0.006, 0.01, 0.015, 0.018, 0.04, 0.06, 0.8} and a pause of $s=0.5$ and a total simulation time of $100 \mu s$ were selected, with $823365$ total samples (QEMC chain length of $535$ per parameter). \texttt{Advantage\_system6.4} QEMC parameters are \emph{0.0002, 0.002, 0.007, 0.014, 0.03, 0.09, 0.1, 0.8} J energy scales, $s=0.2$, a simulation time of $2 \mu s$, and a QEMC chain length of $470$ per parameter, and a total sample count of $669280$.

\clearpage

\twocolumngrid
\bibliographystyle{apsrev4-2-titles}
\bibliography{references}

\end{document}